\documentclass[aps,12pt,showpacs,nofootinbib,amsmath,eqsecnum, floatfix]{revtex4}
\usepackage{graphicx}
\usepackage{amsmath}
\usepackage{amsfonts}
\usepackage{amssymb}
\usepackage{color}
\usepackage{booktabs}

\usepackage{ifpdf}

\ifpdf   
  \RequirePackage[pdftex]{hyperref}
\else    
    \RequirePackage[dvips]{hyperref}
  \fi
\hypersetup{bookmarksnumbered,%
              colorlinks,%
               linkcolor=blue,%
               citecolor=blue,%
              plainpages=false,%
            pdfstartview=FitH}

\def\no{\noindent}

\def\bc{\begin{center}}
\def\nno{\nonumber}
\def\ec{\end{center}}
\def\be{\begin{eqnarray}}
\def\ee{\end{eqnarray}}

\newcommand{\omits}[1]{}

\definecolor{dyellow}{rgb}{1.,0.8,.0}
\definecolor{myblue}{rgb}{.1,.1,.7}
\definecolor{dcyan}{rgb}{.0,.6,.6}
\definecolor{dmagenta}{rgb}{0.6,0.0,0.6}
\definecolor{brown}{rgb}{0.6,0.2,0.}
\definecolor{darkblue}{rgb}{.0,.0,0.5}
\definecolor{darkred}{rgb}{0.75,0.0,0.0}
\definecolor{orange}{rgb}{1.,.6,.0}
\definecolor{dorange}{rgb}{0.8,.4,.0}
\definecolor{green}{rgb}{0.0,1.0,0.0}
\definecolor{lightgrey}{rgb}{0.7,0.7,0.7}
\definecolor{purple}{rgb}{.4,.0,.4}

\def\blue{\color{blue}}

\def\Ga{\Gamma}

\def\Om{\Omega}

\def\dl{\delta}
\def\eps{\epsilon}
\def\ka{\kappa}
\def\la{\lambda}
\def\th{\theta}
\def\si{\sigma}

\def\del{\nabla}
\def\d#1#2{\frac{\displaystyle #1}{\displaystyle #2}}
\def\r{\partial}

\newcommand{\BdS}{${B}d{S}$}
\newcommand{\Riem}{$Riem$}
\newcommand{\Lob}{$Lob$}
\newcommand{\dS}{$dS$}
\newcommand{\AdS}{$AdS$}

\newcommand{\PoR}{principle of relativity}

\newcommand{\vect}[1]{\mbox{\boldmath $#1$}}
\def\P{{\vect P}}
\def\K{{\vect K}}

\def\J{{\vect J}}

\def\NH+{$NH_+$}
\def\NH-{$NH_-$}

\newcommand\btd{\raise 2pt
\hbox{$\hat\bigtriangledown$}\hskip 1.5pt}
\newcommand\bt{\raise 2pt
\hbox{$\bigtriangledown$}\hskip 1.5pt}

\begin{document}



\title{Geometries for Possible Kinematics}%

\author{{Chao-Guang Huang}$^{1}$} 
\author{{Yu Tian}$^2$}
\author{{Xiao-Ning Wu}$^{3}$}
\author{{Zhan Xu}$^4$}
\author{{Bin Zhou}$^5$}

\affiliation{%
${}^1$ Institute of High Energy Physics, and
Theoretical Physics Center for Science Facilities, Chinese Academy of
Sciences, Beijing 100049, China}

\affiliation{%
${}^2$ Graduate University of Chinese Academy of
Sciences, Beijing 100049, China}

\affiliation{%
${}^3$Institute of Mathematics, Academy of Mathematics and System Sciences,
Chinese Academy of Sciences, Beijing
100190, China,}

\affiliation{%
${}^4$Department of Physics, Tsinghua University, Beijing
100084, China.}

\affiliation{%
${}^5$Department of Physics, Beijing Normal University, Beijing
100875, China.}

\date{July 2010}


\begin{abstract}
The algebras for all possible Lorentzian and Euclidean
kinematics with $\frak{so}(3)$ isotropy  except static ones are re-classified.
The geometries for algebras are presented by contraction approach.  The relations
among the geometries are revealed.  Almost all geometries fall into pairs.  There
exists $t \leftrightarrow 1/(\nu^2t)$ correspondence in each pair. In the viewpoint of
differential geometry, there are only 9 geometries, which have right signature
and geometrical spatial isotropy.  They are 3 relativistic geometries,
3 absolute-time geometries, and 3 absolute-space geometries.
\end{abstract}


\maketitle

\tableofcontents

\section{Introduction}

The contraction is a powerful method in mathematical physics,
which reveals the relations among groups and geometries.  By the In\"on\"u-Wigner
contraction method \cite{IW}, Bacry and L\'evy-Leblond show that there are 11
kinematical groups of 8 types according to their algebras under the assumption that
a kinematical group
should possess (i) an $SO(3)$ isotropy generated by $\J$, (ii) parity ($\Pi:
H\to H, \P \to -\P, \K \to -\K, \J \to \J$)
and time-reversal ($\Theta: H\to -H, \P \to \P, \K \to -\K, \J \to \J$)
automorphism, and
(iii) non-compact one-dimensional subgroup generated by each boost $K_i$ \cite{BLL}.
These groups are the Poincar\'e ($P$), de Sitter (\dS), anti-de Sitter (\AdS),
Inhomogenous $SO(4)$ ($E'$\footnote{It is sometimes denoted as $P'_+$ in the
literature.}), para-Poincar\'e ($P'$),
Galilei ($G$), Newton-Hooke ($NH_+$), anti-Newton-Hooke ($NH_-$), para-Galilei ($G'$),
Carroll ($C$), and static ($S$) groups.  Releasing the third condition, three
geometrically kinematical groups --- Euclid ($Euc$), Riemann ($Riem$), and
Lobachevsky ($Lob$) groups --- should be added.  Among these 14 kinematical groups,
the geometries corresponding to \Riem, \Lob, \dS, \AdS, $Euc$, $P$, $NH_\pm$, $G$ and
$C$ groups are clear.  But, the others still need to be clarified though
the correspondence between 2d possible kinematics and 9 Cayley-Klein
geometries \cite{nonEuclGeom}
have been set up \cite{Sanjuan}.  Different from the higher dimension cases,
each direction in 2d case can be identified as the
timelike direction and the other as the spacelike one when the
signature is $(+,-)$, which results in that different kinematics'
correspond to the same Cayley-Klein geometry.

The possible kinematical algebras can also be obtained from very different
approach --- combinatory approach \cite{GHWZ}.  The foundation of the new approach is
the \PoR\ with an invariant speed $c$ and an invariant length $l$, denoted as
the $PoR_{c,l}$ \cite{GHXZ, GWZ, GHWZ}.  It is well known that the principle of relativity
is the foundation of physics, which is closely related to the symmetry of space and
time.  Based on $PoR_{c,l}$, the triality of special relativity with \dS-,
\AdS-, and $P$-invariance can be established \cite{GWZ}.  It has been shown that the
general form of transformations
preserving $PoR_{c,l}$ is the linear fractional transformation \cite{UWFH,GHTXZ}.  All
linear fractional transformations form a group, known as $PGL(5,\mathbb{R})$
(or ``inertial-motion" group $IM(4)$) \cite{GHWZ}.
The group contains 24 possible kinematical or geometrical (sub-)groups
with the same SO(3) isotropy.  In addition to the 11 possible kinematics in
\cite{BLL} and 3 geometrical kinematics, there are additional 10 groups.
A natural question appears: what are the meanings of these additional
possible kinematical groups or what do these additional possible
kinematical groups represent?

The purposes of the present paper are two-fold.  The first is to present the geometrical
structures for all possible kinematics revealed in \cite{BLL, GHWZ} but static ones.
Seeing that the different kinematical algebras are linked
together by the contraction,  their corresponding
geometries should also be linked by the same contraction.  Therefore,
the unknown geometries can be obtained from known geometries by the contraction approach.
The obtained geometrical structures are required to be
invariant under the given 10-parameter transformations and are not identically
vanishing or divergent everywhere.  The second is further to explore the relations
among the geometries with different kinematical groups and to pick out the
geometries for the genuine possible kinematics.

The paper will be organized in the following way.  In the next
section, we shall review all the possible kinematical algebras and
clarify their relation to the $Riem$, $Lob$, \dS\ and \AdS\ algebras.
In sections III, we shall ``derive" the geometries from the geometries for $Riem$,
$Lob$, $dS$, and $AdS$ algebras and/or their contractions.  In section IV, we shall
further explore the relations among the geometries. The concluding remarks
are given in the last section.  In the section, we shall make some comments
on the requirements to select the possible kinematics proposed by Bacry and
L\'evy-Leblond and modify the requirements.  Under the modified requirements,
there are only 9 possible kinematics.  They are 3 relativistic ones, $P$, \dS, and
\AdS, 3 absolute-time ones, $G$ and $NH_\pm$, and 3 absolute-space
ones, $C$, $E_{2-}$, and $P_{2-}$.


\section{Possible Kinematical Algebras}
\begin{table}[thp]
\caption{\quad All possible kinematical and geometrical algebras}
\begin{tabular}{c c c | c c c }
\toprule[2pt]
Algebra & Symbol & Generator set &  Algebra & Symbol & Generator set \\
\hline $\begin{array}{c} {\it Riemann}\\ {\it
Lobachevsky}\end{array}$ & $\begin{array}{c} \mathfrak{r} \\
\mathfrak{l}\end{array}$ & $\begin{array}{c} (H^-, P^+_i,
N_i,J_i) \\ (H^+, P^-_i, N_i,J_i)\end{array} $&
{\it Euclid}&$\begin{array}{c}\mathfrak{e}\\
\mathfrak{e}_2\end{array}$&$\begin{array}{c}(H, P_i, N_i, J_i)\\
(-H', P'_i, N_i, J_i)\end{array}$\\
\hline
$\begin{array}{c} dS \\ AdS\end{array}$ &
$\begin{array}{c}\mathfrak{d}_+\\ \mathfrak{d}_-\end{array}$ &
$\begin{array}{c}(H^+, P^+_i, K_i, J_i) \\ (H^-, P^-_i, K_i, J_i) \end{array}$ &
{\it Poincar\'e} &$\begin{array}{c}\mathfrak{p}\\
\mathfrak{p}_2\footnote{The second versions of algebras here
have different meaning from the one in Ref. \cite{AP}. Say, the second Poincar\'e
group here is the semi-product of Lorentz group with pseudo-translations, while
in \cite{AP} the second Poincar\'e group is the semi-product of Lorentz group
with special conformal transformations.}\end{array}$ & $\begin{array}{c}(H, P_i, K_i, J_i)\\
(H', P'_i, K_i, J_i) \end{array}$\\
${NH}_+$  & $\begin{array}{c}\mathfrak{n_+}\\
\mathfrak{n}_{+2}\end{array}$ & $\begin{array}{c}(H^+, P_i , K^{\frak g}_i,J_i )\\
(H^+, P'_i, K^{\frak c}_i,J_i )\end{array}$&{\it Galilei}&$\begin{array}{c}\mathfrak{g}\\
\mathfrak{g}_2\end{array}$&$\begin{array}{c}(H, P_i, K^{\frak g}_i,J_i)\\
(H', P'_i, K^{\frak c}_i,J_i)\end{array}$  \\
${NH}_-$  &$\begin{array}{c}\mathfrak{n}_-\\
\mathfrak{n}_{-2}\end{array}$ & $\begin{array}{c}(H^-, P_i, K^{\frak g}_i,J_i )\\
(-H^-, P'_i, K^{\frak c}_i,J_i )\end{array}$ &
{\it Carroll}&$\begin{array}{c}\mathfrak{c}\\
\mathfrak{c}_2\end{array}$ & $\begin{array}{c}(H, P_i, K^{\frak c}_i,J_i )\\
(H', P'_i, K^{\frak g}_i,J_i )\end{array}$ \\
$HN_+$\footnote{$HN_\pm$ are historically called the para-Poincar\'e algebras
\cite{BLL}.  They are called Hooke-Newton algebras because they are different
from the Newton-Hooke algebras by the replacement $H^\pm\leftrightarrow H$,
$P_i \leftrightarrow P_i^\pm$, $K_i^g\leftrightarrow K_i^c$.  From the geometrical
point of view (see: Sec. \ref{Sec:HN}), the second version of $HN_\pm$ algebra are called para-Euclid and
para-Poincar\'e algebra, respectively.}
&$\begin{array}{c}\mathfrak{h}_+\\
\mathfrak{e}'\end{array}$&$\begin{array}{c}(H, P^+_i, K^{\frak c}_i,J_i )\\
(H', P^+_i, K^{\frak g}_i,J_i )\end{array}$&{\it para-Galilei}&
$\begin{array}{c}\mathfrak{g}'\\
\mathfrak{g}'_2\end{array}$ & $\begin{array}{c}(H', P, K^{\frak g}, J_i)\\
(H, P'_i, K^{\frak c}_i,J_i )\end{array}$ \\
$HN_-$&$\begin{array}{c}\mathfrak{h_-}\\
\mathfrak{p}'\end{array}$&$\begin{array}{c}(H, P^-_i, K^{\frak c}_i,J_i)\\
(-H', P^-_i, K^{\frak g}_i,J_i )\end{array}$&{\it Static}&$\begin{array}{c}\mathfrak{s}\\
\mathfrak{s}_2\end{array}$& $\begin{array}{c}(H^{\frak s}, P'_i, K^{\frak c}_i,
J_i)\footnote{The generator $H^{\frak s}$ is meaningful only when the central
extension is considered.}
\\
(H^{\frak s}, P_i , K^{\frak g}_i,J_i )\end{array}$ \\
\bottomrule[2pt]
\end{tabular}
\end{table}

In \cite{GHWZ}, we show that there are 24 possible kinematical or geometrical
(sub-)algebras with the same $\frak{so}$(3) isotropy in $\frak{pgl}(5,\mathbb{R})$.
They are listed in the TABLE I, in which the generators are
defined by
\be\begin{cases}
H=\r_t,\ H'=-\nu^{2} t x^\mu\r_\mu,\ H^\pm = \r_t \mp \nu^{2} t x^\mu\r_\mu, &
x^0=ct, \,\nu=c/l;  \\
{P}_i=\r_i,\ {P}_i'=-l^{-2}x_i x^\mu \r_\mu,\
{P}_i^\pm=\r_i \mp l^{-2}x_i x^\mu \r_\mu, &\\
K_i= t\r_i-c^{-2}x_i\r_t,\ {K}_i^g= t\r_i,\ {K}_i^c=-c^{-2}x_i\r_t,\
{N}_i= t\r_i + c^{-2}x_i\r_t &\\
{J}_i=\d 1 2 \eps_{i}^{\ jk} (x_j\r_k-x_k\r_j), &
\end{cases}
\ee
where $c$, $l$ and thus $\nu$ are invariant parameters with dimensions of speed,
length and the inverse of time, respectively.  Hereafter,
the lowercase Greek letters $\mu$, $\nu$, $\ka$, $\rho$, $\si$, $\cdots$ in indices
run from 0 to 3 while the lowercase Latin letters $i$, $j$, $k$, $l$, $m$, $n$,
$\cdots$ in indices run from 1 to 3.
$H$, $H'$, and $H^\pm$ are known as the (algebraic) translation, pseudo-translation,
and Beltrami translation of time, respectively.  $\vect{P}$, $\vect{P}'$, and
$\vect{P}^\pm$ are the (algebraic) translation, pseudo-translation, and Beltrami
translation of space, respectively.  $\vect{K}$, $\vect{K}^g$, $\vect{K}^c$, and
\vect{N} are the Lorentz, Galilei, Carroll, and geometrical boosts, respectively.
The other generators in $\frak{pgl}(5,\mathbb{R})$ may transform the generators
in one kinematical algebra to the ones for another.

According to the classification of Bacry and L\'evy-Leblond, the first four algebras
in TABLE I are purely geometrical ones and the remaining 20 are kinematical algebras.
All possible kinematical and geometrical algebras are
related together in the two extremely different approaches.  One is the combinatory
method \cite{GHWZ} and the other is the contraction method \cite{BLL}.

From $dS$ and $AdS$ algebras $\frak{d}_\pm$, one may obtain the generators of the
Poincar\'{e} algebra $\frak p$ and the second Poincar\'e algebra $\frak{p}^{}_2$
by the simple summation or subtraction,
\be
{\frak p}:&\quad & H= \d 1 2 (H^+ + H^-) ,\  \vect{P}=\d 1 2 (\vect{P}^+ +\vect{P}^-),
\ \vect{K}, \ \vect{J},\\
\mbox{and \qquad}  {\frak p}_2:& \quad & H'= \d 1 2 (H^+ - H^-) ,\
\vect{P}'=\d 1 2 (\vect{P}^+ -\vect{P}^-),\ \vect{K}, \ \vect{J},
\ee
respectively.  One may also obtain them by the contraction of
the generators of $\frak{d}_\pm$, \be {\frak p}:\ &&H= \lim_{l_r
\to \infty} H_r^\pm,\ \vect{P}=\lim_{l_r \to \infty} \vect{P}_r^\pm,
\ \vect{K}, \ \vect{J},\\
\mbox{and \qquad}{\frak p}_2:&\quad &
H'= \pm\lim_{l_r \to 0}\d {l_r^2} {l^2} H_r^\pm,\ \vect{P}'=\pm\lim_{l_r \to 0}
\d {l_r^2}{l^2}\vect{P}_r^\pm,
\ \vect{K}, \ \vect{J}, \label{p2contr}
\ee
respectively.  Here, $l_r$ is a running parameter of dimension $L$.
It replaces $l$ in the definition of generators and thus the generators is denoted
with a subscript $r$.  The contraction prescription used here is
slightly different from that used by Bacry and L\'evy-Leblond, in which
the contraction for ${\frak p}$, for example, is defined by
\be \label{BLLContr}
H \to \varepsilon H,\ \vect{P}\to \varepsilon \vect{P}, \ \vect{K}, \
\vect{J}, \ \varepsilon \to 0.
\ee
In comparison of Eq.(\ref{p2contr}) with Eq. (\ref{BLLContr}), the Bacry-L\'evy-Leblond
contraction gives ${\frak p}_2$ if $\eps = \pm l_r^2/l^2$ is taken.

Similarly, from $Riem$ algebra $\frak r$ and $Lob$ algebra $\frak l$, one may
attain the generators of the $Euc$ algebra $\frak e$
and the second $Euc$ algebra $\frak{e}_2$ by the summation or substraction,
\be
{\frak e}:\ &\quad & H= \d 1 2 (H^+ + H^-) ,\  \vect{P}=\d 1 2 (\vect{P}^+ +\vect{P}^-),
\ \vect{N}, \ \vect{J},\\
\mbox{and \qquad}{\frak e}_2:&\quad & -H'= \d 1 2 (H^- - H^+) ,\
\vect{P}'=\d 1 2 (\vect{P}^+ -\vect{P}^-),\ \vect{N}, \ \vect{J},
\ee
respectively, or by the contraction of the generators of
$\frak r$ and $\frak l$,
\be
{\frak e}:\ &&H= \lim_{l_r \to \infty} H_r^\mp,\ \vect{P}=\lim_{l_r \to \infty}
 \vect{P}_r^\pm, \ \vect{N}, \ \vect{J},\\
\mbox{and \qquad}{\frak e}_2:&\quad &
-H'= \pm \lim_{l_r \to 0}\d {l_r^2} {l^2} H_r^\mp,\ \vect{P}'=\pm \lim_{l_r \to 0}
\d {l_r^2}{l^2}\vect{P}^\pm, \ \vect{N}, \ \vect{J},
\ee
respectively.

From $\frak{d}_+$, $\frak l$ and $\frak{d}_-$, $\frak r$, one may get
the generators of $NH_\pm$ algebras $\frak{n}_\pm$ and the second
$NH_\pm$ algebras $\frak{n}_{\pm 2}$ by the
combinatory method, respectively,
\be
{\frak n}_\pm: &\quad & H^\pm, \ \vect{P}=\d 1 2 (\vect{P}^+ +\vect{P}^-), \
\vect{K}^g=\d 1 2(\vect{K}+\vect{N}), \ \vect{J},\\
\mbox{and \qquad}{\frak n}_{\pm2}:&\quad &\pm H^\pm, \ \vect{P}'=\d 1 2
(\vect{P}^+ -\vect{P}^-),\ \vect{K}^c= \d 1 2(\vect{K}-\vect{N}), \ \vect{J}.
\ee
The generators of $\frak{n}_+$ and $\frak{n}_{+2}$ can also be acquired
by the contraction from $\frak{d}_+$ or ($\Pi$ of) $\frak l$,
\be
{\frak n}_+: &&H^+ =\lim_{\substack{c_r,\,l_r\to \infty \\ \nu {\rm\; fixed}}}H_r^+,
\ \vect{P}=\lim_{\substack{c_r,\,l_r\to \infty \\ \nu {\rm\; fixed}}} \vect{P}_r^\pm,
\vect{K}^g=\lim_{\substack{c_r,\,l_r\to \infty \\ \nu {\rm\; fixed}}}
\vect{K}_r (\mbox{or }\vect{N}_r), \ \vect{J},\\
\mbox{and\quad}{\frak n}_{+2}:&&
H^+= \lim_{\substack{c_r,\,l_r \to 0 \\ \nu\;{\rm fixed}}} H_r^+,\
\vect{P}'=\pm \lim_{\substack{c_r,\,l_r \to 0 \\ \nu\;{\rm fixed}}}
\d {l_r^2}{l^2}\vect{P}_r^\pm, \vect{K}^c=
\lim_{\substack{c_r,\,l_r \to 0 \\ \nu\;{\rm fixed}}}
\d {c_r^2}{c^2}\vect{K}_r(\mbox{or }-\d {c_r^2}{c^2}\vect{N}_r),\ \vect{J}, \quad
\label{NH+Contr}
\ee
where $c_r$ is a running parameter of dimension $LT^{-1}$.
In comparison of Eq.(\ref{NH+Contr}) with Eq. (9a) in \cite{BLL}, the Bacry-L\'evy-Leblond
contraction gives ${\frak n}_{+2}$ if $\varepsilon = l_r^2/l^2= c_r^2/c^2$ is taken.
Similarly, the generators of
$\frak{n}_-$ and $\frak{n}_{-2}$ can also  be acquired
by the contraction from ($\Theta\Pi$ of) $\frak{d}_-$ or
($\Theta$ of) $\frak r$,
\be
{\frak n}_-: &&H^- =\lim_{\substack{c_r,\,l_r\to \infty \\ \nu {\rm\; fixed}}}H_r^-,
\ \vect{P}=\lim_{\substack{c_r,\,l_r\to \infty \\ \nu {\rm\; fixed}}} \vect{P}_r^\mp,\
\vect{K}^g=\lim_{\substack{c_r,\,l_r\to \infty \\ \nu {\rm\; fixed}}}\vect{K}_r
(\mbox{or }\vect{N}_r), \ \vect{J},\\
\mbox{and }{\frak n}_{-2}:&&
- H^-= - \lim_{\substack{c_r,\,l_r \to 0 \\ \nu\;{\rm fixed}}} H_r^-,
\vect{P}'=\mp \lim_{\substack{c_r,\,l_r \to 0 \\ \nu\;{\rm fixed}}} \d {l_r^2}{l^2}
\vect{P}_r^\mp,\
\vect{K}^c= \lim_{\substack{c_r,\,l_r \to 0 \\ \nu\;{\rm fixed}}}
\d {c_r^2}{c^2}\vect{K}_r(\mbox{or }-\d {c_r^2}{c^2}\vect{N}_r),\ \vect{J}.\qquad
\ee

The generators of $HN_+$ algebra $\frak{h}_+$ and para-Euclid [isomorphic
to $\frak{iso}$(4)] algebra $\frak{e}'$ (or $HN_-$ algebra $\frak{h}_-$
and para-Poincar\'e algebra $\frak{p}'$) are obtained
from those of $\frak{d}_+$ and $\frak r$ (or $\frak{d}_-$ and $\frak l$) by the
combinatory method,
\be
{\frak h}_\pm: &\quad & H=\d 1 2(H^++H^-), \ \vect{P}^\pm , \
\vect{K}^c=\d 1 2(\vect{K}-\vect{N}), \ \vect{J},\\
\mbox{and \qquad }{\frak e}',{\frak p}': &\quad &\pm H'=\pm \d 1 2 (H^+-H^-),
\ \vect{P}^\pm,\ \vect{K}^g= \d 1 2(\vect{K}+\vect{N}), \ \vect{J},
\ee
respectively.  They are related to  $\frak{d}_\pm$ by
\be
{\frak h}_\pm: &&H =\lim_{c_r \to 0}H_r^\pm,\ \vect{P}^\pm, \vect{K}^c=\lim_{c_r\to 0}
\d {c_r^2}{c^2}\vect{K}_r, \ \vect{J},\\
\mbox{and \quad }{\frak e}',{\frak p}': &\quad & \pm H'= \lim_{c_r \to \infty} \d{c^2}{c_r^2}
H_r^\pm,\ \vect{P}^\pm, \vect{K}^g=\lim_{c_r \to \infty} \vect{K}_r ,\ \vect{J},
\ee
respectively.  $\frak{h}_\pm$, $\frak{e}'$, and $\frak{p}'$ with $H'$ replaced
by $-H'$ are also the contractions of ($\Theta$ of) $\frak r$ and ($\Theta$ of)
$\frak l$ in the limit $c_r\to 0$ and $c_r\to \infty$,
respectively.

The generators of Galilei/Carroll algebras $\frak{g}/\frak{c}$
are the linear combination of
generators of $\frak{d}_\pm$, $\frak{r}$, and $\frak{l}$\,, respectively,
\be
{\frak g}/{\frak c}: &&H=\d 1 2 (H^++H^-),\ \,\vect{P}=\d 1 2 (\vect{P}^++
\vect{P}^-), \ \,\vect{K}^{g/c}=\d 1 2 (\vect{K}\pm\vect{N}),
\ \vect{J},\\
\mbox{and \qquad}{\frak g}_2/{\frak c}_2:&\quad &
H'=\d 1 2 (H^+-H^-),\ \vect{P}'=\d 1 2 (\vect{P}^+-
\vect{P}^-), \ \vect{K}^{c/g}=\d 1 2 (\vect{K}\mp\vect{N}), \ \vect{J},
\ee
or as the contraction from $\frak{d}_\pm$,
\be
{\frak g}: &&H=\lim_{\substack{c_r,\,l_r\to \infty \\ \nu_r \to 0}}
H_r^\pm,\ \vect{P}=\lim_{\substack{c_r,\,l_r\to \infty \\ \nu_r \to 0}}
\vect{P}^\pm,
\ \vect{K}^g=\lim_{\substack{c_r,\,l_r\to \infty \\ \nu_r \to 0}}\vect{K}_r,
\ \vect{J},\\
{\frak c}:&&H=\lim_{\substack{l_r\to \infty \\ c_r\to 0 }}
H_r^\pm,\; \quad \vect{P}=\lim_{\substack{l_r\to \infty \\ c_r\to 0}}
\vect{P}^\pm,\quad \;
\vect{K}^c=\lim_{\substack{l_r\to \infty \\ c_r\to 0}} \d {c_r^2}{c^2}\vect{K}_r,
 \ \vect{J},\\
{\frak g}_2:&&
H'=\pm \lim_{\substack{c_r,\,l_r\to 0 \\ \nu_r \to \infty}}
\d{\nu^2}{\nu_r^2}H_r^\pm,\ \vect{P}'=\pm \lim_{\substack{c_r,\,l_r\to 0
 \\ \nu_r \to \infty}}
\d {l_r^2}{l^2}\vect{P}^\pm,\ \vect{K}^c=\lim_{\substack{c_r,\,l_r\to 0 \\ \nu_r \to \infty}}
\d {c_r^2}{c^2}\vect{K}_r,  \ \vect{J} ,\\
\mbox{and \qquad}{\frak c}_2:&\quad&
H'=\pm \lim_{\substack{l_r\to 0 \\ c_r\to \infty}}
\d{\nu^2}{\nu_r^2}H_r^\pm,\quad \vect{P}'=
\pm \lim_{\substack{l_r\to 0 \\ c_r\to \infty}}
\d {l_r^2}{l^2}\vect{P}^\pm,\ \
\vect{K}^g=\lim_{\substack{l_r\to 0 \\ c_r\to \infty
}}\vect{K}_r, \qquad \vect{J}.
\ee
Obviously, $\frak{g}$ can also be derived by the contraction from
$\frak{r}$ or $\frak{l}$
in the limit of $c_r,\,l_r \to \infty, \nu_r \to 0$,
\be
{\frak g}: &&H=\lim_{\substack{c_r,\,l_r\to \infty \\ \nu_r \to 0}}
H_r^\mp,\ \vect{P}=\lim_{\substack{c_r,\,l_r\to \infty \\ \nu_r \to 0}}
\vect{P}^\pm,
\ \vect{K}^g=\lim_{\substack{c_r,\,l_r\to \infty \\ \nu_r \to 0}}\vect{N}_r,
\ \vect{J};
\ee
from $\frak{e}$ and $\frak{p}$ in the limit of $c_r\to \infty$; and from
$\frak{n}_\pm$
in the limit of $\nu_r \to 0$.
$\frak{g}_2$ can also be derived by the contraction from ($\Theta$ of)
$\frak r$ or ($\Pi$ of) of $\frak l$ in the limit of $c_r,\,l_r \to 0, \nu_r \to \infty$,
\be
{\frak g}_2:\ H'= \mp\lim_{\substack{c_r,\,l_r \to 0\\ \nu_r \to \infty}}\d{\nu^2}{\nu_r^2}
H_r^\mp,\
\vect{P}'=\pm \lim_{\substack{c_r,\,l_r \to 0\\ \nu_r \to \infty}}\d{l_r^2}{l^2}
\vect{P}_r^\pm,
\,\vect{K}^{\frak c}= -\lim_{\substack{c_r,\,l_r \to 0\\ \nu_r \to \infty}}\d {c_r^2}{c^2}
\vect{N}, \ \vect{J};
\ee
from ($\Theta$ of) $\frak{e}_2$ or $\frak{p}_2$ in the limit of $c_r\to 0$;
and from $\frak{n}_{\pm2}$ in the limit of $\nu_r\to \infty$.
Similarly, $\frak{c}$ and $\frak{c}_2$ are the direct contractions of
$\frak{h}_\pm$ in
the limit $l_r\to \infty$
and $\frak{e}', \frak{p}'$ in the limit $l_r\to 0$, respectively, or
the direct contractions of $\frak p$ in the limit $c_r \to 0$ and
$\frak{p}^{}_2$ in the limit of $c_r\to \infty$, respectively.  $\frak c$
and $\frak{c}_2$ with $H'$
replaced by $-H'$ are also the contraction of ($\Theta$ of) $\frak{e}$
in the limit $c_r\to 0$ and $\frak{e}_2$ in the limit $c_r \to \infty $,
respectively.

\begin{figure}[t]
  \includegraphics[width=160mm,height=120mm]{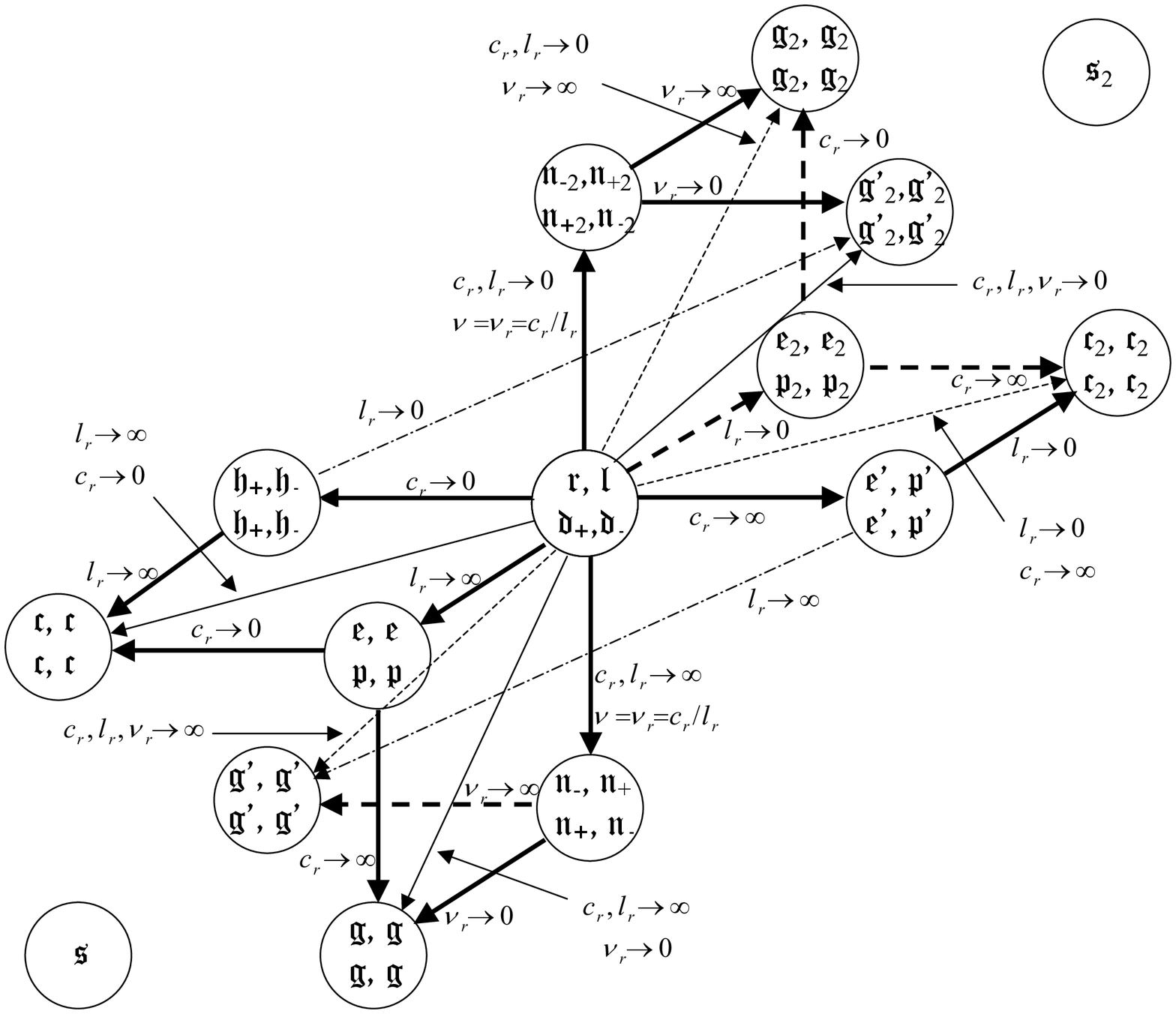}\\
 \caption{Contraction scheme for the possible kinematics.  (The degeneracy
in each algebra and its partner has been released.) } \label{Fig:ar}
\end{figure}

The combination of the generators of $\frak{d}_\pm$, $\frak{r}$ and $\frak{l}$
can define the generators of para-Galilei algebra $\frak{g}'$ in the following way.
\be
{\frak g'}: && H'=\d 1 2(H^+-H^-),\
\vect{P}=\d 1 2 (\vect{P}^+ + \vect{P}^-),\ \vect{K}^g=\d 1 2 (\vect{K}+\vect{N}),
\ \vect{J},\\
\mbox{and \qquad}{\frak g'}_2:&\quad &
H=\d 1 2(H^++H^-),\ \vect{P}'=\d 1 2 (\vect{P}^+ - \vect{P}^-),\
\vect{K}^c=\d 1 4 (\vect{K}-\vect{N}), \ \vect{J}.
\ee
$\frak{g}'$ and $\frak{g}'_2$ can be deduced from the contraction from $\frak{d}_+$,
\be
{\frak g}': &&H'=  \lim_{\substack{l_r,\, c_r\to\infty\\ \nu_r \to \infty}}
 \d {\nu^2}{\nu_r^2} H_r^+,\ \vect{P}=\lim_{\substack{l_r,\,
 c_r\to\infty\\\nu_r\to \infty}}  \vect{P}_r^+,\quad
\vect{K}^g=\lim_{\substack{l_r,\, c_r\to\infty\\\nu_r\to \infty}}\vect{K}_r, \quad
\vect{J},\\
\mbox{and \quad}{\frak g}'_2:& &
H=\lim_{\substack{l_r,\, c_r\to 0\\\nu_r\to 0}} H_r^+,\qquad
\vect{P}'= \lim_{\substack{l_r,\, c_r\to 0\\\nu_r\to 0}}
\frac {l_r^2}{l^2}\vect{P}_r^+,\
\vect{K}^c=\lim_{\substack{l_r,\, c_r\to 0\\\nu_r\to 0}}
\d {c_r^2}{c^2}\vect{K}_r,  \ \vect{J}.
\ee
The contraction from $\frak{l}$ in the limit of $c_r,\, l_r,\, \nu_r \to \infty$
can also give rise to $\frak{g}'$,
\be
{\frak g}': &&H'=  \lim_{\substack{l_r,\, c_r\to\infty\\ \nu_r \to \infty}}
 \d {\nu^2}{\nu_r^2} H_r^+,\ \vect{P}=\lim_{\substack{l_r,\,
 c_r\to\infty\\\nu_r\to \infty}}
  \vect{P}_r^-,
\ \vect{K}^g=\lim_{\substack{l_r,\,
 c_r\to\infty\\\nu_r\to \infty}}
\vect{N}_r
, \ \vect{J}.
\ee
Besides, $\frak{g}'$ is the results of the
contractions from $\frak{n}_{+}$ in the limit $\nu_r\to \infty$
or from $\frak{e}'$ in the limit $l_r\to \infty $, $\frak{g}'$
with $H'$ replaced by $-H'$ is the result of the contraction from $\frak{d}_-$
and $\frak{r}$ in the limit $l_r,\, c_r, \, \nu_r \to \infty$, from
$\frak{n}_{-}$ in the limit $\nu_r\to \infty $, or from
($\Pi$ of) $\frak{p}'$ in the limit $l_r\to \infty$.
Similarly, the contraction from ($\Pi$ of) $\frak{l}$ in the limit of
$c_r,\, l_r,\, \nu_r \to 0$ results in $\frak{g}'_2$,
\be
{\frak g}'_2:& &
H=\lim_{\substack{l_r,\, c_r\to 0\\\nu_r\to 0}} H_r^+,\ \
\vect{P}'= -\lim_{\substack{l_r,\, c_r\to 0\\\nu_r\to 0}}
\frac {l_r^2}{l^2}\vect{P}_r^-,\
\vect{K}^c=-\lim_{\substack{l_r,\, c_r\to 0\\\nu_r\to 0}}
\d {c_r^2}{c^2}\vect{N}_r
,  \ \vect{J}.
\ee
Finally, $\frak{g}'_2$ is the results of the
contractions  from $\frak{n}_{+2}$ in the limit $\nu_r\to 0$
or from $\frak{h}_{+}$ in the limit $l_r\to 0$, $\frak{g}'_2$
with $H'$ replaced by $-H'$ is the result of the contraction from ($\Pi\Theta$
of) $\frak{d}_-$ and ($\Theta$ of) $\frak{r}$ in the limit
$l_r,\, c_r, \, \nu_r \to 0$, from $\frak{n}_{-2}$ in the
limit $\nu_r\to 0$ or from $\Pi$ of
$\frak{h}_{-}$ in the
limit $l_r\to 0$.

\begin{figure}[tb]
  \includegraphics[width=160mm,height=120mm]{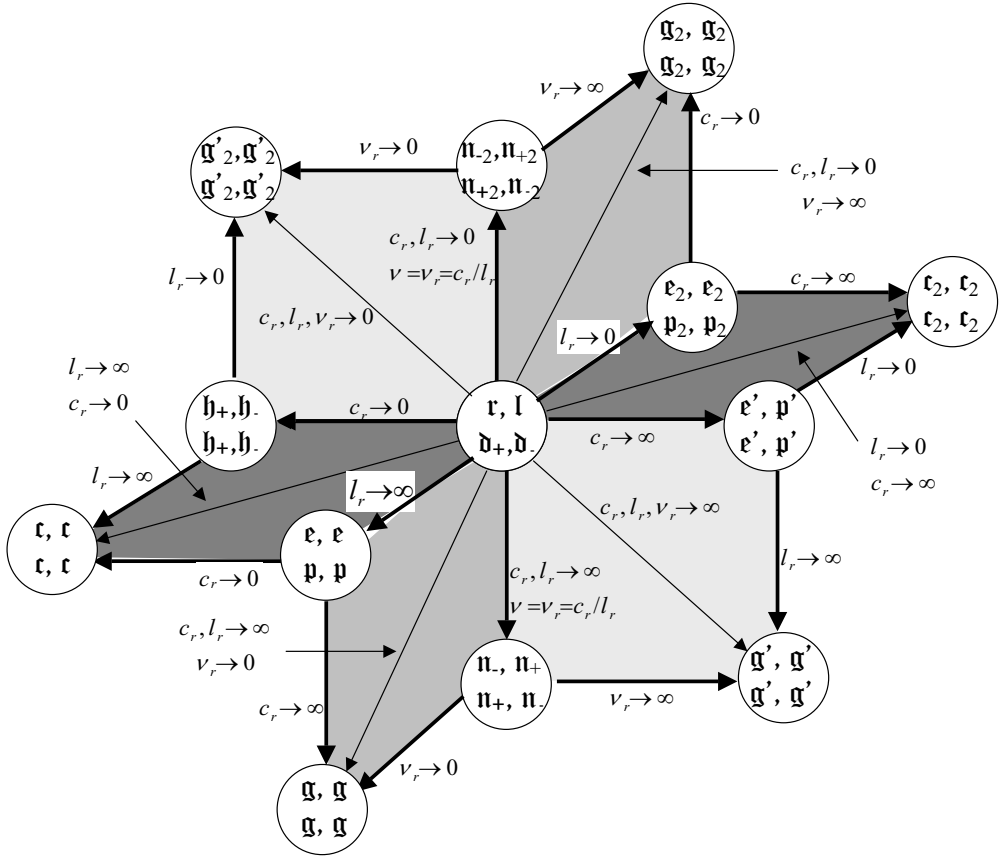}\\
 \caption{Contraction scheme for the possible kinematics in a more symmetric way.
 } \label{Fig:arn}
\end{figure}

The contraction scheme for the possible kinematics are shown in FIG.
\ref{Fig:ar} in the similar way as in Ref.\cite{BLL}.  In the figure the degeneracy
in the each algebra and its partner has been released and the diagram
is not two perfect cubes with a common vertex.
FIG. \ref{Fig:arn}\ presents the contraction scheme for possible kinematics in
a more symmetric way, where the two static algebras as exceptions are ignored,
whose time-translation
generator is meaningful only when the central extension is taken into consideration.

\section{Geometries for the possible kinematical algebras}

\subsection{Geometries for $\frak{r}$, $\frak{l}$, $\frak{e}$, and
$\frak{e}_2$}

The metrics of 4d Riemann and Lobachevsky spaces in a Beltrami coordinate system
are well known,
\be
ds_{E\pm}^2=\d 1 {\si_E^\pm}\left ( \dl_{\mu \nu} \mp\d {\dl_{\mu\ka}
\dl_{\nu \la}x^\ka x^\la} {l^2\si_E^\pm}\right ) dx^\mu dx^\nu,
\ee
where
\be
\si_E^\pm= 1\pm l^{-2} \dl_{\ka\la}x^\ka x^\la >0.
\ee
(If the antipodal identification is not taken, more Beltrami coordinate
charts are needed to cover all 4d Riemann sphere, for example \cite{GHXZ}.
Here, we do not plan to discuss the problem in details in this paper.)
They are invariant under the transformations generated by
$\frak r$ and $\frak l$\,, respectively.  $\si_E^+ > 0$ is automatically satisfied.
$\si_E^- > 0$ puts the constraint on the domain.  Obviously, if $\si_E^- < 0$,
the metric
\be
ds^2= - \d 1 {\si_E^-}\left ( \dl_{\mu \nu} +\d {\dl_{\mu\ka}
\dl_{\nu \la}x^\ka x^\la}
{l^2\si_E^-}\right ) dx^\mu dx^\nu,
\ee
is also invariant under the transformations generated by $\frak l$\,. But,
it has the signature $(-, +, +, +)$ and thus is the alternative
representation of \dS\ space-time though the 1d sub-groups generated by
each boost is compact.  For brevity, it is referred to as
Lobachevsky-Beltrami-de Sitter ($LBdS$) space-time later and the line-element is
denoted by $ds_{LBdS}^{2}$.

Both \Riem\ and \Lob\ geometries contract to the Euclid metric in the limit
of $l_r\to \infty$,
\be
ds_{E}^2=\lim_{l_r\to \infty}\d 1 {\si_{E,\,r}^\pm}
\left ( \dl_{\mu \nu} \mp\d {\dl_{\mu\ka}\dl_{\nu \la}x^\ka x^\la}
{l_r^2\si_{E,\,r}^\pm}\right ) dx^\mu dx^\nu=\dl_{\mu \nu} dx^\mu dx^\nu ,
\ee
where $\si_{E,\,r}^\pm$ is the $\si_E^\pm$ with a running parameter $l_r$, namely,
\be
\si_{E,\,r}^\pm= 1\pm l_r^{-2} \dl_{\ka\la}x^\ka x^\la >0.
\ee
The inequality $\si_{E,\,r}^- <0$ will be violated in the
limiting process of $l_r \to \infty$.  Thus, $LBdS$ space-time is not contractible
in this way\footnote{Please note that ``contractible" here is
different from the usual concept in topology.}.  Obviously,
these metrics contain all local geometrical information
of the given spaces because of non-degeneracy.

On the other hand, in the limiting process of $l_r\to 0$, $l_r^2\si_{E,r}^+ > 0$
and $l_r^2\si_{E,r}^- < 0$
are always hold, while $l_r^2\si_{E,r}^- >0$ is not.  Thus, the \Lob\ geometry is not
contractible in this limit, while both \Riem\ and $LBdS$ geometries give rise to
\be \label{e2metric}
ds_{E_2}^2=\pm
\displaystyle{\lim_{l_r\to 0}}\d {l^2}{l_r^2}\d 1 {\si_{E,\,r}^\pm}\left (
\dl_{\mu \nu} \mp \d {\dl_{\mu\ka}\dl_{\nu \la} x^\ka x^\la}
{l_r^2\si_{E,\,r}^\pm}\right ) dx^\mu dx^\nu
=l^2\d {\dl_{\mu \nu}\dl_{\ka \la}-
\dl_{\mu\ka}\dl_{\nu\la}}{(\dl_{\si\tau}x^\si x^\tau)^2}x^\ka x^\la dx^\mu dx^\nu
=:\vect{g}^{E_2}.
\ee
Unfortunately, Eq.(\ref{e2metric}) defines a degenerate metric.
It does not contain enough geometrical
information to determine a 4d geometry uniquely. Also, it may possess larger
symmetry with more than 10 parameters.  To determine the geometry, one should
consider the limit of the inverse metric and the connection with the limit of
connection coefficients.  For the contraction of \Riem\ geometry,
\be \label{e2inv}
\left (\d {\r}{\r s}\right )_{E_2}^{2}=
\lim_{l_r\to 0}\d {l_r^4}{l^4}\si_{E,\,r}^+
(\dl^{\mu\nu}+ l_r^{-2}x^\mu x^\nu) \d{\r }{\r x^\mu}\otimes\d{\r }{\r x^\nu}
=l^{-4}\dl_{\si\tau}x^\si x^\tau x^\mu x^\nu \d{\r }{\r x^\mu}\otimes
\d{\r }{\r x^\nu} =: \vect{h}_{E_2},
\ee
\be \label{e2conn}
{\Ga_{E_2}}^{\la}_{\mu\nu}=-\lim_{l_r\to 0}\d{(\dl^\la_\mu \dl_{\nu \ka} + \dl^\la_\nu
\dl_{\mu \ka})x^\ka}{l_r^2\si_{E,\,r}^+}=-\d{(\dl^\la_\mu \dl_{\nu \ka} + \dl^\la_\nu
\dl_{\mu \ka})x^\ka}{\dl_{\si\tau}x^\si x^\tau}.
\ee
It can be shown that the connection is compatible to
$g^{E_2}_{\mu\nu}$ and $h_{E_2}^{\mu\nu}$, {\it i.e.}
\be
\del^{E_2}_\la g^{E_2}_{\mu\nu}
=\r_\la g^{E_2}_{\mu\nu}-{\Gamma_{E_2}}^\ka_{\la\nu}g^{E_2}_{\mu\ka}
-{\Gamma_{E_2}}^\ka_{\mu\la}g^{E_2}_{\ka\nu} =0,
\ee
\be
\del_{\la}^{E_2} h_{E_2}^{\mu\nu}= \r_\la {h}_{E_2}^{\mu\nu}+
{\Gamma_{E_2}}^\nu_{\la\ka}{h}_{E_2}^{\mu\ka}
+{\Gamma_{E_2}}^\mu_{\la\ka}{h}_{E_2}^{\ka\nu}=0,
\ee
and that the curvature of the connection has the form for constant curvature space,
\be
{R_{E_2}}^\si_{\ \mu \rho \nu}=\r_\rho{\Gamma_{E_2}}^\si_{\ \mu  \nu }-
\r_\nu{\Gamma_{E_2}}^\si_{\ \mu \rho}
+{\Gamma_{E_2}}^{\si}_{\ \tau \rho}{\Gamma_{E_2}}^\tau_{\ \mu \nu}-
{\Gamma_{E_2}}^{\si}_{\ \tau \nu}{\Gamma_{E_2}}^\tau_{\ \mu \rho}
=-l^{-2}(\dl^\si_\nu g^{E_2}_{\mu\rho}-\dl^\si_\rho g^{E_2}_{\mu\nu}
) \label{e2curvaturecomponents}
\ee
\be \label{e2Ricci}
R^{E_2}_{\mu\nu}={R_{E_2}}^\si_{\ \mu \nu \si}=
-3l^{-2}g^{E_2}_{\mu\nu}.
\ee
It can be checked that $\{M^{E_2},\vect{g}^{E_2},\vect{h}_{E_2},
\del^{E_2}\}$ is invariant under
$E_2$ transformations,
namely, $\forall \vect\xi \in {\frak e}_2\subset TM^{E_2}$
\be
{\cal L}_{\vect \xi} \vect{g}^{E_2} =  g^{E_2}_{\mu\nu,\,\la}\xi^\la
+g^{E_2}_{\mu\la}\r_\nu\xi^\la
+g^{E_2}_{\la\nu}\r_\mu\xi^\la=0,
\ee
\be
{\cal L}_{\vect \xi} {\vect h}_{E_2}={h}_{E_2, \,\la}^{\, \mu\nu}\xi^\la -
h_{E_2}^{\mu\la}\r_\la\xi^\nu
- h_{E_2}^{\la\nu}\r_\la\xi^\mu=0,
\ee
\be
[{\cal L}_{\vect \xi}, \del^{E_2} ] =0,
\ee
or Eqs.(\ref{e2metric}), (\ref{e2inv}), (\ref{e2conn}) are invariant under
the coordinate transformation
\be \label{e2CoordTrans}
x'=\d {S x}{1+l^{-1}b^T x},
\ee
 and its inverse transformation,
\be
x=\d {S^{-1} x'}{1+l^{-1}b^T S^{-1}x'}=\d {S^{-1}
x'}{1+l^{-1}(b')^T x'},
\ee
where $x$, $x'$, and $b$ are the
$4\times 1$ matrixes, $b^T$ is the transpose of $b$, $x$ and $x'$
have the dimension of length while $b$ is dimensionless, $S\in
SO(4)$, $b'=Sb$.  The points satisfying
\be \label{e2Infinitypts}
1+ l^{-1} b^T x =0
\ee
will be transformed to infinity in the new
coordinate system $x'$ under the coordinate transformation
(\ref{e2CoordTrans}).  Therefore, the infinity point should be in
the manifold $M^{E_2}$.  In contrast, the origin $x=0$
is an invariant point under the transformation, and so has to be
detached from the manifold.

In order to see the manifold more transparently, consider the coordinate transformations,
\be
\label{Transf2Riem}
\begin{cases}
x^0=l^2\rho^{-1} \cos \chi & \\
x^1=l^2\rho^{-1} \sin \chi \sin \th \cos\phi &\\
x^2=l^2\rho^{-1} \sin \chi \sin \th \sin \phi&\\
x^3=l^2\rho^{-1} \sin \chi \cos \th ,
\end{cases}
\ee
where $\tilde x^0=\rho \in (-\infty, +\infty)$, $\tilde x^1=\chi \in [0,\pi],\,
\tilde x^2=\th \in [0,\pi]$, $\tilde x^3=\phi \in [0,2\pi)$.
Under the coordinate transformation, Eqs.(\ref{e2metric}), (\ref{e2inv}), and
(\ref{e2conn}) become
\be
{\vect g}^{E_2}
=l^2(d\chi^2 +\sin^2\chi d\Om_2^2) =\tilde g^{E_2}_{ij}d{\tilde x}^i
d{\tilde x}^j,
\ee
\be \label{e2inv2}
{\vect h}_{E_2}=\left (\d {\r }{\r \rho}\right)^{2},
\ee
\be\begin{cases}
\tilde \Gamma^{\ \ 0}_{E_2ij} = l^{-2}\rho g^{}_{ij}, \qquad \, &\\
\tilde \Gamma^{\ \ i}_{E_2jk} =- \sin\chi \cos\chi \dl^i_{1}(\dl_j^2\dl_k^2
+\sin^2\th \dl_j^3\dl_k^3 )+2 \cot\chi ({ \dl^i_{(j}}
- \dl^i_1\dl^1_{(j})\dl_{k)}^1&\\
\qquad \quad {  + 2\cot\th \dl^i_3\dl_{(j}^2\dl^3_{k)}}-
\sin\th\cos\th\dl^i_2\dl^3_j\dl^3_k ,&\\
\mbox{others vanish,} &\end{cases}
\ee
respectively.  Then, Eqs.(\ref{e2curvaturecomponents}) and (\ref{e2Ricci}) become
\be
\tilde R^{\ \ \si}_{E_2\ \mu \rho \nu}
=-l^{-2}(\dl^\si_\nu {\tilde g^{E_2}}_{\mu\rho}-\dl^\si_\rho
{\tilde g^{E_2}}_{\mu\nu} ),
\ee
\be
\tilde R^{E_2}_{\mu\nu}=\tilde R^{\ \, \si}_{E_2\ \mu \nu \si}=-3l^{-2}
{\tilde g^{E_2}}_{\mu\nu}.
\ee
All these equations show that the manifold is $\mathbb{S}_3\times \mathbb{R}$,
where 3d sphere has the radius $l$.  The signature of $\vect{g}^{E_2}$ and
$\vect{h}^{E_2}$ are $(+,+,+)$ and $(+)$, respectively, and denoted by
$(+,+,+;+)$ for brevity.

For the contraction of $LBdS$ space-time,
\be
\lim_{l_r\to 0}\d {l_r^4}{l^4}(-\si_{E,\,r}^-)
(\dl^{\mu\nu} - l_r^{-2}x^\mu x^\nu) \d{\r }{\r x^\mu}\otimes\d{\r }{\r x^\nu}
=-\vect{h}_{E_2}.
\ee
\be
\lim_{l_r\to 0}\d{(\dl^\la_\mu \dl_{\nu \ka} + \dl^\la_\nu
\dl_{\mu \ka})x^\ka}{l_r^2\si_{E,\,r}^-}=-\d{(\dl^\la_\mu \dl_{\nu \ka} + \dl^\la_\nu
\dl_{\mu \ka})x^\ka}{\dl_{\si\tau}x^\si x^\tau}=\Ga_{E_2\ \mu\nu}^{\ \; \la}.
\ee
The resulting space-time possesses the same topology, the same symmetry, and the
same connection and curvature as the $E_2$ geometry.  And the 4d geometry is also
split into 3d and 1d same geometries.  The only difference is the opposite sign in
$\vect{h}$.  Therefore, it may also be denoted by $E_2$.  In order to see
the contraction path and for convenience in the following discussion, however, we
denote it by $E_{2-}$ and the signature by $(+,+,+;-)$.  (We shall treat other
degenerate geometries in the similar way in the following discussion.)

\subsection{Geometries for $\frak{d}_\pm$, $\frak{p}$,
and $\frak{p}_2$}

The metrics of 4d \dS\  and \AdS\ space-times in a Beltrami coordinate system
are \cite{Guo77, GHXZ}
\be
ds_\pm^2=\d 1 {\si^\pm}\left ( \eta_{\mu \nu} \pm \d {\eta_{\mu\ka}\eta_{\nu \la}x^\ka x^\la}
{l^2\si^\pm}\right ) dx^\mu dx^\nu,
\ee
respectively, where
\be \label{dSdomain}
\si^\pm= 1\mp l^{-2} \eta_{\ka\la}x^\ka x^\la >0.
\ee
If the domain condition (\ref{dSdomain}) is changed to
\be
\si^\pm <0,
\ee
the geometries become
\be
ds_{\pm,<}^2=\pm\d 1 {\si^\pm}\left ( \eta_{\mu \nu} \pm \d {\eta_{\mu\ka}
\eta_{\nu \la}x^\ka x^\la} {l^2\si^\pm}\right ) dx^\mu dx^\nu.
\ee
They are also invariant under the transformations generated by $\frak{d}_\pm$.
The metric with $\si_+<0$ has the signature $(+,+,+,+)$.  It is a \BdS\ model of
\Lob\ space.  Hence, it may be named as $BdSL$ space.  The metric with $\si_-<0$
has the signature $(+,+,-,-)$. It may be called the double time \dS\
space-time, denoted as $DTdS$. All these metrics are non-degenerate and thus contain
all local geometrical information.

Form either \dS\ or \AdS\ space-time, take $l_r\to \infty$ limit,
we get the familiar Minkowski ($Min$) metric
\be
ds_{Min}^2=\lim_{l_r\to \infty}\d 1 {\si_{r}^\pm}\left ( \eta_{\mu \nu} \pm
\d {\eta_{\mu\ka}\eta_{\nu \la}x^\ka x^\la}
{l_r^2\si_{r}^\pm}\right ) dx^\mu dx^\nu=\eta_{\mu \nu} dx^\mu dx^\nu ,
\ee
where
\be
\si_{r}^\pm= 1\mp l_r^{-2} \eta_{\ka\la}x^\ka x^\la >0.
\ee
In the limit of $l_r\to \infty$, the conditions $\si_r^\pm <0$ will be violated.
Thus, $BdSL$ space and $DTdS$ space-time are uncontractible.

In the limit of $l_r\to 0$,
\be \label{si_l_to0}
l_{r}^2\si_r^\pm= \mp \eta_{\ka\la}x^\ka x^\la >0,
\ee
the metrics of \dS\ and \AdS\ space-times reduce to the degenerate ones
\be \label{p2metric}
ds_{P^{}_{2\pm}}^2=\lim_{l_r\to 0}\d {l^2}{l_r^2}\d 1 {\si_{r}^\pm}\left ( \eta_{\mu \nu}
\pm
\d {\eta_{\mu\ka}\eta_{\nu \la}x^\ka x^\la}
{l_r^2\si_{r}^\pm}\right ) dx^\mu dx^\nu=\pm l^2\d {\eta_{\mu\ka}\eta_{\nu\la}-
\eta_{\mu \nu}\eta_{\ka \la}}{(\eta_{\si\tau}x^\si x^\tau)^2}x^\ka x^\la dx^\mu dx^\nu
=:\vect{g}^{P^{}_{2\pm}}.
\ee
The upper sign corresponds to the limit of \dS\ space-time while
the lower sign corresponds to the limit of \AdS\ space-time.  Similar to the second
Euclid case, to determine the geometry completely, we should
consider the limit of the inverse metrics
\be \label{p2inv}
\left (\d {\r}{\r s}\right )_{P^{}_{2\pm}}^{2}
=\lim_{l_r\to 0}\d {l_r^4}{l^4}\si_{r}^\pm
(\eta^{\mu\nu} \mp l_r^{-2}x^\mu x^\nu) \d{\r }{\r x^\mu}\otimes\d{\r }{\r x^\nu}
=l^{-4}\eta_{\si\tau}x^\si x^\tau x^\mu x^\nu \d{\r }{\r x^\mu}\otimes\d{\r }{\r x^\nu}
=: \vect{h}_{P^{}_{2\pm}},
\ee
and the connection with the limit of connection coefficients
\be \label{p2conn}
{\Ga_{P^{}_{2\pm}}}^{\la}_{\mu\nu}=\pm \lim_{l_r\to 0}\d{(\dl^\la_\mu \eta_{\nu \ka}
+ \dl^\la_\nu
\eta_{\mu \ka})x^\ka}{l_r^2\si_{r}^\pm}=-\d{(\dl^\la_\mu \eta_{\nu \ka} + \dl^\la_\nu
\eta_{\mu \ka})x^\ka}{\eta_{\si\tau}x^\si x^\tau}.
\ee
It is easy to check that the connection $\del^{P^{}_{2\pm}}$ with coefficients
(\ref{p2conn}) is compatible with Eq.(\ref{p2metric}) and (\ref{p2inv}), {\it i.e.}
\be
\del^{P^{}_{2\pm}}_\la \, g^{P^{}_{2\pm}}_{\mu\nu}= \r_\la g^{P^{}_{2\pm}}_{\mu\nu}
-{\Gamma_{P^{}_{2\pm}}}^\ka_{\la\nu}g^{P^{}_{2\pm}}_{\mu\ka}
-{\Gamma_{P^{}_{2\pm}}}^\ka_{\mu\la}g^{P^{}_{2\pm}}_{\ka\nu} =0
\ee
and
\be
\del^{P^{}_{2\pm}}_\la \, {h}_{P^{}_{2\pm}}^{\mu\nu}= \r_\la {h}^{\mu\nu}_{P^{}_{2
\pm}}+ {\Gamma_{P^{}_{2\pm}}}^\nu_{\la\ka}{h}^{\mu\ka}_{P^{}_{2\pm}}
+{\Gamma_{P^{}_{2\pm}}}^\mu_{\la\ka}{h}^{\ka\nu}_{P^{}_{2\pm}}=0,
\ee
respectively.  It can also be shown that $\{M^{P^{}_{2\pm}},\vect{g}^{P^{}_{2\pm}},
\vect{h}_{P^{}_{2\pm}},\del^{P^{}_{2\pm}}\}$ is invariant under $P_2$ transformations,
namely, $\forall \vect\xi \in {\frak p}_2\subset TM^{P^{}_{2\pm}}$
\be
{\cal L}_{\vect \xi} g^{P^{}_{2\pm}}_{\mu\nu} =  g^{P^{}_{2\pm}}_{\mu\nu,\la}\xi^\la
+g^{P^{}_{2\pm}}_{\mu\la}\r_\nu\xi^\la +g^{P^{}_{2\pm}}_{\la\nu}\r_\mu\xi^\la=0,
\ee
\be
{\cal L}_{\vect \xi} h_{P^{}_{2\pm}}^{\mu\nu}=\xi^\la\r_\la h_{P^{}_{2\pm}}^{\mu\nu}
-h_{P^{}_{2\pm}}^{\mu\la} \r_\la\xi^\nu - h_{P^{}_{2\pm}}^{\la\nu}\r_\la\xi^\mu=0,
\ee
\be
[{\cal L}_{\vect \xi}, \del^{P^{}_{2\pm}} ] =0,
\ee
or Eqs.(\ref{p2metric}), (\ref{p2inv}), and (\ref{p2conn}) are invariant under
the coordinate transformations
\be
x'=\d {L x}{1+l^{-1}b^T x }
\ee
and its inverse transformation,
\be
x=\d {L^{-1} x'}{1+l^{-1}b^T L^{-1}x'}=\d {L^{-1} x'}{1+l^{-1}{b'}^T x'},
\ee
where $L$ is the Lorentz transformation, $b$ is dimensionless,
$b'=Lb$, a superscript $T$ stands for the transpose under the metric $\eta_{\mu\nu}=
{\rm diag}(1,-1,-1,-1)$.

By definition, the curvature tensor is
\be
{R_{P^{}_{2\pm}}}^\si_{\ \mu \rho \nu}
=\pm l^{-2}(\dl^\si_\nu g^{P^{}_{2\pm}}_{\mu\rho}
-\dl^\si_\rho g^{P^{}_{2\pm}}_{\mu\nu}).\quad \label{p2curvaturecomponents}
\ee
The Ricci curvature tensor is then
\be \label{p2riccicurv}
R^{P^{}_{2\pm}}_{\mu\nu}={R_{P^{}_{2\pm}}}^\si_{\ \mu\nu\si}=\pm 3
l^{-2}g^{P^{}_{2\pm}}_{\mu\nu}.
\ee
They are obviously invariant under $P_2$ transformation.

The structure of the space-times has been analyzed in \cite{Huang}.
The manifold is $dS_3\times \mathbb{R}$ for $x\cdot x<0$ and $\mathbb{H}_3\times \mathbb{R}$
for $x\cdot x >0$ and has the signature $(+,-,-;-)$ and $(-,-,-;+)$, respectively.

In the limit of $l_r\to 0$,
the  $BdSL$ space and $DTdS$ space-time contract to
\be
ds_{\substack{EP_{2-}\\DTP_{2+}}}^{2}&=&\pm\lim_{l_r\to 0}\d {l^2}{l_r^2}
\d 1 {\si_{r}^\pm}\left ( \eta_{\mu \nu}\pm
\d {\eta_{\mu\ka}\eta_{\nu \la}x^\ka x^\la}
{l_r^2\si_{r}^\pm}\right ) dx^\mu dx^\nu= l^2\d {\eta_{\mu\ka}\eta_{\nu\la}-
\eta_{\mu \nu}\eta_{\ka \la}}{(\eta_{\si\tau}x^\si x^\tau)^2}x^\ka x^\la dx^\mu dx^\nu
\nno \\
&=&\begin{cases}l^2\left [ \left (\d {d(r/ct)}{1-r^2/c^2t^2}\right )^2+
\d {r^2/c^2t^2}{1-r^2/c^2t^2} d\Om_2^2\right ]
& \qquad  \eta_{\mu\nu} x^\mu x^\nu >0 \smallskip\\
l^2\left [ \left (\d {d(r/ct)}{r^2/c^2t^2-1}\right )^2-
\d {r^2/c^2t^2}{r^2/c^2t^2-1}
d\Om_2^2\right ]& \qquad \eta_{\mu\nu} x^\mu x^\nu <0
\end{cases},
\ee
\be
\left (\d \r {\r s}\right )_{\substack{EP_{2-}\\DTP_{2+}}}^{2}&=&
\pm\lim_{l_r\to 0}\d {l_r^4}{l^4}\si_{r}^\pm
(\eta^{\mu\nu} \mp l_r^{-2}x^\mu x^\nu) \d{\r }{\r x^\mu}\otimes\d{\r }{\r x^\nu}
=\pm l^{-4}\eta_{\si\tau}x^\si x^\tau x^\mu x^\nu \d{\r }{\r x^\mu}\otimes\d{\r }{\r x^\nu}
\nno \\
&=&\begin{cases} \left(\d {\r}{\r (c^2t^2-r^2)^{-1/2}}\right )\otimes
\left(\d {\r}{\r (c^2t^2-r^2)^{-1/2}}\right ) & \qquad \eta_{\mu\nu}x^\mu
x^\nu >0 \\
\left (\d {\r}{\r (r^2-c^2t^2)^{-1/2}}\right )\otimes
\left (\d {\r}{\r (r^2-c^2t^2)^{-1/2}}\right )& \qquad \eta_{\mu\nu}x^\mu
x^\nu <0 \end{cases},
\ee
\be
\Ga_{\substack{EP_{2-}\\DTP_{2+}}\mu\nu}^{\qquad \la}=\pm \lim_{l_r\to 0}
\d{(\dl^\la_\mu \eta_{\nu \ka} + \dl^\la_\nu
\eta_{\mu \ka})x^\ka}{l_r^2\si_{r}^\pm}=-\d{(\dl^\la_\mu \eta_{\nu \ka} + \dl^\la_\nu
\eta_{\mu \ka})x^\ka}{\eta_{\si\tau}x^\si x^\tau}
\ee
where $r^2:=\dl_{ij}x^i x^j$ and $d\Om_2^2$ is the line-element of 2d unit sphere.
They have the forms of $\vect{g}^{P^{}_{2+}}$,
$\pm \vect{h}_{P^{}_{2\pm}}$,
and $\Gamma_{P_{2\pm}}$ but with
\be \label{si_l_to0<}
l_{r}^2\si_r^\pm \to \mp \eta_{\ka\la}x^\ka x^\la <0.
\ee
The contraction of $BdSL$ space is $\mathbb{H}_3\times \mathbb{R}$ with
the signature $(+,+,+;+)$.  It is the Euclidean version of $P_{2-}$ space-time,
denoted by $EP_{2-}$. The contraction of $DTdS$ space is $dS_3\times \mathbb{R}$ with
the signature $(+,-,-;+)$.  It is the double time version of $P_{2+}$ space-time,
denoted by $DTP_{2+}$.

\subsection{Geometries for $\frak{n}_\pm$ and $\frak{n}_{\pm 2}$}

The $NH_\pm$ space-times can be obtained from \dS\ and \AdS\ space-times by the
contraction in the limit of $c_r, l_r\to \infty$ but $\nu=c_r/l_r$ fixed \cite{NH}.
When  $c_r, l_r\to \infty$ and $\nu=c_r/l_r=c/l$,
\be
\si_{\frak{n}}^\pm:=\lim_{\substack{c_r,\, l_r\to \infty \\ \nu\ {\rm fixed}}}
\si_{r}^\pm=\lim_{\substack{c_r,\, l_r\to \infty \\ \nu\ {\rm fixed}}}
[1\mp l_r^{-2} (c_r^2t^2-\dl_{ij}x^i x^j)]=1\mp l^{-2} (x^0)^2 >0,
\ee
the metrics, inverse metrics, and connection coefficients become
\be \label{n_pm-metric}
ds^{2}_{NH_\pm} = \lim_{\substack{c_r,\, l_r\to \infty \\ \nu\ {\rm fixed}}}
\d {c^2}{c_r^2}
\d 1 {\si_r^\pm}\left ( \eta_{\mu \nu} \pm \d {\eta_{\mu\ka}\eta_{\nu \la}x_r^\ka x_r^\la}
{l_r^2\si_r^\pm}\right ) dx_r^\mu dx_r^\nu = \d {c^2}{(\si_{\frak{n}}^\pm)^2}dt^2 =:
\vect{g}^{NH_\pm},
\ee
\be \label{n_pm-inv}
\left (\d {\r}{\r s}\right )_{NH_\pm}^{2}=\lim_{\substack{c_r,\,l_r\to \infty\\
\nu\ {\rm fixed}}}\si_{r}^\pm
(\eta^{\mu\nu} \mp l_r^{-2}x_r^\mu x_r^\nu) \d{\r }{\r x_r^\mu}\otimes\d{\r }{\r x_r^\nu}
=-\si_{\frak{n}}^\pm\dl^{ij}\d{\r }{\r x^i}\otimes\d{\r }{\r x^j}
=: \vect{h}_{NH_\pm},
\ee
\be \label{n_pm-conn}
{\Ga_{NH_\pm}}^{\la}_{\mu\nu}=\pm \lim_{\substack{c_r,\,l_r\to \infty \\
\nu\ {\rm fixed}}}\d {c_r}{c}\d{(\dl^\la_\mu \eta_{\nu \ka} + \dl^\la_\nu
\eta_{\mu \ka})x_r^\ka}{l_r^2\si_{r}^\pm}
=\pm\d{(\dl^\la_\mu \eta_{\nu 0} + \dl^\la_\nu
\eta_{\mu 0})x^0}{l^2\si_{\frak{n}}^\pm},
\ee
where $x_r^\mu$ means $(c_r t, x^i)$.  The nonzero connection coefficients
are\footnote{${\Ga_{NH_\pm}}^{t}_{tt}
=c{\Ga_{NH_\pm}}^{0}_{00}, {\Ga_{NH_\pm}}^{i}_{tj}=
c{\Ga_{NH_\pm}}^{i}_{0j}$ are used  in \cite{NH}.}
\be
{\Ga_{NH_\pm}}^{0}_{00}=\pm\d{2 x^0}{l^2 \si_{\frak{n}}^\pm}, \qquad
{\Ga_{NH_\pm}}^{i}_{0j}={\Ga_{NH_\pm}}^{i}_{j0}=\pm\d{ x^0}
{l^2 \si_{\frak{n}}^\pm}\dl^i_j .
\ee
The non-zero components of curvature tensor and Ricci tensor are given by
\be
R^{\ \ \la}_{{\frak n}_\pm\ 0\mu \nu}=\pm l^{-2}(\dl^\la_\nu g^{NH_\pm}_{\ 0\mu}
-\dl^\la_\mu g^{NH_\pm}_{0\nu}),
\ee
\be
R_{00}=\pm 3 l^{-2} g^{NH_\pm}_{\ \ 00},
\ee
respectively.  The $NH_\pm$ space-times are the non-relativistic space-times.  They are the
generalization of Galilei space-time and have absolute-time.  In terms of
Beltrami coordinates, the time is ``curved"  and
the space is conformally flat.  The conformal factor is independent of
point in the space.  It has been shown that $\{M^{NH_\pm},\vect{g}^{NH_\pm},
\vect{h}_{NH_\pm}, \del^{NH_\pm}\}$ is invariant under $NH_\pm$
transformations \cite{NH}.  The mechanics on the $NH_+$ space-times has been studied
in details in \cite{NH, NHmore}.

The contractions of 4d \Lob\ and 4d \Riem\ spaces in the same limit give
the same domain conditions
\be
\si_{\frak n}^\pm = \lim_{\substack{c_r,\, l_r\to \infty \\ \nu\ {\rm fixed}}}\si_{E,r}^\mp
=1 \mp l^{-2} (x^0)^2 >0,
\ee
the same covariant degenerate metrics
\be
ds_{ENH_\pm}^{2}=\lim_{\substack{c_r,\, l_r\to \infty \\ \nu\ {\rm fixed}}}\d 1
 {\si_{E,r}^\mp}\left ( \dl_{\mu \nu} \pm \d {\dl_{\mu\ka}\dl_{\nu \la}x_r^\ka x_r^\la}
{l^2\si_{E,r}^\mp}\right ) dx_r^\mu dx_r^\nu = \d {c^2} {(\si_{\frak n}^\pm)^2} dt^2
=:\vect{g}^{ENH_\pm} =\vect{g}^{NH_\pm},
\ee
and the same connection coefficients
\be
{\Ga_{ENH_\pm}}^{\la}_{\mu\nu}=\pm \lim_{\substack{c_r,\,l_r\to \infty \\
\nu\ {\rm fixed}}}\d {c_r}{c}\d{(\dl^\la_\mu \dl_{\nu \ka} + \dl^\la_\nu
\dl_{\mu \ka})x_r^\ka}{l_r^2\si_{E,r}^\mp}
=\pm\d{(\dl^\la_\mu \dl_{\nu 0} + \dl^\la_\nu
\dl_{\mu 0})x^0}{l^2\si_{\frak{n}}^\pm}={\Ga_{NH_\pm}}^{\la}_{\mu\nu}.
\ee
The contravariant degenerate metrics are
\be
\lim_{\substack{c_r,\,l_r\to \infty\\
\nu\ {\rm fixed}}}
\si_{E,\,r}^\mp
(\dl^{\mu\nu} \mp l_r^{-2}x_r^\mu x_r^\nu) \d{\r }{\r x_r^\mu}\otimes\d{\r }{\r x_r^\nu}
=\si_{\frak{n}}^\pm\dl^{ij}\d{\r }{\r x^i}\otimes\d{\r }{\r x^j} =:\vect{h}_{ENH_\pm}
=-\vect{h}_{NH_\pm}.
\ee
Since the signatures of $\vect{g}^{NH_\pm}$ and $-\vect{h}_{NH_\pm}$
are $(+)$ and $(+,+,+)$, respectively,
the geometries obtained from the contraction of \Lob\ and \Riem\ geometries are
called Euclidean $NH_\pm$ ($ENH_\pm$) space-times.

The domain conditions
\be
\left . \begin{array}{c}\displaystyle \lim_{\substack{c_r,\, l_r\to \infty \\
\nu\ {\rm fixed}}}\si_{E,r}^-\\
\displaystyle \lim_{\substack{c_r,\, l_r\to \infty \\ \nu\ {\rm fixed}}}\si_{r}^+
\end{array}\right \}
=1 - l^{-2} (x^0)^2 =\si_{\frak n}^+<0
\ee
means $ t^2>\nu^{-2}$.  In this case, the $LBdS$ space-time contracts to
\be
ds_{NH_+'}^{2}&=&-\lim_{\substack{c_r,\, l_r\to \infty \\ \nu\ {\rm fixed}}}
\d 1 {\si_{E,r}^-}\left ( \dl_{\mu \nu} +\d {\dl_{\mu\ka}
\dl_{\nu \la}x_r^\ka x_r^\la}
{l_r^2\si_{E,r}^-}\right ) dx_r^\mu dx_r^\nu = -\d {c^2 dt^2}
{(\si_{\frak n}^+)^2}\omits{=:-\vect{g}^{\frak{n}_+'}} , \\
\left(\d {\r}{\r s}\right )_{NH_+'}^{2}&=&-\lim_{\substack{c_r,\, l_r\to \infty \\ \nu\ {\rm fixed}}}\si_{E,\,r}^-
(\dl^{\mu\nu} - l_r^{-2}x_r^\mu x_r^\nu) \d{\r }{\r x_r^\mu}\otimes\d{\r }{\r x_r^\nu}
=-\si_{\frak{n}}^+\dl^{ij}\d{\r }{\r x^i}\otimes\d{\r }{\r x^j},\\
{\Ga_{NH_+'}}^\la_{\mu\nu}&=& \lim_{\substack{c_r,\,l_r\to \infty \\
\nu\ {\rm fixed}}}\d {c_r}{c}\d{(\dl^\la_\mu \dl_{\nu \ka} + \dl^\la_\nu
\dl_{\mu \ka})x_r^\ka}{l_r^2\si_{E,r}^-}
=\d{(\dl^\la_\mu \dl_{\nu 0} + \dl^\la_\nu
\dl_{\mu 0})x^0}{l^2\si_{\frak{n}}^+}\omits{=:{\Ga_{\frak{n}_+'}}^{\la}_{\mu\nu}}.
\ee
Except an overall minus, the degenerate covariant and contravariant metrics have the
same form as $ENH_+$ but the same signature as $NH_+$ because $\si_{\frak n}^+<0$.
In order to distinguish it from $NH_+$ space-time, it may be called para-$NH_+$ ($NH_+'$)
 space-time.
The $BdSL$ space-time contracts to
\be
ds_{ENH'_+}^{2}&=&\lim_{\substack{c_r,\, l_r\to \infty \\ \nu\ {\rm fixed}}}
\d 1 {\si_{r}^+}\left ( \eta_{\mu \nu} +\d {\eta_{\mu\ka}
\eta_{\nu \la}x_r^\ka x_r^\la}
{l_r^2\si_{r}^+}\right ) dx_r^\mu dx_r^\nu = \d {c^2 dt^2}
{(\si_{\frak n}^+)^2}\omits{=\vect{g}^{\frak{n}_+'}} , \\
\left(\d {\r}{\r s}\right )_{ENH_+'}^{2}&=&\lim_{\substack{c_r,\, l_r\to \infty \\ \nu\ {\rm fixed}}}\si_{r}^+
(\eta^{\mu\nu} - l_r^{-2}x_r^\mu x_r^\nu) \d{\r }{\r x_r^\mu}\otimes\d{\r }{\r x_r^\nu}
=-\si_{\frak{n}}^+\dl^{ij}\d{\r }{\r x^i}\otimes\d{\r }{\r x^j}\omits{=\vect{h}_{\frak{n}_+'}},\\
\Ga_{NH_+' \mu \nu}^{\quad\ \la}&=& \lim_{\substack{c_r,\,l_r\to \infty \\
\nu\ {\rm fixed}}}\d {c_r}{c}\d{(\dl^\la_\mu \eta_{\nu \ka} + \dl^\la_\nu
\eta_{\mu \ka})x_r^\ka}{l_r^2\si_{r}^+}
=\d{(\dl^\la_\mu \eta_{\nu 0} + \dl^\la_\nu
\eta_{\mu 0})x^0}{l^2\si_{\frak{n}}^+}\omits{={\Ga_{\frak{n}_+'}}^{\la}_{\mu\nu}}.
\ee
The signature is $(+;+,+,+)$.  It is the Euclidean version of para-$NH_+$ space-time,
denoted by $ENH_+'$.  Finally, $DTdS$ is uncontractible in this case because
$\si_r^- <0$ is not preserved in the limit.

On the other hand, in the limit $c_r, l_r\to 0$ but $\nu=c_r/l_r$
fixed,
\be
\left . \begin{array}{l}
\displaystyle\lim_{\substack{c_r,\, l_r\to 0 \\
\nu\ {\rm fixed}}} \d {l_r^2}{l^2}\si_{r}^\pm=\displaystyle\lim_{\substack{c_r,\, l_r\to 0
\\ \nu\ {\rm fixed}}} \d {l_r^2}{l^2}[1\mp l_r^{-2}
(c_r^2t^2-\dl_{ij}x^i x^j)] \\
\displaystyle\lim_{\substack{c_r,\, l_r\to 0 \\
\nu\ {\rm fixed}}} \d {l_r^2}{l^2}\si_{E,r}^\pm=\displaystyle\lim_{\substack{c_r,\, l_r\to 0
\\ \nu\ {\rm fixed}}} \d {l_r^2}{l^2}[1\pm l_r^{-2}
(c_r^2t^2+\dl_{ij}x^i x^j)] \end{array} \right \}= \pm l^{-2}\dl_{ij}x^i x^j=
\si_{\frak{n}_2}^\pm .
\ee
The inequalities $\si^+_r > 0$, $\si_{E,r}^- < 0$, $\si_{E,r}^+>0$,
and $\si^-_r<0$ are always valid in the limit.  Thus, the second $NH$ geometry ($NH_{2}$),
the second para-$NH$ geometry ($NH_2'$), the second Euclidean $NH$ geometry ($ENH_{2}$),
and the second double time $NH$ space-time ($DTNH_2$)
can be obtained from the contraction of $dS$, $LBdS$, $Riem$, and $DTdS$ geometries,
respectively.

For the $NH_{2}$ geometry, the covariant degenerate metric, contravariant
degenerate metric, and connection coefficients are
\be
\label{n_2-metric}
ds_{NH_{2}}^{2} = \lim_{\substack{c_r,\,
l_r\to 0 \\ \nu\ {\rm fixed}}}\d {l^2}{l_r^2} \d 1 {\si_r^+}\left (
\eta_{\mu \nu} + \d {\eta_{\mu\ka}\eta_{\nu \la}x_r^\ka x_r^\la}
{l_r^2\si_r^+}\right ) dx_r^\mu dx_r^\nu = l^2\d
{(\dl_{ik}\dl_{jl}-\dl_{ij}\dl_{kl})x^k x^l}
{(\dl_{mn}x^m x^n)^2}dx^i dx^j =:\vect{g}^{NH_{2}},
\ee
\be \label{n_2-inv}
\left (\d {\r}{\r s}\right )_{NH_{2}}^{2}&=&\lim_{\substack{c_r,\,l_r\to 0\\
\nu\ {\rm fixed}}}\d {l_r^4}{l^4}\si_{r}^+
(\eta^{\mu\nu} - l_r^{-2}x_r^\mu x_r^\nu) \d{\r }{\r x_r^\mu}\otimes\d{\r }{\r x_r^\nu}\nno \\
&=&\d {\dl_{mn}x^m x^n}{l^4}\left (\nu^{-2}(1-\nu^2t^2)\d{\r }{\r t}\otimes\d{\r }{\r t} -
x^i x^j \d{\r }{\r x^i}\otimes\d{\r }{\r x^j}-2tx^i\d{\r }{\r t}\otimes\d{\r }{\r x^j}\right )
\omits{\nno \\
&=:&h_{NH_{2}}^{\mu\nu}\d{\r }{\r x^\mu}\d{\r }{\r x^\nu}} \nno \\
&=:& \vect{h}_{NH_{2}},
\ee
\be \label{n_2-conn}
{\Ga_{NH_{2}}}^{\la}_{\mu\nu}= \lim_{\substack{c_r,\,l_r\to 0 \\
\nu\ {\rm fixed}}}\d{(\dl^\la_\mu \eta_{\nu \ka} + \dl^\la_\nu
\eta_{\mu \ka})x^\ka}{l_r^2\si_{r}^+}
=\d{(\dl^\la_\mu \eta_{\nu i} + \dl^\la_\nu
\eta_{\mu i})x^i}{\dl_{mn}x^m x^n},
\ee
respectively.
The rank of the degenerate metric (\ref{n_2-metric}) is 2, which may serve as the
non-degenerate metric of the sphere $\mathbb{S}_2$ with the radius $l$.  The rank of
the degenerate ``inverse" metric (\ref{n_2-inv}) is also 2.  Their signatures are
$(-,-)$ and $(+,-)$, respectively, denoted by $(-,-;+,-)$.  The coordinate $c^2t^2$
serves as a timelike coordinate.  The non-zero connection
coefficients are
\be
{\Ga_{NH_{2}}}^{0}_{0i}={\Ga_{NH_{2}}}^{0}_{i0}=
-\d{\dl_{ij}x^j}{\dl_{mn}x^m x^n},\qquad {\Ga_{NH_{2}}}^{i}_{jk}=-\d{(\dl^i_j \dl_{kl}
+ \dl^i_k \dl_{jl})x^l}{\dl_{mn}x^m x^n}.
\ee
The nonzero components of curvature tensor and Ricci tensor are
\be
{R_{NH_{2}}}^{0}_{\ i 0 j }=-{R_{NH_{2}}}^{0}_{\ ij 0}
= - l^{-2}g^{NH_{2}}_{ij}, \qquad
{R_{NH_{2}}}^{k}_{\ i l j }=  l^{-2}(\dl^k_j g^{NH_{2}}_{i l}
-\dl^k_l g^{NH_{2}}_{ij}),
\ee
\be
R^{NH_{2}}_{ij}=3l^{-2}g^{NH_{2}}_{ij},
\ee
respectively.

Under the coordinate transformation
\be
\begin{cases}
x^0= \psi r/l & \\
x^1=r \sin \th \cos\phi &\\
x^2=r \sin \th \sin \phi&\\
x^3=r \cos \th ,
\end{cases}
\ee
Eqs.(\ref{n_2-metric}) and (\ref{n_2-inv}) become
\be
ds_{NH_{2}}^{2}=-l^2(d\th^2+\sin^2 d\phi^2),
\ee
\be
\left (\d {\r}{\r s}\right )_{NH_{2}}^{2}
=\d{\r}{\r \psi}\otimes \d{\r}{\r \psi} - \d{r^4}{l^4} \d{\r}{\r r}\otimes
\d{\r}{\r r},
\ee
respectively.  Obviously, in $NH_2$ space-time, there is no geometrical
SO(3) isotropy with respect to any fixed point in $M^{NH_{2}}$.
The $NH_2$ space-time describes the Beltrami coordinate region
$\delta_{ij} x^i x^j>0$.   It implies that the spatial point $x^1=x^2=x^3=0$
is not in the manifold.

The contraction from the $LBdS$ gives rise to the $NH_{2}'$ geometry:
\be
ds_{NH_{2}'}^{2}=- \lim_{\substack{c_r,\,
l_r\to 0 \\ \nu\ {\rm fixed}}}\d {l^2}{l_r^2} \d 1 {\si_{E,r}^-}\left (
\dl_{\mu \nu} + \d {\dl_{\mu\ka}\dl_{\nu \la}x_r^\ka x_r^\la}
{l_r^2\si_{E,r}^-}\right ) dx_r^\mu dx_r^\nu = l^2\d
{(\dl_{ij}\dl_{kl}-\dl_{ik}\dl_{jl})x^k x^l}
{(\dl_{mn}x^m x^n)^2}dx^i dx^j =-\vect{g}^{NH_{2}}, \nno \\
\ee
\be
\left (\d {\r}{\r s}\right )_{NH_{2}'}^{2}&=&-\lim_{\substack{c_r,\,l_r\to 0\\
\nu\ {\rm fixed}}}\d {l_r^4}{l^4}\si_{E,r}^-
(\dl^{\mu\nu} - l_r^{-2}x_r^\mu x_r^\nu) \d{\r }{\r x_r^\mu}\otimes\d{\r }{\r x_r^\nu}\nno \\
&=&\d {\dl_{mn}x^m x^n}{l^4}\left (\nu^{-2}(1-\nu^2t^2)\d{\r }{\r t}\otimes\d{\r }{\r t} -
x^i x^j \d{\r }{\r x^i}\otimes\d{\r }{\r x^j}-2tx^i\d{\r }{\r t}\otimes\d{\r }{\r x^j}
\right )
= \vect{h}_{NH_{2}},\qquad\
\ee
\be \label{n_-2-conn}
{\Ga_{NH_{2}'}}^{\la}_{\mu\nu}= \lim_{\substack{c_r,\,l_r\to 0 \\
\nu\ {\rm fixed}}}\d{(\dl^\la_\mu \dl_{\nu \ka} + \dl^\la_\nu
\dl_{\mu \ka})x^\ka}{l_r^2\si_{E,r}^-}
=-\d{(\dl^\la_\mu \dl_{\nu i} + \dl^\la_\nu
\dl_{\mu i})x^i}{\dl_{mn}x^m x^n}.
\ee
The resulting $-\vect{g}^{NH_{2}}$ and $\vect{h}_{NH_{2}}$ have rank 2
each and signature $(+,+)$ and $(+,-)$, respectively, denoted by $(+,+;+,-)$.
In this case, the coordinate $\rho=(\dl_{\mu\nu}x^\mu x^\nu)^{1/2}$ rather than
$x^0=ct$ serves as a timelike coordinate.

The contraction of $Riem$ geometry defines the $ENH_2$ geometry:
\be
ds_{ENH_{2}}^{2}= \lim_{\substack{c_r,\,
l_r\to 0 \\ \nu\ {\rm fixed}}}\d {l^2}{l_r^2} \d 1 {\si_{E,r}^+}\left (
\dl_{\mu \nu} - \d {\dl_{\mu\ka}\dl_{\nu \la}x_r^\ka x_r^\la}
{l_r^2\si_{E,r}^+}\right ) dx_r^\mu dx_r^\nu = l^2\d
{(\dl_{ij}\dl_{kl}-\dl_{ik}\dl_{jl})x^k x^l}
{(\dl_{mn}x^m x^n)^2}dx^i dx^j =-\vect{g}^{NH_{2}},\qquad
\ee
\be
\left (\d {\r}{\r s}\right )_{ENH_{2}}^{2}&=&\lim_{\substack{c_r,\,l_r\to 0\\
\nu\ {\rm fixed}}}\d {l_r^4}{l^4}\si_{E,r}^+
(\dl^{\mu\nu} + l_r^{-2}x_r^\mu x_r^\nu) \d{\r }{\r x_r^\mu}\d{\r }{\r x_r^\nu}\nno \\
&=&\d {\dl_{mn}x^m x^n}{l^4}\left (\nu^{-2}(1+\nu^2t^2)\d{\r }{\r t}\otimes \d{\r }{\r t} +
x^i x^j \d{\r }{\r x^i}\otimes \d{\r }{\r x^j}+2tx^i\d{\r }{\r t}\otimes
\d{\r }{\r x^j}\right )\nno \\
&=:&\omits{h_{NH_{2}}^{\mu\nu}\d{\r }{\r x^\mu}\d{\r }{\r x^\nu}}
 \vect{h}_{ENH_{2}},
\ee
\be \label{n_-2-conn}
{\Ga_{ENH_{2}}}^\la_{\mu\nu}= -\lim_{\substack{c_r,\,l_r\to 0 \\
\nu\ {\rm fixed}}}\d{(\dl^\la_\mu \dl_{\nu \ka} + \dl^\la_\nu
\dl_{\mu \ka})x^\ka}{l_r^2\si_{E,r}^\pm}
=-\d{(\dl^\la_\mu \dl_{\nu i} + \dl^\la_\nu
\dl_{\mu i})x^i}{\dl_{mn}x^m x^n}.
\ee
The resulting $-\vect{g}^{NH_{2}}$ and $\vect{h}_{ENH_{2}}$ have rank 2
and signature $(+,+)$ and
$(+,+)$, respectively.  It describes a pure geometry
and may serve as the Euclidean version
of both $NH_2$ and $NH_2'$ geometries.

The similar contraction of $DTdS$ space-time gives
\be
ds_{DTNH_{2}}^{2}&=&-\lim_{\substack{c_r,\,
l_r\to 0 \\ \nu\ {\rm fixed}}}\d {l^2}{l_r^2} \d 1 {\si_{r}^-}\left (
\eta_{\mu \nu} - \d {\eta_{\mu\ka}\eta_{\nu \la}x_r^\ka x_r^\la}
{l_r^2\si_{r}^-}\right ) dx_r^\mu dx_r^\nu = l^2\d
{(\dl_{ik}\dl_{jl}-\dl_{ij}\dl_{kl})x^k x^l}
{(\dl_{mn}x^m x^n)^2}dx^i dx^j \nno \\
&=&\vect{g}^{NH_{2}},
\ee
\be
\left (\d {\r}{\r s}\right )_{DTNH_{2}}^{2}&=&-\lim_{\substack{c_r,\,l_r\to 0\\
\nu\ {\rm fixed}}}\d {l_r^4}{l^4}\si_{r}^-
(\eta^{\mu\nu} + l_r^{-2}x_r^\mu x_r^\nu) \d{\r }{\r x_r^\mu}\otimes\d{\r }{\r x_r^\nu}\nno \\
&=&\d {\dl_{mn}x^m x^n}{l^4}\left (\nu^{-2}(1+\nu^2t^2)\d{\r }{\r t}\otimes \d{\r }{\r t} +
x^i x^j \d{\r }{\r x^i}\otimes\d{\r }{\r x^j}+2tx^i\d{\r }{\r t}\otimes \d{\r }{\r x^j}
\right ) \nno \\
&=& \vect{h}_{DTNH_{2}},
\ee
\be
{\Ga_{DTNH_{2}}}^\la_{\mu\nu}= -\lim_{\substack{c_r,\,l_r\to 0 \\
\nu\ {\rm fixed}}}\d{(\dl^\la_\mu \eta_{\nu \ka} + \dl^\la_\nu
\eta_{\mu \ka})x^\ka}{l_r^2\si_{r}^-}
=\d{(\dl^\la_\mu \eta_{\nu i} + \dl^\la_\nu
\eta_{\mu i})x^i}{\dl_{mn}x^m x^n}.
\ee
The resulting $\vect{g}^{NH_{2}}$ and $\vect{h}_{DTNH_{2}}$ have rank 2 each
and the signature $(-,-)$ and
$(+,+)$, respectively.  It describes a double time space-time, named $DTNH_2$ space-time.

\subsection{Geometries for $\frak{h}_\pm$, $\frak{e}'$, and $\frak{p}'$
\label{Sec:HN}}

The $HN_\pm$ space-times can be obtained from the \dS\ and \AdS\ space-times
by the
contraction in the limit of $c_r\to 0$.
When  $c_r\to 0$,
\be
\lim_{c_r\to 0}
\si_{r}^\pm=\lim_{c_r\to 0 }
[1\mp l^{-2} (c_r^2t^2-\dl_{ij}x^i x^j)]=1\pm l^{-2}\dl_{ij}x^i x^j=
\si_{E,\,3}^\pm (x^i),
\ee
which is 3d $\si_E^\pm$.  The metrics and ``inverse" metrics become
\be \label{gh'1}
ds_{HN_\pm}^{2}=\lim_{c_r\to 0} \d 1 {\si_r^\pm} \left (\eta_{\mu\nu} \pm \d
{\eta_{\mu\ka}\eta_{\nu\la}x_r^\ka x_r^\la} {l^2\si_r^\pm}\right ) dx_r^\mu dx_r^\nu
=-\d 1 {\si_{E,\,3}^\pm}\left (\dl_{ij} \mp \d
{\dl_{ik}\dl_{jl}x^k x^l} {l^2\si_{E,\,3}^\pm}\right ) dx^i dx^j=:{\vect g}^{HN_\pm},\quad
\ee
\be \label{gh'2}
\left (\d {\r}{\r s}\right)_{HN_\pm}^{2}=\lim_{c_r\to 0} \frac {c_r^2}{c^2}
\left (\si_r^\pm
(\eta^{\mu\nu}\mp l^{-2}x_r^\mu x_r^\nu)\d {\r}{\r x_r^\mu}\otimes \d {\r}{\r x_r^\nu}\right )
=\si_{E,\,3}^\pm \d {\r}{\r x^0}\otimes \d {\r}{\r x^0}=:\vect{h}_{HN_\pm},
\ee
or
\be \label{gh'2-2}
\left (\d {\r}{\r \tau}\right)_{HN_\pm}^{2} ={\si_{E,\,3}^\pm}
\left (\d {\r}{\r t}\right )^{ 2}. %
\ee
Namely, the space-time is split into 3d space and 1d time. The
3d space is a  Riemann sphere or a Lobachevsky space. The scaling
factor of the 1d time depends on the position in space.

In comparison, $NH_\pm$ space-times are split into 1d time and 3d
space.  The 3d space is conformally
flat. The conformal factor depends on time.  From this point of
view, the geometries for $HN_\pm $ is in contrast to the
geometries for ${\frak n}^\pm$.  It is one of the reasons that the
geometries are referred to as (anti-)Hooke-Newton geometries.

The connection are
\be
{\Ga_{HN_\pm}}^{\la}_{\mu\nu}=\pm \lim_{c_r\to 0}
\d{(\dl^\la_\mu \eta_{\nu \ka} + \dl^\la_\nu
\eta_{\mu \ka})x_r^\ka}{l^2\si_{r}^\pm}
=\pm\d{(\dl^\la_\mu \eta_{\nu l} + \dl^\la_\nu
\eta_{\mu l})x^l}{l^2 \si_{E,\,3}^\pm}.
\ee
The nonzero coefficients are
\be
{\Ga_{HN_\pm}}^{0}_{0i}={\Ga_{HN_\pm}}^{0}_{i0}
=\mp\d{\dl_{il} x^l}{l^2 \si_{E,\,3}^\pm}, \qquad
{\Ga_{HN_\pm}}^{i}_{jk}
=\mp\d{(\dl^i_j \dl_{k l} + \dl^i_k
\dl_{j l})x^l}{l^2 \si_{E,\,3}^\pm}.
\ee
The nonzero components of curvature tensor and Ricci tensor are
\be
{R_{HN_\pm}}^{0}_{\ i 0 j }=-{R_{HN_\pm}}^{0}_{\ ij 0}
= \pm l^{-2}g^{HN_\pm}_{ij}, \qquad
{R_{HN_\pm}}^{k}_{\ i l j }= \mp l^{-2}(\dl^k_j g^{HN_\pm}_{i l}
-\dl^k_l g^{HN_\pm}_{ij}),
\ee
\be
R^{HN_\pm}_{ij}= \mp 3l^{-2}g^{HN_\pm}_{ij},
\ee
respectively.

When  $c_r\to 0$, $\si_{E,r}^\pm$ also tend to $
\si_{E,\,3}^\pm (x^i)$.  The $Riem$ and $Lob$ spaces tend to
\be
ds_{EHN_\pm}^{2}=\lim_{c_r\to 0} \d 1 {\si_{E,r}^\pm} \left (\dl_{\mu\nu}
\mp \d
{\dl_{\mu\ka}\dl_{\nu\la}x_r^\ka x_r^\la} {l^2\si_{E,r}^\pm}\right )
dx_r^\mu dx_r^\nu
=\d 1 {\si_{E,\,3}^\pm}\left (\dl_{ij} \mp \d
{\dl_{ik}\dl_{jl}x^k x^l} {l^2\si_{E,\,3}^\pm}\right ) dx^i dx^j=-{\vect g}^{HN_\pm}, \quad
\ee
\be
\left (\d {\r}{\r s}\right)_{EHN_\pm}^{2}=\lim_{c_r\to 0} \frac {c_r^2}{c^2}\left (\si_{E,r}^\pm
(\dl^{\mu\nu}\pm l^{-2}x_r^\mu x_r^\nu)\d {\r}{\r x_r^\mu} \otimes \d {\r}{\r x_r^\nu}\right )
=\si_{E,\,3}^\pm \d {\r}{\r x^0} \d {\r}{\r x^0}=\vect{h}_{HN_\pm}
\ee
with $\si_{E,3}^\pm >0$.
The signatures of $-\vect{g}^{HN_\pm}$ and $\vect{h}_{HN_\pm}$ are
$(+,+,+)$ and $(+)$, respectively.  Therefore, they describe the Euclidean version
of $HN_\pm$ space-times.

For $\si_{E,r}^-<0$ and $\si_r^- <0$, the $LBdS$ and $DTdS$ space-times
contract to
\be
\left . \begin{array}{l}
ds_{HN_-'}^{2}=-\displaystyle
\lim_{c_r\to 0} \d 1 {\si_{E,r}^-} \left (\dl_{\mu\nu}
+ \d
{\dl_{\mu\ka}\dl_{\nu\la}x_r^\ka x_r^\la} {l^2\si_{E,r}^-}\right )
dx_r^\mu dx_r^\nu \\
ds_{DTHN}^{2}=-\displaystyle
\lim_{c_r\to 0}\d 1 {\si_r^-}\left (\eta_{\mu\nu}-\d
{\eta_{\mu\ka}\eta_{\nu\la}x_r^\ka x_r^\la}{l_r^2\si_r^-}\right )
dx_r^\mu dx_r^\nu\end{array} \right \}
&=&\mp \d 1 {\si_{E,\,3}^-}\left (\dl_{ij} + \d
{\dl_{ik}\dl_{jl}x^k x^l} {l^2\si_{E,\,3}^-}\right ) dx^i dx^j \nno \\
&=&\pm {\vect g}^{LBdS}_3,
\ee
\be
\left . \begin{array}{l}
\left (\d {\r}{\r s}\right)_{HN_-'}^{2}=-\displaystyle\lim_{c_r\to 0} \frac {c_r^2}{c^2}\left (\si_{E,r}^-
(\dl^{\mu\nu}- l^{-2}x_r^\mu x_r^\nu)\d {\r}{\r x_r^\mu} \otimes
\d {\r}{\r x_r^\nu}\right ) \smallskip \\
\left (\d {\r}{\r s}\right)_{DTHN}^{2}=-\displaystyle\lim_{c_r\to 0}\frac {c_r^2}{c^2}\left (\si_{r}^-
(\eta^{\mu\nu} + l^{-2}x_r^\mu x_r^\nu)\d {\r}{\r x_r^\mu} \otimes
\d {\r}{\r x_r^\nu}\right )\end{array} \right \}
=-\si_{E,\,3}^- \d {\r}{\r x^0} \d {\r}{\r x^0}=\vect{h}_{HN'_-},
\ee
\be
\left . \begin{array}{l}
{\Ga_{HN_-'}}^{\la}_{\mu\nu}=\displaystyle\lim_{c_r\to 0}\d{(\dl^\la_\mu \dl_{\nu \ka}
+ \dl^\la_\nu \dl_{\mu \ka})x_{r}^\ka}{l^2\si_{E,\,r}^-}\\
{\Ga_{DTHN}}^{\la}_{\mu\nu}=-\displaystyle\lim_{c_r\to 0}\d{(\dl^\la_\mu \eta_{\nu \ka}
+ \dl^\la_\nu \eta_{\mu \ka})x_{r}^\ka}{l_r^2\si_{r}^-}\end{array}\right \}
=\d{(\dl^\la_\mu \dl_{\nu k} + \dl^\la_\nu
\dl_{\mu k})x^k}{l^2\si_{E,3}^-},
\ee
with $\si_{E,3}^-<0$, where ${\vect g}^{LBdS}_3$ is the metric of the 3d
$LBdS$ space-time.  $(\pm{\vect g}^{LBdS}_3, \vect{h}_{HN'_-})$
have the signatures $(\mp,\pm, \pm; +)$ and thus are called
para-$HN_-$ space-time and double time $HN$ space-time,
respectively. $\si_r^+ < 0$ is not valid in the limit and thus $BdSL$ space-time
is uncontractible in this way.

On the other hand, the para-Euclid space (or $HN_{+ 2}$ space) can be obtained
from the 4d Riemann sphere by the contraction in the limit of $c_r\to \infty$.  As $c_r$ is
running,
\be
\si_{E,\,r}^{+}=\omits{1 \mp \eta_{\mu \nu} l^{-2} x^\mu x^\nu=} \nu_r^2 t^2 + \si_{E,\,3}^+
 >0.
\ee
The metric becomes
\be \label{ge'1}
ds_{E'}^2&=&
\lim_{c_r \to \infty}\d {c_r^2}{c^2} \d 1 {\nu_r^2 t^2}
\left[ \left (c_r^2dt^2 +\dl_{ij}dx^idx^j  - \d
{1} { c_r^2 t^2}(1- \frac{\si_{E,\,3}^+}{\nu_r^2 t^2})(c_r^2tdt+\dl_{ij}x^idx^j)^2
\right )\right ] \nno \\
&=&\d 1 { \nu^2 t^2}
\left (\frac{l^2\si_{E,\,3}^+}{ t^2}dt^2+ \dl_{ij}dx^idx^j  - \d{2} { t}\dl_{ij}x^idt dx^j
\right )=:{\vect g}^{E'},
\ee
which is non-degenerate.  It provides all local information of the space-time.
For example, the inverse metric
\be
{\vect h}_{E'}:=\vect{g}^{-1}_{E'}=l^{-2}\nu^2t^2\left ( t^2\d {\r}{\r t} \otimes \d {\r}{\r t}
+(l^{2}\dl^{ij}+x^i x^j) \d {\r}{\r x^i} \otimes  \d {\r}{\r x^j}+2{tx^i}\d{\r}{\r t}
 \otimes \d{\r}{\r x^i}\right ),
\ee
and the connection coefficients
\be
{\Ga_{E'}}^0_{00}=-\frac {2}{ct}, \qquad {\Ga_{E'}}^i_{j0}=
{\Ga_{E'}}^i_{0j}=-\frac 1 {ct}\dl^i_j, \qquad \mbox{others vanish,}
\ee
which can also be obtained from the limit
\be
\left (\d {\r}{\r s}\right)_{E'}^{ 2} &=& \lim_{c_r\to \infty} \frac{c^2}{c_r^2}
\nu_r^2t^2(\dl^{\mu\nu}+ l^{-2}x_r^\mu x_r^\nu)\d {\r}
{\r x_r^\mu}\otimes \d {\r}{\r x_r^\nu}
\ee
and
\be
{\Ga_{E'}}^{\la}_{\mu\nu}=- \lim_{c_r\to \infty}\d{c_r}{c}
\d{(\dl^\la_\mu \dl_{\nu \ka} + \dl^\la_\nu
\dl_{\mu \ka})x_r^\ka}{l^2\si_{E,\,r}^+},
\ee
respectively.  The space (\ref{ge'1}) has the signature
$(+,+,+,+)$ and vanishing curvature.  Therefore, it is better to be
referred to as para-Euclid geometry rather than the $HN_{+2}$ or $P_+'$
geometry.

Similarly, the para-Poincar\'e  space-time  can be obtained from the
AdS space-times by the contraction in the limit of $c_r\to \infty$.  As $c_r$ is
running,
\be
\si_r^{-}= \nu_r^2 t^2 + \si_{E,\,3}^-
\omits{=\lim_{c_r\to \infty }\d {c^2}{c_r^2}
[1\mp l^{-2} (c_r^2t^2-\dl_{ij}x^i x^j)]=\mp \nu^2 t^2}>0.
\ee
The metric reads
\be \label{gp'1}
ds_{P'}^2&=&\lim_{c_r \to \infty}\d {c_r^2}{c^2}\d 1 {\nu_r^2 t^2}
\left[ \left (c_r^2dt^2 -\dl_{ij}dx^idx^j  - \d
{1} { c_r^2 t^2}(1- \frac{\si_{E,\,3}^-}{\nu_r^2 t^2})(c_r^2tdt-\dl_{ij}x^idx^j)^2
\right )\right ] \nno \\
&=&
\d 1 { \nu^2 t^2}
\left (\frac{l^2\si_{E,\,3}^-}{ t^2}dt^2 - \dl_{ij}dx^idx^j + \d{2} { t}\dl_{ij}x^idt dx^j
\right )=:{\vect g}^{P'},
\ee
which is also non-degenerate.  The inverse metric is
\be
\left (\d {\r}{\r s}\right)_{P'}^{ 2} &=& \lim_{c_r\to \infty} \frac{c^2}{c_r^2}
\nu_r^2t^2(\eta^{\mu\nu}+ l^{-2}x_r^\mu x_r^\nu)\d {\r}{\r x_r^\mu} \otimes
\d {\r}{\r x_r^\nu}\nno \\
&=&l^{-2}\nu^2 t^2 \left ( t^2\d {\r}{\r t} \otimes  \d {\r}{\r t}
-(l^{2}\dl^{ij}-x^i x^j) \d {\r}{\r x^i} \otimes  \d {\r}{\r
x^j}+2{tx^i}\d{\r}{\r t} \otimes  \d{\r}{\r x^i}\right )\nno \\
&=&\vect{g}^{-1}_{P'}=:{\vect h}_{P'}.
\ee %
The nonzero connection coefficients are %
\be
{\Ga_{P'}}^{0}_{00} =-\d{2}{ct}, \qquad
{\Ga_{P'}}^{i}_{j0} ={\Ga_{P'}}^{i}_{0j}
=-\d{1}{ct}\dl^i_j .
\ee
The geometry (\ref{gp'1}) has the signature
$(+,-,-,-)$ and vanishing curvature. It is better to be referred to as
para-Poincar\'e geometry as the nomenclature  in Ref. \cite{BLL}.

When $c_r \to \infty$,
\be
\si_{E,\,r}^{-} &\to& - \nu_r^2 t^2 + \si_{E,\,3}^- <0, \\
\si_r^+ &\to& - \nu_r^2 t^2 + \si_{E,\,3}^+ <0.
\ee
The $LBdS$ space-time and $BdSL$ space also contract to, respectively,
the $P'$ space-time,
\be
&&-\lim_{c_r\to \infty }
\d {c_r^2}{c^2}\d 1 {\si_{E,r}^-}\left ( \dl_{\mu \nu} +\d {\dl_{\mu\ka}
\dl_{\nu \la}x_r^\ka x_r^\la}
{l^2\si_{E,r}^-}\right ) dx_r^\mu dx_r^\nu \nno \\
&=&\lim_{c_r \to \infty}\d {c_r^2}{c^2} \d 1 {\nu_r^2 t^2}
\left[ \left (c_r^2dt^2 +\dl_{ij}dx^idx^j  - \d
{1} { c_r^2 t^2}(1+ \frac{\si_{E,\,3}^-}{\nu_r^2 t^2})(c_r^2tdt+\dl_{ij}x^idx^j)^2
\right )\right ] \nno  \\
&=&\d 1 { \nu^2 t^2}
\left (-\frac{l^2\si_{E,\,3}^-}{ t^2}dt^2+ \dl_{ij}dx^idx^j  - \d{2} { t}\dl_{ij}x^idt dx^j
\right )=-\vect{g}^{P'}
\ee
and the $E'$ space,
\be
&&\lim_{c_r\to \infty }
\d {c_r^2}{c^2}\d 1 {\si_{r}^+}\left ( \eta_{\mu \nu} +\d {\eta_{\mu\ka}
\eta_{\nu \la}x_r^\ka x_r^\la}
{l^2\si_{r}^+}\right ) dx_r^\mu dx_r^\nu \nno \\
&=&\lim_{c_r \to \infty}\d {c_r^2}{c^2} \d 1 {\nu_r^2 t^2}
\left[ \left (c_r^2dt^2 -\dl_{ij}dx^idx^j  - \d
{1} { c_r^2 t^2}(1+ \frac{\si_{E,\,3}^+}{\nu_r^2 t^2})(c_r^2tdt-\dl_{ij}x^idx^j)^2
\right )\right ] \nno  \\
&=&\d 1 { \nu^2 t^2}
\left (-\frac{l^2\si_{E,\,3}^+}{ t^2}dt^2 - \dl_{ij}dx^idx^j  - \d{2} { t}\dl_{ij}x^idt dx^j
\right )=-\vect{g}^{E'}.
\ee

\subsection{Geometries for $\frak{g}$, $\frak{c}$, $\frak{g}_2$, and $\frak{c}_2$}

It is well known that the $G$ geometry can be obtained from the $Min$ space-time:
\be \label{gg1}
ds_G^2\omits{=\lim_{c_r \to \infty}\d {c^2}{c_r^2}ds_r^2}=
\lim_{c_r \to \infty}\d {c^2}{c_r^2}(c_r^2dt^2-\dl_{ij}dx^idx^j)
=c^2 dt^2 =:{\vect g}^G.
\ee
\be
\left (\d {\r}{\r s}\right )_G^{ 2} =\lim_{c_r\to \infty}\left (
\d 1 {c_r^2}\d {\r}{\r t} \otimes \d {\r}{\r t}-\dl^{ij}\d{\r}{\r x^i} \otimes
\d{\r}{\r x^j}\right )
=-\dl^{ij}\d{\r}{\r x^i} \otimes \d{\r}{\r x^j}=:{\vect h}_G.
\ee
Similarly, the $C$ geometry can also be obtained from the $Min$ space-time:
\be \label{gc1}
ds_C^2\omits{=\lim_{c_r \to 0}ds_r^2}=
\lim_{c_r \to 0}(c_r^2dt^2-\dl_{ij}dx^idx^j)
=-\dl_{ij}dx^idx^j =:{\vect g}^C,
\ee
\be \label{gc2}
\left (\d {\r}{\r s}\right )_C^{ 2} =\lim_{c_r\to 0}\d {c_r^2}{c^2}
\left (\d 1 {c_r^2}\d {\r}{\r t} \otimes \d {\r}{\r t}-\dl^{ij}\d{\r}{\r x^i} \otimes
\d{\r}{\r x^j}\right )
=\d 1 {c^2}\d{\r}{\r t} \otimes \d{\r}{\r t}=:{\vect h}_C.
\ee
For both $G$ and $C$ geometries, the spaces and times are flat:
\be
\Ga^{\quad \ \la}_{G/C\,\mu\nu} =0, \qquad
R^{\quad \ \la}_{G/C\ \mu\si\nu}=0.
\ee
The $G$ and $C$ geometries can also be obtained from \dS\ and \AdS\ space-times directly.
In the contraction procedure to obtain $G$ geometry from \dS\ and \AdS\ space-times, both
$c_r$ and $l_r$ tend to infinity but $\nu_r=c_r/l_r \to 0$.  Without loss of generality,
we may suppose $c_r^2/l_r$ keeps finite in the limit.
Then
\be
ds_G^2
=\lim_{\substack{c_r,\, l_r \to \infty \\ c_r^2/l_r\ \mbox{fixed}}}
\d {c^2}{c_r^2}\d 1 {\si_r^\pm}\left (\eta_{\mu\nu} \pm
\d {\eta_{\mu \ka}\eta_{\nu \la}x_r^\ka x_r^\la}{l_r^2\si_r^\pm}\right )dx_r^\mu dx_r^\nu,
\ee
\be
\left (\d {\r}{\r s}\right )_G^{ 2} =\lim_{\substack{c_r,\, l_r \to \infty \\
c_r^2/l_r\ \mbox{fixed}}}
\si_r^\pm (\eta^{\mu\nu}-l_r^{-2}x_r^\mu x_r^\nu)\d{\r}{\r x_r^\mu} \otimes \d{\r}{\r x_r^\nu}.
\ee
The contraction from \dS\ and \AdS\ space-times to $C$ space-time
is easier:
\be
ds_C^2
=\lim_{\substack{l_r \to \infty \\ c_r \to 0}}
\d 1 {\si_r^\pm}\left (\eta_{\mu\nu} \pm
\d {\eta_{\mu \ka}\eta_{\nu \la}x_r^\ka x_r^\la}{l_r^2\si_r^\pm}\right )dx_r^\mu dx_r^\nu,
\ee
\be
\left (\d {\r}{\r s}\right )_C^{ 2} =\lim_{\substack{l_r \to \infty \\
c_r\to 0}}
\d {c_r^2}{c^2}\si_r^\pm (\eta^{\mu\nu} \mp l_r^{-2}x_r^\mu x_r^\nu)\d{\r}{\r x_r^\mu}
 \otimes \d{\r}{\r x_r^\nu}.
\ee
In addition, the direct contractions from \Riem\ and \Lob\ geometries can give the
Euclidean $G$ space-time (denoted by $EG$)  and Euclidean $C$ space-time (
denoted by $EC$).
Since the inequalities
\be
\si_{E,r}^-<0, \qquad \si_r^\pm <0
\ee
are not valid either in the limit $c_r, l_r \to \infty,\ \nu_r\to 0$ or in the
limit of $l_r \to \infty, \ c_r\to 0$, the $LBdS$, $BdSL$ and $DTdS$
geometries are not contractible in the two ways.

In the similar way, the $G_2$ and $C_2$ geometries
are related to the $P_{2\pm}$ space-time.

If the inequality (\ref{si_l_to0}) is
always valid ({\it i.e.} $\mp\eta_{\ka\la}x_r^\ka x_r^\la>0$) when $c_r\to 0$,
only the upper sign ({\it i.e.} $P_{2+}$) is meaningful.  Therefore,
\be \label{gg21}
ds_{EG_{2}}^2=\lim_{c_r \to 0} l^2\d {\eta_{\mu\ka}\eta_{\nu\la}-
\eta_{\mu \nu}\eta_{\ka \la}}{(\eta_{\si\tau}x_{r}^\si x_{r}^\tau)^2}
x_{r}^\ka x_{r}^\la dx_{r}^\mu dx_{r}^\nu
= l^2\d {\dl_{ik}\dl_{jl}-\dl_{ij}\dl_{kl}}{(\dl_{mn}x^m x^n)^2}x^k x^l dx^i dx^j
=:\vect{g}^{EG_{2}}=\vect{g}^{NH_{2}}.\qquad
\ee
\be \label{gg22}
\left (\d {\r}{\r s}\right )_{EG_{2}}^{ 2}&=&\lim_{c_r\to 0}
l^{-4}\eta_{\si\tau}x_r^\si x_r^\tau x_{r}^\mu x_{r}^\nu
\d{\r }{\r x_{r}^\mu} \otimes \d{\r }{\r x_{r}^\nu}
=
-l^{-4}\dl_{mn}x^m x^n x^\mu x^\nu \d{\r }{\r x^\mu} \otimes \d{\r }{\r x^\nu}\nno \\
&=:& \vect{h}_{EG_{2}}
= \lim_{\nu_r\to \infty} \vect{h}_{NH_{2}},
\ee
\be
{\Ga_{EG_{2}}}^{\la}_{\mu\nu}\omits{=\lim_{c_r\to 0}{\Ga_{\frak{p}^{}_2,r}}^{\la}_{\mu\nu}}=
-\lim_{c_r\to 0}\d{(\dl^\la_\mu \eta_{\nu \ka} + \dl^\la_\nu
\eta_{\mu \ka})x_r^\ka}{\eta_{\si\tau}x_r^\si x_r^\tau}
=\d{(\dl^\la_\mu \eta_{\nu k} + \dl^\la_\nu
\eta_{\mu k})x^k}{\dl_{mn}x^m x^n}={\Ga_{NH_{2}}}^{\la}_{\mu\nu}.
\ee
The curvature tensor and Ricci tensor are the same as those for the $NH_2$ geometry.
It should be noted that the sum of ranks of $\vect{g}^{G_2}$ and
$\vect{h}_{G_2}$ is only 3.  They have the signature $(-, -)$ and $(-)$, respectively.
Therefore, they describe a 3d Euclidean geometry with the free parameter $\psi=lx^0/
\sqrt{\dl_{ij}x^ix^j}$, denoted by $EG_{2}$ geometry.
The $EG_2$ geometry can also be obtained from the direct contraction of
$dS$ space-time.  Since both $c_r$ and $l_r$ tend
to 0 but $\nu_r=c_r/l_r \to \infty$, we may suppose $c_r^2/l_r$ keeps finite in the
limit without loss of generality. Then,
\be
 ds_{EG_{2}}^2
=\displaystyle
\lim_{\substack{c_r,\, l_r \to 0 \\ c_r^2/l_r\ \mbox{fixed}}}
\d {l^2}{l_r^2}\d 1 {\si_r^+}\left (\eta_{\mu\nu} +
\d {\eta_{\mu \ka}\eta_{\nu \la}x_r^\ka x_r^\la}{l_r^2\si_r^+}\right )dx_r^\mu dx_r^\nu,
\ee
\be
\left (\d {\r}{\r s}\right )_{EG_2}^{ 2}
=\displaystyle \lim_{\substack{c_r,\, l_r \to 0 \\ c_r^2/l_r\
\mbox{fixed}}}\d{l_r^4}{l^4}
\si_r^+ (\eta^{\mu\nu}-l_r^{-2}x_r^\mu x_r^\nu)\d{\r}{\r x^\mu} \otimes \d{\r}{\r x^\nu} .
\ee

The direct contraction from the \Riem\ geometry in the limit of $c_r, l_r\to 0$
and $c_r^2/l_r =c^2/l$ gives rise to $-\vect{g}^{EG_{2}}$ and
$-\vect{h}_{EG_{2}}$ and the free parameter $\psi=lx^0/\sqrt{\dl_{ij}x^ix^j}$.
The contractions of $LBdS$ and $DTdS$ space-times are, respectively,
\be
ds_{G_{2}}^2
= -\displaystyle \lim_{\substack{c_r,\, l_r \to 0 \\ c_r^2/l_r\ \mbox{fixed}}}
\d {l^2}{l_r^2}\d 1 {\si_{E,r}^-}\left ( \dl_{\mu \nu} +\d {\dl_{\mu\ka}
\dl_{\nu \la}x_r^\ka x_r^\la}
{l_r^2\si_{E,r}^-}\right ) dx_r^\mu dx_r^\nu=-\vect{g}^{NH_2},
\ee
\be
\left (\d {\r}{\r s}\right )_{G_2}^{ 2}
= -\displaystyle\lim_{\substack{c_r,\, l_r \to 0 \\ c_r^2/l_r\ \mbox{fixed}}}
\frac {l_r^4}{l^4}\si_{E,r}^-
(\dl^{\mu\nu}- l_r^{-2}x_r^\mu x_r^\nu)\d {\r}{\r x_r^\mu} \otimes
\d {\r}{\r x_r^\nu}= \vect{h}_{EG_2},
\ee
and
\be
ds_{G_2}^{2}=-\displaystyle
\lim_{\substack{c_r,\, l_r \to 0 \\ c_r^2/l_r\ \mbox{fixed}}}
\d {l^2}{l_r^2}\d 1 {\si_r^-}\left (\eta_{\mu\nu}-\d
{\eta_{\mu\ka}\eta_{\nu\la}x_r^\ka x_r^\la}{l_r^2\si_r^-}\right )
dx_r^\mu dx_r^\nu =\vect{g}^{EG_2},
\ee
\be
\left (\d {\r}{\r s}\right )_{G_2}^{2}=-\displaystyle\lim_{\substack{c_r,\, l_r \to 0 \\ c_r^2/l_r\ \mbox{fixed}}}
\frac {l_r^4}{l^2}\si_{r}^-
(\eta^{\mu\nu} + l^{-2}x_r^\mu x_r^\nu)\d {\r}{\r x_r^\mu} \otimes
\d {\r}{\r x_r^\nu}=-\vect{h}_{EG_2}
\ee
with the free parameter $\psi=lx^0/\sqrt{\dl_{ij}x^i x^j}$.
The domain condition is again $\dl_{ij}x^i x^j>0$.
The \Lob,  \AdS\ and $BdSL$ geometries are not contractible in this limit.

If the inequality (\ref{si_l_to0}) is
always valid as $c_r\to \infty$, only the $P_{2-}$ geometry
is taken.
\be \label{gc21}
ds_{C_2}^2=-\lim_{c_r \to \infty} l^2\d {c_r^2}{c^2}\d {\eta_{\mu\ka}\eta_{\nu\la}-
\eta_{\mu \nu}\eta_{\ka \la}}{(\eta_{\si\tau}x^\si x^\tau)^2}
x_r^\ka x_r^\la dx_r^\mu dx_r^\nu
=- l^2 \dl_{ij}d(\frac {x^i} {ct})d(\frac {x^j} {ct}) =:{\vect g}^{C_2},
\ee
\be \label{gc22}
\left (\d {\r}{\r s}\right )_{C_2}^2&=&\lim_{c_r\to \infty}\d {c^2}{c_r^2}
l^{-4}\eta_{\si\tau}x_r^\si x_r^\tau x^\mu x^\nu \d{\r }{\r x^\mu}\d{\r }{\r x^\nu}
=l^{-4}c^2t^2 x^\mu x^\nu \d{\r }{\r x^\mu}\otimes\d{\r }{\r x^\nu}
=: \vect{h}_{C_2}.
\ee
\be
\Ga^{\ \ 0}_{C_200}=-\d 2 {ct}, \qquad  \Ga^{\ \ i}_{C_20j}=
\Ga^{\ \ i}_{C_2j0}=-\d 1 {ct}\dl^i_j, \qquad
\mbox{others vanish}.
\ee
\be
R^{\ \ \la}_{C_2\ \,\mu\nu\rho}=0.
\ee
The $C_2$ geometry can also be directly obtained from the contraction of
$AdS$ and $LBdS$ space-times:
\be
ds_{C_2}^2
=\left \{\begin{array}{l}\displaystyle \lim_{\substack{l_r \to 0 \\ c_r\to \infty}}
\d {\nu_r^2}{\nu^2}\d 1 {\si_r^-}\left (\eta_{\mu\nu} -
\d {\eta_{\mu \ka}\eta_{\nu \la}x_r^\ka x_r^\la}{l_r^2\si_r^-}\right )dx_r^\mu dx_r^\nu
 =-l^2\dl_{ij} d(\d {x^i}{ct})d(\d {x^j}{ct})
\\
 -\displaystyle \lim_{\substack{l_r \to 0 \\ c_r \to \infty}}
\d {\nu_r^2}{\nu^2}\d 1 {\si_{E,r}^-}\left ( \dl_{\mu \nu} +\d {\dl_{\mu\ka}
\dl_{\nu \la}x_r^\ka x_r^\la}
{l_r^2\si_{E,r}^-}\right ) dx_r^\mu dx_r^\nu =l^2\dl_{ij} d(\d {x^i}{ct})d(\d {x^j}{ct})  \end{array}\right .,
\ee
\be
\left (\d {\r}{\r s}\right )_{C_2}^{ 2}
=\left \{\begin{array}{l}\displaystyle \lim_{\substack{l_r \to 0 \\ c_r\to \infty}}
\d{l_r^2\nu^2}{l^2\nu_r^2}
\si_r^- (\eta^{\mu\nu}+l_r^{-2}x_r^\mu x_r^\nu)\d{\r}{\r x_r^\mu}\otimes\d{\r}{\r x_r^\nu}
=l^{-4} c^2t^2x^\mu x^\nu\d{\r}{\r x^\mu}\otimes\d{\r}{\r x^\nu}\\
-\displaystyle\lim_{\substack{l_r \to 0 \\ c_r \to \infty}}
\d{l_r^2\nu^2}{l^2\nu_r^2}\si_{E,r}^-
(\dl^{\mu\nu}- l_r^{-2}x_r^\mu x_r^\nu)\d {\r}{\r x_r^\mu} \otimes
\d {\r}{\r x_r^\nu}=-l^{-4} c^2t^2x^\mu x^\nu\d{\r}{\r x^\mu}\otimes\d{\r}{\r x^\nu}
\end{array}\right . ,
\ee
for
\be
\left . \begin{array}{l} 0<\si_r^- \\
0>\si_{E,r}^-\end{array} \right \}\to \pm \nu_r^2t^2(1 \mp \d {\dl_{ij}x^i x^j}
{c_r^2 t^2}).
\ee
The \Riem\ space and $BdSL$ space contract to $- \vect{g}^{C_2}, \vect{h}_{C_2}$,
which
has the signature $(+, +, +; +)$ and thus is named $EC_2$ space.
\Lob\ and $DTdS$ as well as \dS\ are not contractible in this limit.

\no
\begin{table}[thbp]
\hspace{-1cm}
\caption{Algebras and their corresponding geometries.}
\begin{tabular}{ccccccc}
\toprule[2pt]
  Alg & $\begin{array}{c} \mbox{Geom.}\\
   \mbox{name}\end{array}$ & $\begin{array}{c}\mbox{Geometrical}\\
  \mbox{variables}\footnote{For brevity, the following shorthands are used.
  $(x\cdot x)_E:=\dl_{\mu\nu}x^\mu x^\nu$; $x\cdot x:=
  \eta_{\mu\nu}x^\mu x^\nu$; $\vect{x}\cdot\vect{x}:=\dl_{ij}x^i x^j$;
  $\r_\mu\r_\nu:=\d \r {\r x^\mu}\otimes\d \r {\r x^\nu}$, etc.}
  \end{array}$ &$\begin{array}{c}(\vect g,\ \vect h)\\
  {\rm Ranks} \end{array}$ & Signature & $\begin{array}{c}\mbox{Contrac-}\\
  \mbox{tion}\end{array}$   & Domain \\
\hline
  ${\frak r}$ & {\it Riem}  & $\vect{g}^{Riem}=\d 1 {\si_E^+}\left( (dx\cdot dx)_E
  -\d {(x\cdot dx)_E^2}{l^2\si_E^+}\right )$& (4, 4)
  &$(+,+,+,+)$&No& $\si_E^+>0$ 
  \smallskip \\
  ${\frak l}$ & {\it Lob}  & $\vect{g}^{Lob}=\d 1 {\si_E^-}\left((dx\cdot dx)_E
  +\d {(x\cdot dx)_E^2}{l^2\si_E^-}\right )$& (4, 4) &
  $(+,+,+,+)$&No & $\si_E^->0$ 
  \smallskip \\
 ${\frak l}$ & {\it LBdS}  & $\vect{g}^{LBdS}=-\d 1 {\si_E^-}\left((dx\cdot dx)_E
  +\d {(x\cdot dx)_E^2}{l^2\si_E^-}\right )$& (4, 4) &
  $(-,+,+,+)$&No & $\si_E^-<0$ 
   \smallskip \\
   \hline
  ${\frak e}$ & {\it Euc}  & $\vect{g}^{Euc}=(dx\cdot dx)_E$& (4, 4) &$(+,+,+,+)$
      & $l_r\to \infty$& arbitrary 
  \smallskip \\
${\frak e}_2$ & $E_2$ &
$\begin{array}{l} \vect{g}^{E_2}=l^2\d {(x\cdot x)_E(dx \cdot dx)_E
-(x\cdot dx)_E^2}{(x\cdot x)_E^2}\\
\vect{h}_{E_2}=l^{-4}(x\cdot x)_E\;(x^\mu \r_\mu)^2\\
\Ga_{E_2\mu\nu}^{\ \la}=
-\d {(\dl^\la_\mu\dl_{\nu\ka}+\dl^\la_\nu\dl_{\mu\ka})x^\ka}{(x\cdot x)_E}
\end{array}$& (3, 1) &$(+,+,+;+)$
      & $l_r\to 0$& $(x\cdot x)_E^{}>0$ 
      \\
${\frak e}_2$ & $E_{2-}$ &
$\begin{array}{l} \vect{g}^{E_{2-}}=l^2\d {(x\cdot x)_E(dx \cdot dx)_E
-(x\cdot dx)_E^2}{(x\cdot x)_E^2}\\
\vect{h}_{E_{2-}}=-l^{-4}(x\cdot x)_E\;(x^\mu \r_\mu)^2\\
{\Ga_{E_{2-}}}^{\la}_{\mu\nu}=
-\d {(\dl^\la_\mu\dl_{\nu\ka}+\dl^\la_\nu\dl_{\mu\ka})x^\ka}{(x\cdot x)_E}
\end{array}$& (3, 1) &$(+,+,+;-)$
      & $l_r\to 0$& $(x\cdot x)_E^{}>0$ 
      \\
\hline
  ${\frak d}_+$ & $dS$  & ${\vect g}^{dS}=\d 1 {\si^+}\left(dx\cdot dx
  +\d {(x\cdot dx)^2}{l^2 \si^+}\right )$
  &(4, 4)&$(+,-,-,-)$& No & $\si^+>0$ 
  \smallskip \\
  ${\frak d}_+$ & $BdSL$  & ${\vect g}^{BdSL}=\d 1 {\si^+}\left(dx\cdot dx
  +\d {(x\cdot dx)^2}{l^2 \si^+}\right )$
  &(4, 4)&$(+,+,+,+)$& No &$\si^+<0$ 
  \smallskip \\
  ${\frak d}_-$ & $AdS$  & ${\vect g}^{AdS}=\d 1 {\si^-}\left(dx\cdot dx
  -\d {(x\cdot dx)^2}{l^2\si^-}\right )$&(4, 4)
  &$(+,-,-,-)$& No & $\si^->0$ 
  \smallskip \\
  ${\frak d}_-$ & $DTdS$  & ${\vect g}^{DTdS}=-\d 1 {\si^-}\left(dx\cdot dx
  -\d {(x\cdot dx)^2}{l^2\si^-}\right )$&(4, 4)
  &$(+,+,-,-)$& No & $\si^-<0$ 
  \smallskip \\
  \hline
  ${\frak p}$ & {\it Min} & ${\vect g}^{Min}= dx \cdot dx$& (4, 4) &$(+,-,-,-)$
      & $l_r\to \infty$& arbitrary 
      \smallskip \\
 ${\frak p}_2$ & $P_{2\pm}$ & $\begin{array}{l}\vect{g}^{P_{2\pm }}
 = \pm l^2\d {(x\cdot dx)^2-(x\cdot x)(dx \cdot dx)}{(x\cdot x)^2}\\
 \vect{h}_{P_{2\pm}}=l^{-4}(x\cdot x)(x^\mu \r_\mu)^2\\
 {\Ga_{P_{2\pm}}}^{\la}_{\mu\nu}=
 -\d {(\dl^\la_\mu\eta_{\nu\ka}+\dl^\la_\nu\eta_{\mu\ka})x^\ka}{x\cdot x}
 \end{array}$ & (3, 1) &$\begin{array}{c}
 (+,-,-;-)
\\ (-,-,-;+)\end{array}
$
      & $l_r\to 0$& $\begin{array}{c} x\cdot x<0 \\
      x\cdot x>0\end{array}$
\\
 ${\frak p}_2$ & $EP_{2-}$ & $\begin{array}{l}\vect{g}^{EP_{2-}}
 =l^2\d {(x\cdot dx)^2-(x\cdot x)(dx \cdot dx)}{(x\cdot x)^2}\\
 \vect{h}_{EP_{2-}}=l^{-4}(x\cdot x)(x^\mu \r_\mu)^2\\
 \Ga_{EP_{2-} \mu\nu}^{\qquad \la}=
 -\d {(\dl^\la_\mu\eta_{\nu\ka}+\dl^\la_\nu\eta_{\mu\ka})x^\ka}{x\cdot x} \end{array}
 $ & (3, 1) &$ (+,+,+;+)
$
      & $l_r\to 0$& $x\cdot x >0$ 
      \\
 ${\frak p}_2$ & $DTP_{2+}$ & $\begin{array}{l}\vect{g}^{DTP_{2+}}
 =l^2\d {(x\cdot dx)^2-(x\cdot x)(dx \cdot dx)}{(x\cdot x)^2}\\
 \vect{h}_{DTP_{2{\blue +}}}=-l^{-4}(x\cdot x)(x^\mu \r_\mu)^2\\
 \Ga_{DTP_{2+} \mu\nu}^{\qquad \la}=
 -\d {(\dl^\la_\mu\eta_{\nu\ka}+\dl^\la_\nu\eta_{\mu\ka})x^\ka}{x\cdot x} \end{array}
 $ & (3, 1) &$ (+,-,-;+)$
      & $l_r\to 0$& $x\cdot x<0$ 
      \\
       \hline
\bottomrule[2pt]
\end{tabular}
\end{table}

\begin{table}[t]
\setcounter{table}{1}
{\vspace{-1.5cm}\caption{Algebras and their corresponding geometries (Cont.).}
\begin{tabular}{ccccccc}
\toprule[2pt]
  Alg &  $\begin{array}{c} \mbox{Geom.}\\
   \mbox{name}\end{array}$ & $\begin{array}{c}\mbox{Geometrical}\\
  \mbox{variables}\end{array}$ &$\begin{array}{c}(\vect g,\ \vect h)\\
  {\rm Ranks} \end{array}$ & Signature &$\begin{array}{c}\mbox{Contrac-}\\
  \mbox{tion}\end{array}$& Domain 
  \\
\hline
${\frak n}_\pm$& $NH_\pm$  & $\begin{array}{l}
                       \vect{g}^{NH_\pm}= (\si_{\frak n}^\pm)^{-2}c^2dt^2 \\
                       \vect{h}_{NH_\pm} = -\si_{\frak n}^\pm \dl^{ij}\r_i \r_j\\
                       {\Ga_{NH_\pm}}^{0}_{00}=\pm\d{2\nu^2t}{c \si_{\frak{n}}^\pm}\\
{\Ga_{NH_\pm}}^{i}_{0j}={\Ga_{NH_\pm}}^{i}_{j0}=\pm \d{\nu^2 t}
{c \si_{\frak{n}}^\pm}\dl^i_j
                       \end{array}$& (1, 3) &$ (+;-,-,-)$&
$ \begin{array}{l}l_r,c_r\to \infty \\ \nu=c_r/l_r \\{\rm finite}\end{array}$&
$\si_{\frak{n}}^\pm>0$ 
\smallskip\\
${\frak n}_\pm$& $ENH_\pm$  & $\begin{array}{l}
                       \vect{g}^{ENH_\pm}=(\si_{\frak n}^\pm)^{-2}c^2dt^2 \\
                       \vect{h}_{ENH_\pm} =  \si_{\frak n}^\pm\dl^{ij}\r_i \r_j\\
                       {\Ga_{ENH_\pm}}^{0}_{00}=\pm\d{2\nu^2t}{c \si_{\frak{n}}^\pm}\\
{\Ga_{NH_+}}^{i}_{0j}={\Ga_{NH_+}}^{i}_{j0}=\pm\d{\nu^2 t}
{c \si_{\frak{n}}^\pm}\dl^i_j
                       \end{array}$& (1, 3) &$  (+;+,+,+ )  $&
$ \begin{array}{l}l_r,c_r\to \infty \\ \nu=c_r/l_r \\{\rm finite}\end{array}$&
$\si_{\frak{n}}^\pm >0$ 
\smallskip\\
${\frak n}_+$& $NH_+'$  & $\begin{array}{l}
                       \vect{g}^{NH_+'}=- (\si_{\frak n}^+)^{-2}c^2dt^2 \\
                       \vect{h}_{NH_+'} = - \si_{\frak n}^+  \dl^{ij}\r_i \r_j\\
                       {\Ga_{NH_+'}}^{0}_{00}=\d{2\nu^2t}{c \si_{\frak{n}}^+}\\
{\Ga_{NH_+'}}^{i}_{0j}={\Ga_{NH_+'}}^{i}_{j0}=\d{\nu^2 t}
{c \si_{\frak{n}}^+}\dl^i_j
                       \end{array}$& (1, 3) &$ (-;+,+,+) $&
$ \begin{array}{l}l_r,c_r\to \infty \\ \nu=c_r/l_r \\{\rm finite}\end{array}$&
$\si_{\frak n}^+<0$ 
\smallskip\\
${\frak n}_+$& $ENH_+'$& $\begin{array}{l}
                       \vect{g}^{ENH_+'}= (\si_{\frak n}^+)^{-2}c^2dt^2\\
                       \vect{h}_{ENH_+'}= - \si_{\frak n}^+  \dl^{ij}\r_i \r_j\\
                       {\Ga_{ENH_+'}}^{0}_{00}=\d{2\nu^2t}{c \si_{\frak{n}}^+}\\
{\Ga_{ENH_+'}}^{i}_{0j}={\Ga_{ENH_+'}}^{i}_{j0}=\d{\nu^2 t}
{c \si_{\frak{n}}^+}\dl^i_j\end{array}$
                       & (1, 3) &$  (+;+, +, + )  $&
$ \begin{array}{l}l_r,c_r\to \infty \\ \nu=c_r/l_r \\{\rm finite}\end{array}$&
$\si_{\frak n}^+<0$
\smallskip \\
\hline
${\frak n}_{+2}$& $NH_{2}$  & $\begin{array}{l}
                       \vect{g}^{NH_2}= l^2 \d {(\vect x\cdot \vect{dx})^2-(\vect x
                                      \cdot \vect x)(\vect{dx} \cdot \vect{dx})}
                                      {(\vect x\cdot \vect x)^2}\\
                       \vect{h}_{NH_2} = l^{-4} {\vect x \cdot \vect x}
                       \left[\nu^{-2}\r_t^2-(x^\mu\r_\mu)^2\right]\\
{\Ga_{NH_{2}}}^{0}_{0i}={\Ga_{NH_{2}}}^{0}_{i0}=
-\d{x^i}{\vect x \cdot \vect x}\\
{\Ga_{NH_{2}}}^{i}_{jk}=-\d{\dl^i_j x^k
+ \dl^i_k x^j}{\vect x \cdot \vect x}
                       \end{array}$&   (2, 2)& $(-,-;+,-)$&
$ \begin{array}{l}l_r,c_r\to 0 \\ \nu=c_r/l_r \\{\rm finite}\end{array}$&
           $|\vect{x}|>0$ 
           \smallskip \\
${\frak n}_{+2}$& $NH_{2}'$  & $\begin{array}{l}
                       \vect{g}^{NH_2'}=- l^2 \d {(\vect x\cdot \vect{dx})^2-(\vect x
                                      \cdot \vect x)(\vect{dx} \cdot \vect{dx})}
                                      {(\vect x\cdot \vect x)^2}\\
                       \vect{h}_{NH_2'} = l^{-4} {\vect x \cdot \vect x}
                       \left[\nu^{-2}\r_t^2-(x^\mu\r_\mu)^2\right]\\
{\Ga_{NH_{2}'}}^{0}_{0i}={\Ga_{NH_{2}'}}^{0}_{i0}=
-\d{x^i}{\vect x \cdot \vect x}\\
{\Ga_{NH_{2}'}}^{i}_{jk}=-\d{\dl^i_j x^k
+ \dl^i_k x^j}{\vect x \cdot \vect x}
                       \end{array}$&   (2, 2)& $(+,+;+,-)$&
$ \begin{array}{l}l_r,c_r\to 0 \\ \nu=c_r/l_r \\{\rm finite}\end{array}$&
         $|\vect{x}|>0$ 
         \smallskip \\
${\frak n}_{-2}$& $ENH_{2}$  &$\begin{array}{l}
                       \vect{g}^{ENH_2}= l^2 \d {(\vect x \cdot \vect x)(\vect{dx} \cdot \vect{dx})-
                                      (\vect x\cdot \vect{dx})^2}
                                      {(\vect x\cdot \vect x)^2}\\
                       \vect{h}_{ENH_2} = l^{-4} {\vect x \cdot \vect x}
                       \left[\nu^{-2}\r_t^2+(x^\mu\r_\mu)^2\right]\\
{\Ga_{ENH_2}}^{0}_{0i}={\Ga_{ENH_2}}^{0}_{i0}=
-\d{x^i}{\vect x \cdot \vect x}\\
{\Ga_{ENH_2}}^{i}_{jk}=-\d{\dl^i_j x^k
+ \dl^i_k x^j}{\vect x \cdot \vect x}
                       \end{array}$ &  (2, 2)& $(+,+;+,+)$&
$ \begin{array}{l}l_r,c_r\to 0 \\ \nu=c_r/l_r \\{\rm finite}\end{array}$&
        $|\vect{x}|>0$
        \smallskip \\
${\frak n}_{-2}$& $DTNH_{2}$  &$\begin{array}{l}
                       \vect{g}^{DTNH}= l^2 \d {(\vect x\cdot \vect{dx})^2-
                       (\vect x \cdot \vect x)(\vect{dx} \cdot \vect{dx})}
                                      {(\vect x\cdot \vect x)^2}\\
                       \vect{h}_{DTNH} = l^{-4} {\vect x \cdot \vect x}
                       \left[\nu^{-2}\r_t^2+(x^\mu\r_\mu)^2\right]\\
{\Ga_{DTNH_2}}^{0}_{0i}={\Ga_{DTNH_2}}^{0}_{i0}=
-\d{x^i}{\vect x \cdot \vect x}\\
{\Ga_{DTNH_2}}^{i}_{jk}=-\d{\dl^i_j x^k
+ \dl^i_k x^j}{\vect x \cdot \vect x}
                       \end{array}$ &   (2, 2)&$(-,-;+,+)$&
$ \begin{array}{l}l_r,c_r\to 0 \\ \nu=c_r/l_r \\{\rm finite}\end{array}$&
      $|\vect{x}|>0$
      \smallskip \\
\bottomrule[2pt]
\end{tabular}}
\end{table}
\begin{table}[thb]
\setcounter{table}{1}
\caption{Algebras and their corresponding geometries (Cont.).}
\begin{tabular}{ccccccc}
\toprule[2pt]
  Alg & $\begin{array}{c} \mbox{Geom.}\\
   \mbox{name}\end{array}$ & $\begin{array}{c}\mbox{Geometrical}\\
  \mbox{variables}\end{array}$ &$\begin{array}{c}(\vect g,\ \vect h)\\
  {\rm Ranks} \end{array}$ & Signature &$\begin{array}{c}\mbox{Contrac-}\\
  \mbox{tion}\end{array}$& Domain 
  \\
\hline
$\frak{h}_\pm$& $HN_\pm$  & $\begin{array}{l}
                       \vect{g}^{HN_\pm}= -\d {1}{\si_{E,\,3}^\pm}\left [
                       d\vect{x}\cdot d\vect{x} \mp \d {(\vect{x}\cdot d\vect{x})^2}
                       {l^2\si_{E,\,3}^\pm} \right ]\\
                       \vect{h}_{HN_\pm} =\si_{E,\,3}^\pm \r_{ct}^2\\
                       {\Ga_{HN_\pm}}^{0}_{0i}={\Ga_{HN_\pm}}^{0}_{i0}
=\mp\d{ x^i}{l^2 \si_{E,\,3}^+}\\
{\Ga_{HN_\pm}}^{i}_{jk} =\mp \d{\dl^i_j x^k + \dl^i_k
x^j}{l^2 \si_{E,\,3}^+} \end{array}$& (3, 1)
                       &$(-,-,-;+)$&
$ c_r\to 0 $& $\si_{E,3}^\pm >0 $
\smallskip \\
$\frak{h}_\pm$& $EHN_\pm$  & $\begin{array}{l}
                       \vect{g}^{EHN_\pm}= \d {1}{\si_{E,\,3}^\pm}\left [
                       d\vect{x}\cdot d\vect{x} \mp \d{ (\vect{x}\cdot d\vect{x})^2}
                       {l^2\si_{E,\,3}^\pm} \right ]\\
                       \vect{h}_{EHN_\pm} =\si_{E,\,3}^\pm \r_{ct}^2\\
                       {\Ga_{EHN_\pm}}^{0}_{0i}={\Ga_{EHN_\pm}}^{0}_{i0}
=\mp\d{ x^i}{l^2 \si_{E,\,3}^{\pm}}\\
{\Ga_{EHN_\pm}}^{i}_{jk}
=\mp \d{\dl^i_j x^k + \dl^i_k
x^j}{l^2 \si_{E,\,3}^{\pm}} \end{array}$& (3, 1)
                       &$(+,+,+;+)$&
$ c_r\to 0 $& $\si_{E,3}^\pm>0 $
\smallskip \\
${\frak h}_-$& $HN_-'$  & $\begin{array}{l}
                       \vect{g}^{HN_-'}= -\d {1}{\si_{E,\,3}^-}\left [
                       d\vect{x}\cdot d\vect{x} +\d{(\vect{x}\cdot d\vect{x})^2}
                       {l^2\si_{E,\,3}^-} \right ]\\
                       \vect{h}_{HN_-'} = - \si_{E,\,3}^- \r_{ct}^2\\
                  {\Ga_{HN_-'}}^{0}_{0i}={\Ga_{HN_-'}}^{0}_{i0}
                 =\d{x^i}{l^2 \si_{E,\,3}^-}\\
      {\Ga_{HN_-'}}^{i}_{jk}=\d{\dl^i_j x^k + \dl^i_kx^j}{l^2 \si_{E,\,3}^-}
      \end{array}$& (3, 1)
                       &$(-,+,+;+)$& $ c_r \to 0 $ &$\si_{E,3}^-<0$
                       \\
${\frak h}_-$& $DTHN$  & $\begin{array}{l}
                       \vect{g}^{DTHN}= \d {1}{\si_{E,3}^-}\left [
                       d\vect{x} \cdot d\vect{x} + \d{(\vect{x}\cdot d\vect{x})^2}
                       {l^2\si_{E,3}^-} \right ]\\
                       \vect{h}_{DTHN} = - \si_{E,3}^- \r_{ct}^2\\
                  {\Ga_{DTHN}}^{0}_{0i}={\Ga_{DTHN}}^{0}_{i0}
=\d{x^i}{l^2 \si_{3}^-}\\
{\Ga_{DTHN}}^{i}_{jk}=\d{\dl^i_j x^k + \dl^i_k
x^j}{l^2 \si_{E,\,3}^-}     \end{array}$ & (3, 1)
                       &$(+,-,-;+)$& $ c_r \to 0 $ & $\si_{E,3}^-<0$
                       \\
${\frak e}'$& $E'$ & $\begin{array}{ll}
                       \vect{g}^{E'}&= \d {1}{\nu^2t^2}\left [
                       \d {l^2\si_3^+}{t^2}dt^2 +d\vect{x}\cdot d\vect{x}\right .\\
                       & \left .- \d 2 t\vect{x}\cdot d\vect{x} dt\right ] \end{array}$
                       & (4, 4) &$         (+,+,+,+)$&
$ c_r\to \infty $ &$t^2>0$
\smallskip \\
${\frak p}'$& $P'$  & $\begin{array}{ll}
                       \vect{g}^{P'}&= \d {1}{\nu^2t^2}\left [
                       \d {l^2\si_3^-}{t^2}dt^2- d\vect{x}\cdot d\vect{x}\right .\\
                       & \left . +
                       \d 2 t\vect{x}\cdot d\vect{x}dt\right ] \end{array}$ & (4, 4) &$
                                      (+,-,-,-)$&
$ c_r\to \infty $& $t^2>0$ 
\smallskip \\
\bottomrule[2pt]
\end{tabular}
\end{table}
\begin{table}[thb]
\vspace{-1.cm}
\setcounter{table}{1}
\caption{Algebras and their corresponding geometries (Cont.).}
\begin{tabular}{ccccccc}
\toprule[2pt]
  Alg & $\begin{array}{c} \mbox{Geom.}\\
   \mbox{name}\end{array}$  & $\begin{array}{c}\mbox{Geometrical}\\
  \mbox{variables}\end{array}$ &$\begin{array}{c}(\vect g,\ \vect h)\\
  {\rm Ranks} \end{array}$ & Signature & $\begin{array}{c}\mbox{Contrac-}\\
  \mbox{tion}\end{array}$& Domain 
  \\
\hline
${\frak g}$& {\it G} & $\begin{array}{l}
                                      \vect{g}^{G}=c^{2}dt^2\\
                                      \vect{h}_{G}=- \dl^{ij}\r_i \r_j\\
                                      \Ga_{G\mu\nu}^{\ \la}=0\end{array} $
      & (1, 3)                                &$(+; -, -, -)$ &
$\begin{array}{c}l_r, c_r \to \infty\\
\nu_r \to 0\end{array}$ & arbitrary  
\smallskip \\
${\frak g}$& {\it EG} & $\begin{array}{l}
                                      \vect{g}^{EG}=c^{2}dt^2\\
                                      \vect{h}_{EG}= \dl^{ij}\r_i \r_j\\
                                      \Ga_{G\mu\nu}^{\ \la}=0\end{array} $
      & (1, 3)                                &$(+; +, +, +)$ &
$\begin{array}{c}l_r, c_r \to \infty\\
\nu_r \to 0\end{array}$ & arbitrary 
\smallskip \\
${\frak c}$& {\it C} & $\begin{array}{l}
                                      \vect{g}^{C}=-d\vect{x}\cdot d\vect{x}\\
                                      \vect{h}_{C}=\r_{ct}^2\\
                                      \Ga_{C\mu\nu}^{\ \la}=0\end{array} $& (3, 1)&$
                                      (-,-,-;+)$&
$\begin{array}{c}l_r \to \infty\\
c_r\to 0\end{array}$& arbitrary 
\smallskip  \\
${\frak c}$& {\it EC} & $\begin{array}{l}
                                      \vect{g}^{EC}=d\vect{x}\cdot d\vect{x}\\
                                      \vect{h}_{EC}=\r_{ct}^2\\
                                      \Ga_{C\mu\nu}^{\ \la}=0\end{array} $& (3, 1)&$
                                      (+,+,+;+)$&
$\begin{array}{c}l_r \to \infty\\
c_r\to 0\end{array}$& arbitrary 
\smallskip  \\
${\frak c}_2$& $C_2$ & $\begin{array}{l}
\vect{g}^{C_2}=- d(\vect{x}/\nu t)\cdot d (\vect{x}/\nu t)\, \footnote{An overall minus
in both degenerate covariant metric and degenerate contravariant metric has been ignored.}\\
\vect{h}_{C_2}= l^{-2}\nu^2t^2x^\mu x^\nu \r_\mu \r_\nu\\
\Ga_{C_2 00}^{\ \ 0}=-\d 2 {ct} \\
\Ga_{C_2 j0}^{\ \ i}=\Ga_{C_2 0j}^{\ \ i}=-\d 1 {ct}\dl^i_j
  \end{array}$                            & (3, 1) &$(- ,- , - ; +)$&
$\begin{array}{l}l_r \to 0\\ c_r \to \infty \end{array}$& $t^2>0$
\smallskip \\
${\frak c}_2$& $EC_2$ & $\begin{array}{l}
\vect{g}^{EC_2}=d(\vect{x}/(\nu t)) \cdot d (\vect{x}/(\nu t))\\
\vect{h}_{EC_2}=l^{-4}c^2t^2x^\mu x^\nu \r_\mu \r_\nu\\
{\Ga_{EC_2}}_{00}^{0}=-\d 2 {ct} \\
{\Ga_{EC_2}}_{j0}^{i}={\Ga_{EC_2}}_{0j}^{i}=-\d 1 {ct}\dl^i_j
  \end{array}$                            & (3, 1) &$(+,+,+;+)$&
$\begin{array}{l}l_r \to 0\\ c_r \to \infty \end{array}$& $t^2>0$
\smallskip \\
${\frak g}_2$& $EG_{2} $ & $\begin{array}{l}
\vect{g}^{EG_{2}}=l^2\d {(\vect x\cdot \vect{dx})^2-(\vect x
   \cdot \vect x)(\vect{dx} \cdot \vect{dx})}
   {(\vect x\cdot \vect x)^2}       \\
\vect{h}_{EG_{2}}= - l^{-4}\vect{x}\cdot\vect{x}(x^\mu \r_\mu)^2\\
\mbox{Free parameter: }  lx^0/\sqrt{\vect{x}\cdot\vect{x}} \\
{\Ga_{G_{2}}}^{0}_{0i}={\Ga_{G_{2}}}^{0}_{i0}
=-\d{ x^i}{\vect x\cdot \vect x} \\ {\Ga_{G_{2}}}^{l}_{ij}=-\d{\dl^l_i x^j +
\dl^l_j x^i}{\vect x\cdot \vect x}\end{array} $& (2, 1) &
                                       $(- ,-; - )$ &
$\begin{array}{l}l_r, c_r \to 0\\ \nu_r \to \infty \end{array}$&$|\vect{x}|>0$
\smallskip \\
${\frak g}_2$& $G_2$ & $\begin{array}{l}
\vect{g}^{G_2}= l^2\d {(\vect x
\cdot \vect x)(\vect{dx} \cdot \vect{dx})-(\vect x\cdot \vect{dx})^2}
{(\vect x\cdot \vect x)^2}\\
 \vect{h}_{G_2}= -l^{-4}\vect{x}\cdot\vect{x}(x^\mu \r_\mu)^2\\
 \mbox{Free parameter: } l x^0/\sqrt{\vect{x}\cdot\vect{x}} \\
{\Ga_{G_2}}^{0}_{0i}={\Ga_{G_2}}^{0}_{i0}
=-\d{ x^i}{\vect x\cdot \vect x} \\ {\Ga_{G_2}}^{l}_{ij}=-\d{\dl^l_i x^j +
\dl^l_j x^i}{\vect x\cdot \vect x}\end{array} $& (2, 1) &
                                       $(+,+; -)$ &
$\begin{array}{l}l_r, c_r \to 0\\ \nu_r \to \infty \end{array}$&$|\vect{x}|>0$
\smallskip \\
\bottomrule[2pt]
\end{tabular}
\end{table}

\begin{table}[thb]
\setcounter{table}{1}
\caption{Algebras and their corresponding geometries (Cont.).}
\begin{tabular}{ccccccc}
\toprule[2pt]
  Alg & $\begin{array}{c} \mbox{Geom.}\\
   \mbox{name}\end{array}$   & $\begin{array}{c}\mbox{Geometrical}\\
  \mbox{variables}\end{array}$ &$\begin{array}{c}(\vect g,\ \vect h)\\
  {\rm Ranks} \end{array}$ & Signature & $\begin{array}{c}\mbox{Contrac-}\\
  \mbox{tion}\end{array}$& Domain 
  \\
\hline
${\frak g}'$& $G'$ & $\begin{array}{l}
             \vect{g}^{G'}=- l^2(d\frac 1 {\nu t})^{2} \\
             \vect{h}_{G'}= (\nu t)^{-2}\dl^{ij}\r_i \r_j\\
             \Ga^{\ \ 0}_{{\frak g}'00}=-\dfrac 2 {ct} \\
             \Ga^{\ \ i}_{{\frak g}'j0}
             =\Ga^{\ \ i}_{{\frak g}'0j}=-\dfrac 1 {ct}\dl^i_j
                                      \end{array}$& (1, 3) & $ (-; +, + ,+)$&
$\begin{array}{l}c_r, l_r\to \infty, \\ \nu_r \to \infty \end{array}$&
$t^2>0$
\smallskip  \\
${\frak g}'$& $EG'$ & $\begin{array}{l}
             \vect{g}^{EG'}=l^2(d\frac 1 {\nu t})^{2} \\
             \vect{h}_{EG'}=(\nu t)^{-2}\dl^{ij}\r_i \r_j\\
             \Ga^{\ \ 0}_{{\frak g}'00}=-\dfrac 2 {ct} \\
             \Ga^{\ \ i}_{{\frak g}'j0}
             =\Ga^{\ \ i}_{{\frak g}'0j}=-\dfrac 1 {ct}\dl^i_j
                                      \end{array}$& (1, 3) & $
                                      (+; +, +,+)$&
$\begin{array}{l}c_r, l_r\to \infty, \\ \nu_r \to \infty \end{array}$&
$t^2>0$ 
\smallskip  \\
${\frak g}'_2$& $G_2'$ & $\begin{array}{l}
 \vect{g}^{G_2'}= l^2 \d {(\vect x\cdot \vect{dx})^2-(\vect x \cdot \vect x)
 (\vect{dx} \cdot \vect{dx})}{(\vect x\cdot \vect x)^2} \\
\vect{h}_{G_2'}=c^{-2}l^{-2}(\vect x\cdot \vect x) (\r_t)^2\\
\mbox{Free parameter: } l^2(\vect{x}\cdot\vect{x})^{-1/2}\\
\Ga^{\ \ 0}_{G_2' 0i}=\Ga^{\ \ 0}_{G_2' i0}=-\dfrac {x^i} {\vect x\cdot \vect x} \\
             \Ga^{\ \ l}_{G_2'ij}
             =-\dfrac {\dl^l_i x^j+\dl^l_j x^i} {\vect x\cdot \vect x}
                                      \end{array} $& (2, 1) &$(-,-; +)$&
$\begin{array}{l}c_r, l_r\to 0, \\ \nu_r \to 0 \end{array}$& $|\vect{x}|>0$
\\
${\frak g}'_2$& $EG_2'$  & $\begin{array}{l}
 \vect{g}^{EG_2'}=l^2\d {(\vect x \cdot \vect x)(\vect{dx} \cdot \vect{dx})
 -(\vect x\cdot \vect{dx})^2} {(\vect x\cdot \vect x)^2} \\
\vect{h}_{EG_2'}=c^{-2}l^{-2}(\vect x\cdot \vect x) (\r_t)^2\\
\mbox{Free parameter: } l^2(\vect{x}\cdot\vect{x})^{-1/2}\\
\Ga^{\ \ 0}_{EG_2' 0i}=\Ga^{\ \ 0}_{EG_2' i0}=-\dfrac {x^i} {\vect x\cdot \vect x} \\
             \Ga^{\ \ l}_{G_2'ij}
             =-\dfrac {\dl^l_i x^j+\dl^l_j x^i} {\vect x\cdot \vect x}
                                      \end{array} $& (2, 1) &$(+,+; +)$&
$\begin{array}{l}c_r, l_r\to 0, \\ \nu_r \to 0 \end{array}$&$|\vect{x}|>0$
\smallskip  \\
\bottomrule[2pt]
\end{tabular}
\end{table}

\subsection{Geometries for $\frak{g}'$ and $\frak{g}'_2$}

The inequalities $\si_r^+ >0$, $\si_r^-<0$, and
$\si_{E,r}^->0$ do not keep valid in the limiting process of
$l_r,\, c_r,\, \nu_r\to
\infty$, and the inequalities $\si_r^+ <0$, $\si_r^->0$, and
$\si_{E,r}^->0$  do not keep valid in the
limiting process of $l_r,\, c_r,\, \nu_r\to
0$.  Therefore, \dS, $DTdS$, and \Lob\ geometries and
$BdSL$, \AdS, and \Lob\ geometries cannot define new geometries by the contraction approach
in the two ways, respectively.

The inequalities $\si_{E,r}^+>0$, $\si_{E,r}^-<0$,
$\si_r^+ <0$, and $\si_r^->0$ in the limit of $l_r,\, c_r,\, \nu_r\to \infty$ require
$\nu_r^2 t^2(1\pm \frac {1}{\nu_r^2 t^2}) >0$.  It means that
the hypersurface at $t=0$ should be removed from the manifold.
Without loss of generality, $c_r/l_r^2$ is supposed to be fixed when
$c_r,\, l_r \to \infty$ in these cases.  Then,
\Riem, $LBdS$, $BdSL$,
and \AdS\ geometries contract to, respectively,
\be
\begin{array}{l}
ds_{\substack{EG'\\ G'}}^{2}=\displaystyle
\lim_{\substack{l_r, c_r\to \infty \\ c_r/l_r^2 =c/l^2}}
\d {\pm 1} {\si_{E,r}^\pm}\left ( \dl_{\mu \nu} \mp \d {\dl_{\mu\ka}
\dl_{\nu \la}x_r^\ka x_r^\la}
{l_r^2\si_{E,r}^\pm}\right ) dx_r^\mu dx_r^\nu = \pm l^2 [d(\d 1{\nu t})]^2=:
\pm {\vect g}^{EG'}\\
ds_{\substack{EG'\\ G'}}^{2}=\displaystyle \lim_{\substack{l_r, c_r\to \infty \\ c_r/l_r^2 =c/l^2}}
\d 1 {\si_{r}^\pm}\left ( \eta_{\mu \nu} \pm \d {\eta_{\mu\ka}
\eta_{\nu \la}x_r^\ka x_r^\la}
{l_r^2\si_{r}^\pm}\right ) dx_r^\mu dx_r^\nu= l^2 [d(\d 1{\nu t})]^2= {\vect g}^{G\,'}
\end{array}
\ee
\be
\begin{array}{l}
\left (\d {\r}{\r s}\right )_{\substack{EG'\\ G'}}^{2}=
\displaystyle \lim_{\substack{l_r, c_r\to \infty \\ c_r/l_r^2 =c/l^2}}\d {\nu^2}{\nu_r^2}
(\pm \si_{E,\,r}^\pm)
(\dl^{\mu\nu} \pm l_r^{-2}x_r^\mu x_r^\nu) \d{\r }{\r x_r^\mu}\otimes\d{\r }{\r x_r^\nu}
=
\dl^{ij}\d{\r }{\r (x^i/\nu t)}\otimes\d{\r }{\r (x^j/\nu t)}=:{\vect h}_{EG'}\\
\left (\d {\r}{\r s}\right )_{\substack{EG'\\ G'}}^{2}=
\displaystyle \lim_{\substack{l_r, c_r\to \infty \\ c_r/l_r^2 =c/l^2}}\d {\nu^2}{\nu_r^2}\si_{r}^\pm
(\eta^{\mu\nu} \mp l_r^{-2}x_r^\mu x_r^\nu) \d{\r }{\r x_r^\mu}\otimes\d{\r }{\r x_r^\nu}=
\pm \dl^{ij}\d{\r }{\r (x^i/\nu t)}\otimes\d{\r }{\r (x^j/\nu t)}=\pm{\vect h}_{EG'}
\end{array}
\ee
\be
\begin{array}{l}
{\Ga_{\substack{EG'\\ G'}}}^\la_{\mu\nu}=\mp\displaystyle \lim_{\substack{l_r, c_r\to \infty \\ c_r/l_r^2 =c/l^2}}\d{c_r}{c}
\d{(\dl^\la_\mu \dl_{\nu \ka} + \dl^\la_\nu
\dl_{\mu \ka})x_r^\ka}{l_r^2\si_{E,\,r}^\pm}=-\d{\dl^\la_\mu \dl_{\nu 0} + \dl^\la_\nu
\dl_{\mu 0}}{ct}\\
{\Ga_{\substack{EG'\\ G'}}}^\la_{\mu\nu}=\pm\displaystyle \lim_{\substack{l_r, c_r\to \infty \\ c_r/l_r^2 =c/l^2}}\d{c_r}{c}
\d{(\dl^\la_\mu \eta_{\nu \ka} + \dl^\la_\nu
\eta_{\mu \ka})x^\ka}{l_r^2\si_{r}^\pm}=- \d{\dl^\la_\mu \eta_{\nu 0} + \dl^\la_\nu
\eta_{\mu 0}}{ct}.\end{array}
\ee
The contraction of $LBdS$ and \AdS\ geometries is the para-Galilei geometry, denoted
by $G'$.  The contraction of \Riem\ and $BdSL$ geometries is the Euclidean version of
$G'$ geometry, denoted by $EG'$.  They can also be obtained by the contraction from $P'$
and $EP'$ geometries in the limit of $l_r \to \infty$, respectively.  The curvature of
the $G'$ and $EG'$ geometries are zero.

\begin{figure}[b]
  \includegraphics[width=160mm,height=120mm]{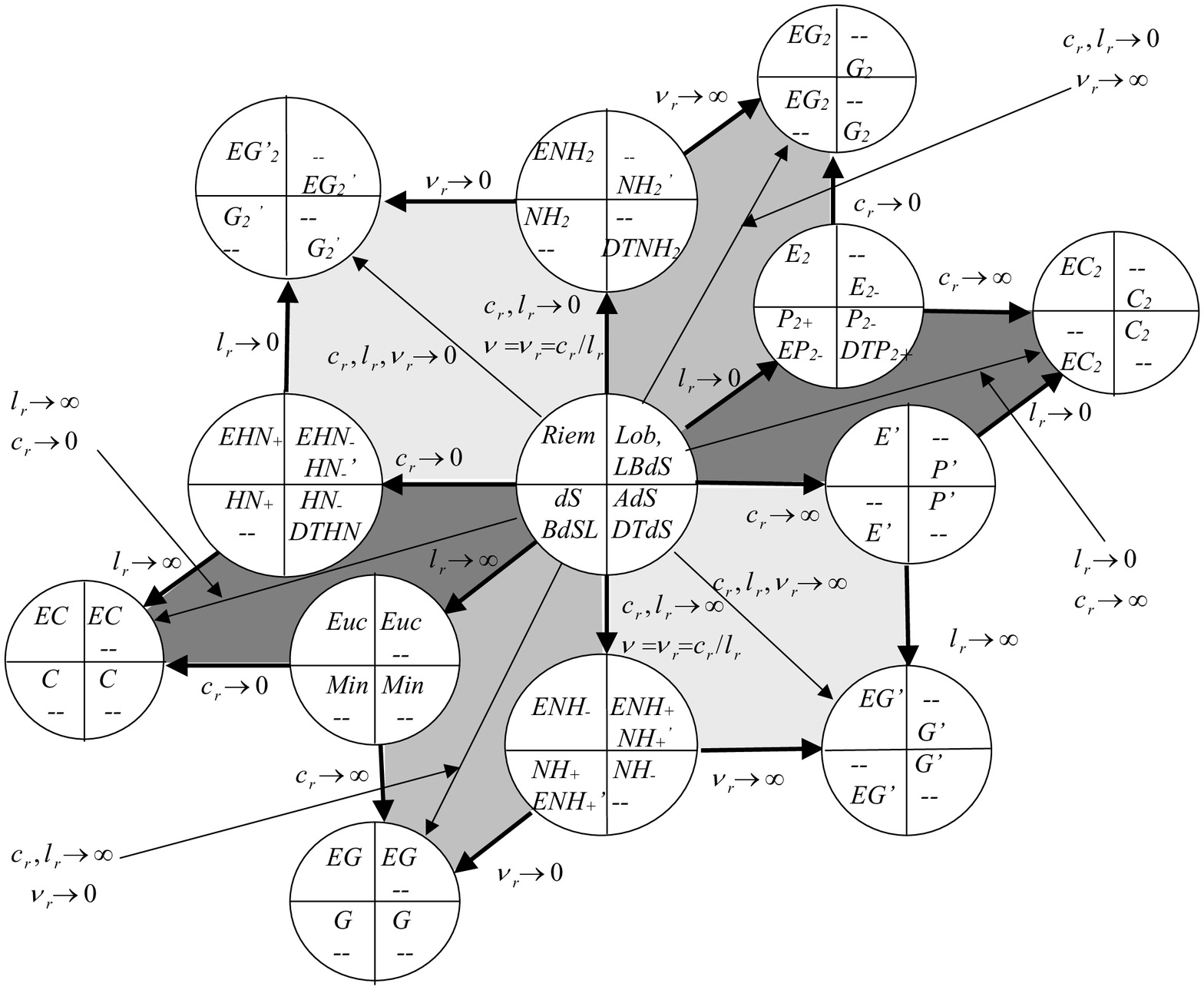}\\
 \caption{Contraction scheme for the geometries for the possible kinematics.
} \label{Fig:gr}
\end{figure}

\begin{figure}[t]
  \includegraphics[width=160mm,height=120mm]{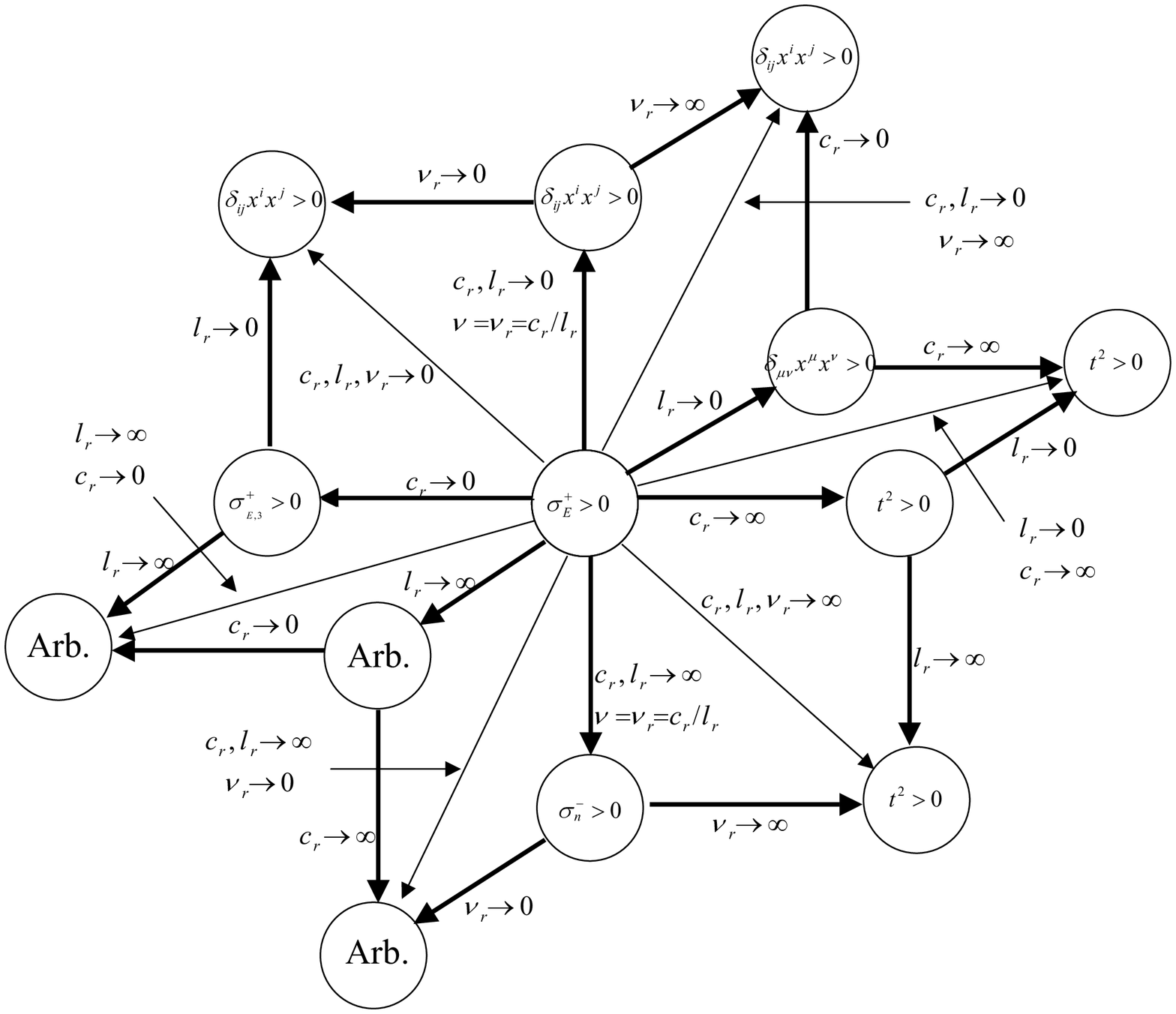}\\
 \caption{Domain condition $\si_E^+>0$ and its contraction.
 The inequalities $\si_E^+>0$, $\si_{E,3}^+>0$ and $\si_{\frak n}^-$
 do not give any constraints on coordinates.
 } \label{Fig:D1}
\end{figure}

On the other hand, the inequalities $\si_{E,r}^+>0$, $\si_{E,r}^-<0$,
$\si_r^+ >0$, and $\si_r^-<0$ in the limit of $l_r,\, c_r,\, \nu_r\to 0$ require
$\dl_{ij}x^i x^j>0$.  It means that
the spatial point $x^i=0\ (i=1, 2, 3)$ should be removed from the manifold
as for $G'_2$ manifold.  Again,
 $c_r/l_r^2$ is supposed to be fixed without loss of generality, when
$c_r,\, l_r \to 0$.  Then,
\Riem, $LBdS$, \dS\ and $DTdS$,
geometries contract to, respectively,
\be
\begin{array}{l}\displaystyle \lim_{\substack{l_r, c_r\to 0 \\ c_r/l_r^2 =c/l^2}}
\d {l^2}{l_r^2}\d {\pm 1} {\si_{E,r}^\pm}\left ( \dl_{\mu \nu} \mp \d {\dl_{\mu\ka}
\dl_{\nu \la}x_r^\ka x_r^\la}
{l_r^2\si_{E,r}^\pm}\right ) dx_r^\mu dx_r^\nu ={l^2}\d
{(\dl_{ij}\dl_{kl}-\dl_{ik}\dl_{jl})x^k x^l} {(\dl_{mn}x^m x^n)^2} dx^i dx^j
=:-{\vect g}^{G'_2}=-{\vect g}^{NH_{2}}
\\
\displaystyle \lim_{\substack{l_r, c_r\to 0 \\ c_r/l_r^2 =c/l^2}}
\d {l^2}{l_r^2}\d {\pm1} {\si_{r}^\pm}\left ( \eta_{\mu \nu} \pm \d {\eta_{\mu\ka}
\eta_{\nu \la}x_r^\ka x_r^\la}
{l_r^2\si_{r}^\pm}\right ) dx_r^\mu dx_r^\nu=  l^2\d
{(\dl_{ik}\dl_{jl}-\dl_{ij}\dl_{kl})x^k x^l} {(\dl_{mn}x^m x^n)^2} dx^i dx^j
= {\vect g}^{G_2'} = {\vect g}^{NH_2}
\end{array}
\ee
\be
\begin{array}{l}\displaystyle \lim_{\substack{l_r, c_r\to 0 \\ c_r/l_r^2
=c/l^2}}\d {c_r^2l_r^2}{c^2l^2}
(\pm \si_{E,\,r}^\pm)
(\dl^{\mu\nu} \pm l_r^{-2}x_r^\mu x_r^\nu) \d{\r }{\r x_r^\mu}\otimes\d{\r }{\r x_r^\nu}
=\d{\dl_{ij}x^i x^j}{c^2l^2}
\d{\r }{\r t}\otimes\d{\r }{\r t}=:{\vect h}_{G'_2}\\
\displaystyle \lim_{\substack{l_r, c_r\to \infty \\ c_r/l_r^2 =c/l^2}}
\d {c_r^2 l_r^2}{c^2 l^2}(\pm\si_{r}^\pm)
(\eta^{\mu\nu} \mp l_r^{-2}x_r^\mu x_r^\nu) \d{\r }{\r x_r^\mu}\otimes\d{\r }{\r x_r^\nu}=
\d {\dl^{ij}x^ix^j}{c^2l^2}\d{\r }{\r t}\otimes\d{\r }{\r t}={\vect h}_{G'_2}
\end{array}
\ee
\be
\begin{array}{l}
\mp\displaystyle \lim_{\substack{l_r, c_r\to 0 \\ c_r/l_r^2 =c/l^2}}
\d{(\dl^\la_\mu \dl_{\nu \ka} + \dl^\la_\nu
\dl_{\mu \ka})x_r^\ka}{l_r^2\si_{E,\,r}^\pm}
=-\d{(\dl^\la_\mu \dl_{\nu k} + \dl^\la_\nu
\dl_{\mu k})x^k}{\dl_{mn}x^m x^n}=:\Ga^{\ \ \la}_{G_2'\mu\nu}\\
\pm\displaystyle \lim_{\substack{l_r, c_r\to \infty \\ c_r/l_r^2 =c/l^2}}
\d{(\dl^\la_\mu \eta_{\nu \ka} + \dl^\la_\nu
\eta_{\mu \ka})x^\ka}{l_r^2\si_{r}^\pm}= \d{(\dl^\la_\mu \eta_{\nu k} + \dl^\la_\nu
\eta_{\mu k})x^k}{\dl_{mn}x^m x^n}=\Ga^{\ \ \la}_{G_2'\mu\nu}.\end{array}
\ee
The curvature is the same as that for $NH_2$ geometry.
The sum of ranks of $\vect{g}^{G_2'}$ and
$\vect{h}_{G_2'}$ is again only 3 and they have the signature $(-, -)$ and $(+)$,
respectively.  Thus, the contractions of \Riem, $LBdS$, \dS, and $DTdS$ geometries
in this limit have the signatures
$(+,+;+)$, $(+,+;+)$, $(-,-;+)$, $(-,-;+)$, respectively.  They have the
free parameter $l^2/r=l^2(\dl_{ij}x^i x^j)^{-1/2}$.

All these results can be easily obtained from the contraction of $NH_2$ geometries.

\begin{figure}[t]
  \includegraphics[width=160mm,height=120mm]{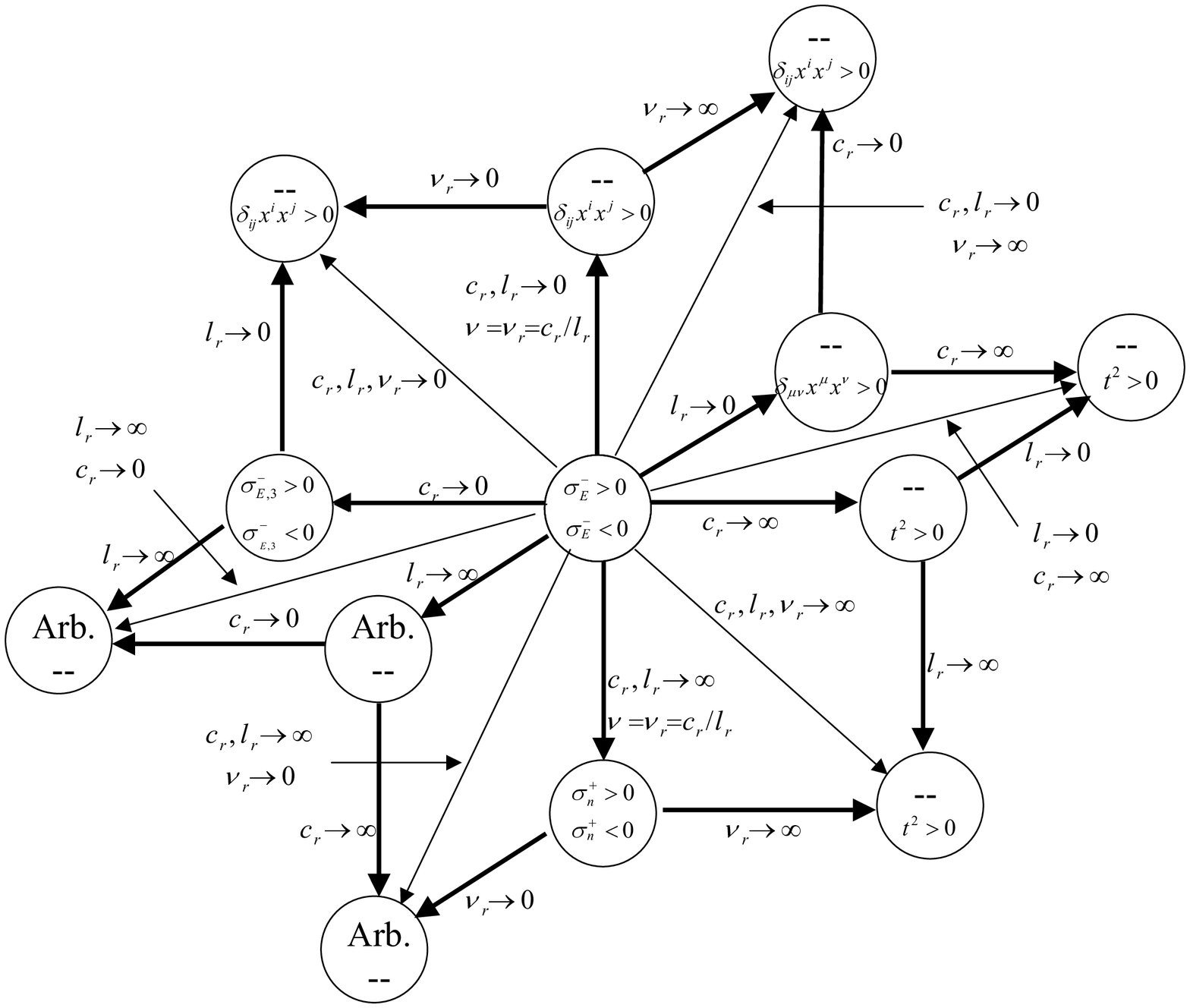}\\
 \caption{Domain condition $\si_E^- \gtrless 0$ and its contraction.
 } \label{Fig:D2}
\end{figure}
\begin{figure}[t]
  \includegraphics[width=160mm,height=120mm]{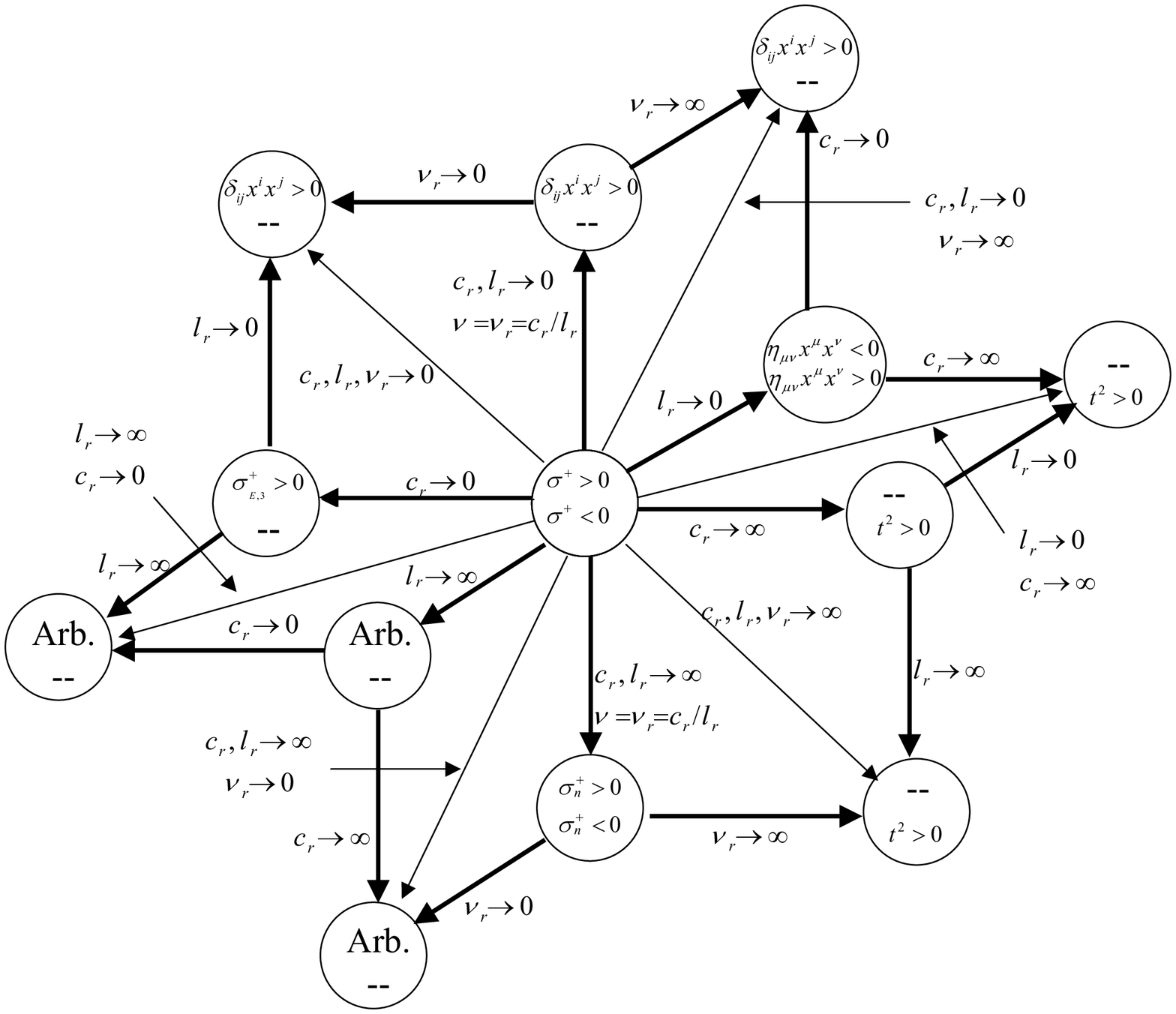}\\
 \caption{Domain condition $\si^+\gtrless 0$ and its contraction.
 } \label{Fig:D3}
\end{figure}
\begin{figure}[t]
  \includegraphics[width=160mm,height=120mm]{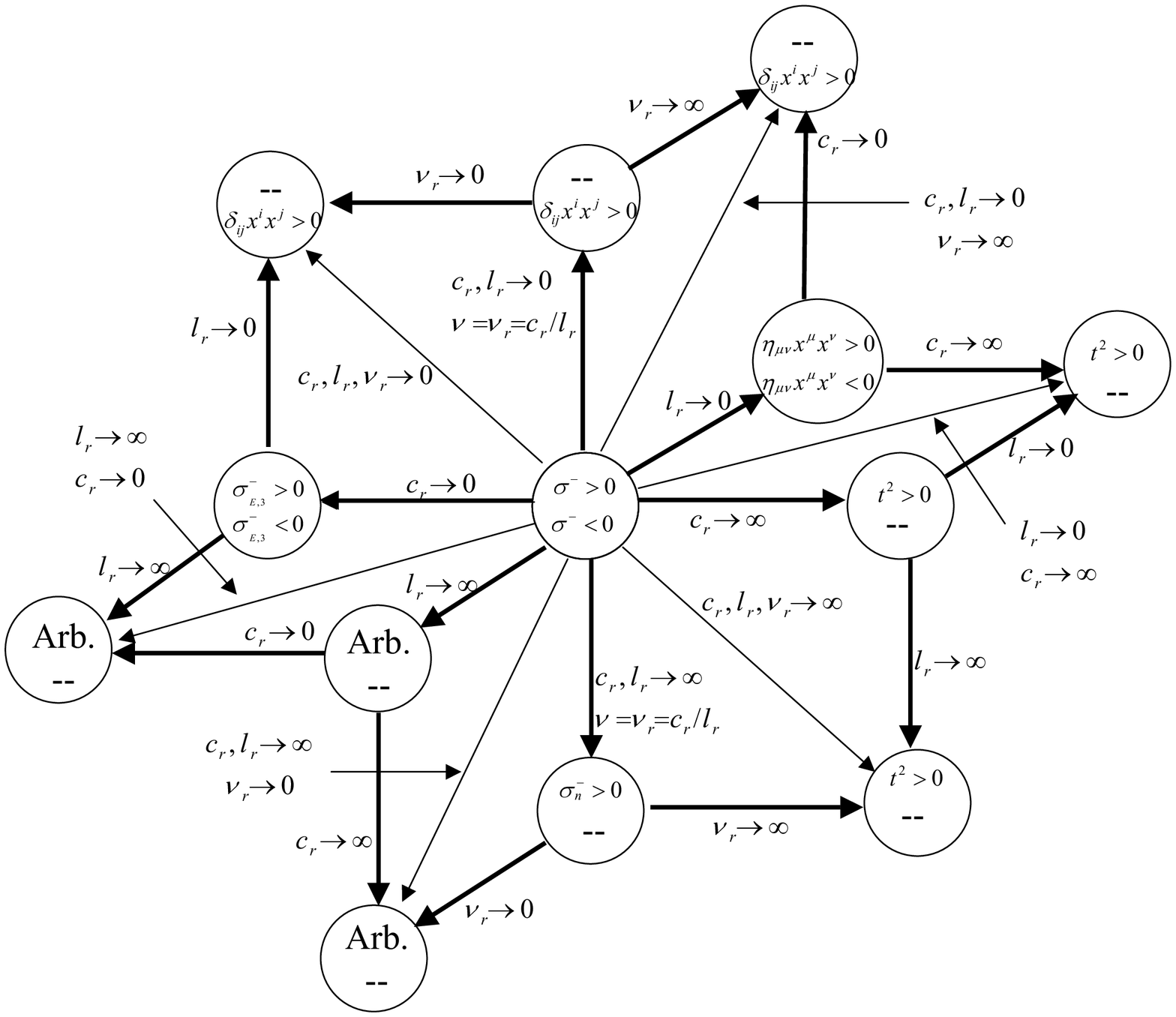}\\
 \caption{Domain condition $\si^-\gtrless 0$ and its contraction.
 } \label{Fig:D4}
\end{figure}

\begin{table}[tb]\label{Tab:pairs}
\setcounter{table}{2}
\caption{Duality of the present and the time infinity}
\begin{tabular}{ccc||ccc}
  \toprule[2pt]
 $t,\ x^i$&  $\Leftrightarrow $\quad &  $1/\nu^2 t,\  x^i/\nu t$
  & $t,\ x^i$&  $\Leftrightarrow $ &  $1/\nu^2 t,\  x^i/\nu t$ \\
  \hline
  $Min$   &  $\Leftrightarrow $ &   $P'$
&  $Euc$ &   $\Leftrightarrow $ &  $E'$  \\
$G$ &       $\Leftrightarrow$ & $G'$
& $EG$ & $\Leftrightarrow$ & $EG'$\\
 $C$   &  $\Leftrightarrow$ &  $C_2$ &
$EC$&     $\Leftrightarrow$ &  $EC_2$\\
$G_2$   & $\Leftrightarrow$ & $G_2'$
 & $EG_{2}$   & $\Leftrightarrow$ &
 $EG_2'$ \\
$HN_+$  & $\Leftrightarrow$ &  $E_{2-}$  &
$EHN_+$   & $\Leftrightarrow$ &
$E_2$   \\
$HN_-$ & $\Leftrightarrow$ &  $P_{2-}$  &
$EHN_-$ & $\Leftrightarrow$ &
$EP_{2-}$ \\
$HN_-'$ & $\Leftrightarrow$ &  $P_{2+}$
   & $DTHN_-$ & $\Leftrightarrow$ &$DTP_{2+}$\\
 $NH_+$\omits{\quad($|\nu t| < 1$)} & $\Leftrightarrow$ &
$NH_+'$\omits{\quad($|\nu t| >1$)}&
$ENH_+$\omits{\ ($|\nu t| < 1$)} & $\Leftrightarrow$ &
$ENH_+'$\omits{\ ($|\nu t| >1$)} \\
\omits{ $NH_+$\quad ($|\nu t| < 1$) & $\Leftrightarrow$ &
$ENH_+$\;($|\nu t| < 1$) &
$NH_+'$\quad ($|\nu t| > 1$) & $\Leftrightarrow$ &
$ENH_+'$\;($|\nu t| > 1$) \\ }
 $NH_2$  & $\Leftrightarrow$ &
 $NH_2'$ & & &  \\
\hline
 $NH_-$\omits{\quad ($|\nu t| \lessgtr 1$)\  }& $\Leftrightarrow $ &   $NH_-$
\omits{  ($|\nu t| \gtrless1$)} &
{\it $ENH_-$}\omits{\ ($|\nu t| \lessgtr 1$) } & $\Leftrightarrow$ &
 $ENH_-$\omits{ \ ($|\nu t| \gtrless 1$)}  \\
$ENH_2$  & $\Leftrightarrow$ &
  $ENH_2$ &
 $DTNH_2$  & $\Leftrightarrow$ & $DTNH_2$ \\
\bottomrule[2pt]
\end{tabular}
\end{table}

\subsection{Summary}

The algebras and their corresponding 45 geometries are listed in TABLE II.
All of them have the same vanishing Weyl projective curvature tensor \cite{Gibbons}
\be
W^\la_{\ \mu \si \nu} := R^\la _{\ \mu \si \nu}+\d 1 3 (\dl^\la_\si R_{\mu\nu}
-\dl^\la_{\nu}R_{\mu \si})
\ee
are projective equivalent to each other.  In fact, each geometry is defined
on a portion of a 4d real projective manifold.

From the viewpoint of differential
geometry, these geometries are not all independent.  For example, $dS$ and $LBdS$
describe the same space-time in different coordinate systems.  Therefore, the
number of independent geometries will be less than 45.

\section{Relations among space-times}

In the previous section, the geometries for all possible kinematics except static ones
are presented
by contraction procedure.  The contraction scheme for these geometries is shown
in FIG \ref{Fig:gr}.  The domains for the geometries are shown in
FIG. \ref{Fig:D1}--FIG. \ref{Fig:D4}.

In addition to the contraction relation, there exist more relations
among the geometries.  In this section, we shall discuss these relations.

\subsection{Duality between the present and the time infinity}

The geometries for the possible kinematics almost appear in pairs.  Each pair
are invariant under the transformation
\be \label{tdual}
\frac 1 {\nu^2 t} \Leftrightarrow t, \quad \frac {x^i} {\nu t} \Leftrightarrow x^i.
\ee
It may be interpreted as the duality between the ``present time"
and ``the time infinity".  For example, ${\frak p}'$ is isomorphic
to the ordinary Poincar\'e algebra but is regarded as a physically different
one in \cite{BLL}.  The analysis in the previous section shows that the
two geometries describe the different portions in the $RP^4$ manifold and
have the time duality.  The duality relation among the geometries for the
possible kinematics are summarized in TABLE III.  It is remarkable
that the last 4 geometries in the TABLE are all self dual.  In particular,
the $NH_-$ geometry is one of them.  Thus, the physics in
the $NH_-$ space-time should share the same property.

It should be noted that in the viewpoint of
differential geometry, the Minkowski space-time and $P'$ geometry actually describe
the same space-time because they have the same metric tensor and topology.  If the
transformation Eq.(\ref{tdual}) is considered as a coordinate transformation, the
independent geometries remain 26.  They are \Riem, \Lob, \dS, \AdS, $DTdS$,
$Euc$, $Min$, $EG$, $G$, $EC$, $C$, $ENH_\pm$, $NH_\pm$, $E_2$, $EP_{2-}$,
$P_{2\pm}$, $E_{2-}$, $DTP_{2+}$, $ENH_2$, $NH_2$, $DTNH_2$, $EG_{2}$, and
$G_2$.

\subsection{Time geometry versus space geometry}

The forms of the degenerate metrics present a striking contrast between
$NH_\pm$ ($ENH_\pm$) geometries and $HN_\pm$ ($EHN_\pm$) geometries
in addition to the algebra consideration in \cite{GHWZ}.  The
TABLE IV shows the contrast when the ``metrics" are expressed in terms of
`inertial' coordinates.

Similarly, $G$ ($EG$) geometry and $C$ ($EC$) geometry also present a contrast
between the time geometry and space geometry though they are both flat.

\begin{table}[h]
\caption{Time geometry versus space geometry}
\begin{tabular}{lccc}
  \toprule[2pt]
      & $\begin{array}{c}\mbox{Covariant}\\
      \mbox{degenerate metric}\end{array}$ &  $\begin{array}{c}\mbox{Contravariant}\\
      \mbox{degenerate metric}\end{array}$ & $\begin{array}{c}\mbox{Conformal factor}\\
      (C^{-2})\end{array}$ \\
  \hline
$\begin{array}{c}NH_\pm\\
(ENH_\pm)\end{array}$ & $\begin{array}{c}\mbox{Beltrami model for 1d time}
\\ \vect{g}
=\d 1 {\si_{\frak n}^\pm}(1\pm\d{c^2t^2}{l^2\si_{\frak n}^\pm})c^2dt^2\end{array}$ &
$\begin{array}{c}\mbox{Conformal to}\\
\mbox{3d flat space}\end{array}$ & $\si_{\frak n}^\pm =1 \mp \nu^2t^2$  \\
$\begin{array}{c}HN_\pm\\
(EHN_\pm)\end{array}$& $\begin{array}{c}\mbox{Beltrami model for 3d space}\\
\vect{g}=-\d 1 {\si_{E,\,3}^\pm}
\left(\dl_{ij}\mp \d{\dl_{ik}\dl_{jl}x^kx^l}{l^2\si_{E,\,3}^\pm}\right)dx^idx^j
\end{array} $ &  $\begin{array}{c}\mbox{Conformal to} \\
\mbox{1d flat time}\end{array}$ & $\si_{E,\,3}^\pm(x^i)=1\pm l^{-2}
\dl_{ij}x^i x^j$\\
\hline
$G$ ($EG$) & 1d flat time & 3d flat space & 1\\
$C$ ($EC$) & 3d flat space & 1d flat time & 1 \\
  \bottomrule[2pt]
\end{tabular}
\end{table}

\subsection{Relation among $(E)G_2$, $(E)NH_2$, and $(E)G_2'$ geometries}

The algebras $\frak{n}_{+2}$, $\frak{g}_2$ and $\frak{g}'_2$ share the same
generators ${\vect P}'$, ${\vect K}^c$,
and ${\vect J}$.  Their only difference is at the ``time translation" generator.
The sum of the `time translation' generators for $\frak{g}_2$
and for $\frak{g}'_2$ gives rise to the ``time translation"
generator --- Beltrami time translation --- for $\frak{n}_{+2}$.

Their corresponding geometries have the same covariant degenerate metric
\be
\omits{{\vect g}_{NH_2}^{}=-{\vect g}_{NH_2'}^{}=}\pm l^2\d {(\vect{x}\cdot\vect{dx})^2
 -(\vect{x}\cdot\vect{x})(\vect{dx}\cdot\vect{dx})}{(\vect{x}\cdot \vect{x}
 )^2}, \nno
\ee
which is the metric of 2d sphere, and the same connection and curvature tensors.
Their contravariant degenerate
metrics also possess the simple additivity\omits{ if the free parameters are ignored}.
Namely, the algebraic sum of contravariant
degenerate metrics of $G_2$ (or $EG_{2}$) and $G'_2$ (or $EG'_2$) gives the
contravariant degenerate metrics of $NH_2$ (or $ENH_2$).

\begin{table}[th]
\caption{Contravariant degenerate metrics of $NH_{2}$, $G_2$ and $G_2'$ geometries}
\begin{tabular}{ll||ll}
\toprule[2pt]
&  \qquad${\vect h}$ &  &\qquad${\vect h}$\\
\hline
$G_2$ & \qquad$ l^{-4}(\vect{x}\cdot\vect{x})
\left (x^\mu\r_\mu\right)^2$&$EG_{2}$ & \qquad$ l^{-4}(\vect{x}\cdot\vect{x})
\left (x^\mu\r_\mu\right)^2$ \\
$NH_{2}$ & \qquad  $l^{-4}(\vect{x}\cdot\vect{x})
\left [\nu^{-2}\r_t^2 - \left (x^\mu\r_\mu \right)^2\right ]$
&$ENH_{2}$ & \qquad  $l^{-4}(\vect{x}\cdot\vect{x})
\left [\nu^{-2}\r_t^2 + \left (x^\mu\r_\mu \right)^2\right ]$\\
$G'_2$  &  \qquad $l^{-4} (\vect{x}\cdot\vect{x}) (\nu^{-2}
\r_t^2) $&$EG'_2$  &  \qquad $l^{-4} (\vect{x}\cdot\vect{x}) (\nu^{-2}
\r_t^2) $\\
\bottomrule[2pt]
\end{tabular}
\end{table}

\subsection{Geometries with spatial isotropy and Lorentz-like signature}

Although all the 22 algebras possess $\frak{so}(3)$ isotropy, the geometries
$DTdS$, $P_{2+}$, $DTP_{2+}$, $ENH_2$, $NH_2$, $DTNH_2$, $G_{2}$, $EG_{2}$,
and their time dualities if exist do not have the spatial isotropy with
respect to any point on the manifolds.  Therefore, they cannot serve as the
geometries for the genuine possible kinematics.

In addition, the geometries for genuine possible kinematics should have the right
signature.  Then, only 9 geometries remains. They are 3 relativistic
geometries, \dS, \AdS, and $Min$, 3 absolute-time geometries,
$NH_\pm$ and $G$, and 3 absolute-space geometries $E_{2-}$, $P_{2-}$ and
$C$.

In order to obtain possible kinematics, Bacry and L\'evy-Leblond require that
the transformations generated by boost in any given direction form a noncompact
subgroup.  However, the requirement cannot guarantee the geometry has suitable
Lorentz-like signature.  For example, the $E'$ geometry is diffeomorphic to
$Euc$ geometry.  On the contrary, even when the transformations generated by
boost form a compact subgroup, geometries still possibly have right
signature.  $E_{2-}$ geometry is one of examples.  Therefore, the
requirement to pick up the possible kinematics should be the right signature.


\section{Concluding remarks}

Except for the static ones, there are 22 different possible kinematical and
geometrical algebras with $\frak{so}$(3) isotropy and 10 parameters.  Their
generators belong to the 4d ``inertial-motion algebra" $\frak{im}(4)$.  Among
these algebras, $\frak r$, $\frak l$ and $\frak d^\pm$ algebras are basic ones.
The generators of others can be obtained by the linear combinations of the
generators of the 4 algebras or by the contraction from the 4 algebras.

The geometries corresponding these algebras are all presented from the
contraction of the Beltrami models of Riemann space, Lobachevsky space,
(anti-)de Sitter space-times, and double time de Sitter space-time,
Lobachevsky-Beltrami model of de Sitter space-time, and \BdS\ model of
Lobachevsky space, in the similar way to obtain the Euclid space, Minkowski
space-time, and (anti-)Newton-Hooke space-times from Riemann space,
Lobachevsky space and $(A)dS$ space-times.  It should be emphasized that the
conditions $\si_E^+ >0$, $\si_E^- \gtrless 0$, $\si^\pm \gtrless0$ are
important in the Beltrami models.  They specify the domains of the geometries.
In the limiting process, they should be always valid.  The requirement implies
that not all geometries are contractible.

The geometries can be classified in several ways.  By the determinant of the
metric, the geometries are classified into two categories.  One is non-degenerate,
and the other is degenerate.  It is well known that for the non-degenerate
geometries the (covariant) metric is enough to determine their local properties.
For degenerate geometries the covariant degenerate metric is not enough to
determine the local properties.  One has to supplement the contravariant
degenerate metric and the connection. Many new geometries belong to the second
category.  $G$, $C$, and $NH_\pm$ space-times are all familiar examples of
4d degenerate space-times.

Each geometry is defined in a suitable portion in the $RP^4$ manifold.  On the
$RP^4$ there exist a set of the coordinate systems, which are called
`inertial' coordinate systems.  For a given ``inertial" coordinate system $x^\mu$,
the geometries fall into 3 categories according to whether $x^0=ct=0$ and
$x^0=ct=\infty$ are in the geometries.  For the first category, $t=0$ belongs to
the geometries, while $t=\infty$ does not belong to the geometries.  $Min$ and
$G$ space-times are the most familiar representatives of the categories.  For
the second one, $t=\infty$ belongs to the geometries while $t=0$ does not.  $P'$
space-times and $G'$ space-times belong to the categories, which  are regarded
as the physically different ones even though their algebras are isomorphic to
$\frak{p}$ and $\frak{g}$, respectively \cite{BLL}.  There exist the
correspondences between the geometries of the first and the second categories.
They are linked by Eq.(\ref{tdual}).  For example, $Min$ and $P'$ space-times, $G$
and $G'$ space-times, $C$ and $C_2$ space-times, $HN_-$ and $P_{2-}$, are linked
together, respectively.   The relation can be interpreted as the time
dualities of the present time and time infinity.  This behavior might be useful
in the study of the space-time structure near the time infinity for asymptotically
flat space-times.  For the third category, both $t=0$ and $t=\infty$ belong to
the geometries.  The $NH_-$ space-time is such a geometry.  It is self dual under
Eq.(\ref{tdual}).  Thus, the physics in the $NH_-$ space-time should share the
same property.  By the way, there is no space duality in these geometries.  The
reason is that the three ``spatial" Beltrami coordinates $x^i$ are required to be
on equal footing so that $\frak{so}(3)$ algebra is preserved.

The geometries can be casted into three categories according to the signature of
metric tensors.  For the non-degenerate cases, the signature of the metric tensor
$\vect{g}$ is well-defined.  The non-degenerate geometries are immediately
classified into Euclidean, Lorentzian, double-time geometries.  For the degenerate
cases, the definition is somewhat obscure.  Now, a 4d geometry is split into one
3d and one 1d geometries, or two 2d geometries, or even one 2d, one 1d geometries
plus one free parameter. Obviously, the signature is meaningless if only the
covariant (or contravariant) degenerate metric of 1d geometry is concerned.
However, since the degenerate geometries are obtained in the contraction approach,
$\vect{g}$ and $\vect{h}$ have the imprints of the non-degenerate progenitor which
has well-defined signature.  Therefore, the signature of a degenerate geometry may
be defined by the imprints of the non-degenerate progenitor in $\vect{g}$ and
$\vect{h}$. In this way, the degenerate geometries can also be classified into
Euclidean, Lorentzian, and double-time geometries.

The aim of the third requirement in \cite{BLL}, the transformations generated by
boost in any given direction form a noncompact subgroup, is to rule out the pure
geometrical kinematics.  However, the geometries shows that the possible kinematics
satisfying the requirement may define a pure geometry (i.e. $E'$ space with
$\frak{e}'\cong \frak{iso}(4)$) on one hand, and that the possible kinematics
violating the requirement may define a Lorentz-like-signature geometries ({\it i.e.}
$E_{2-}$ space with $\frak{e}_2$) on the other.

Similarly, the first requirement in \cite{BLL} is ``space is isotropic
(rotation invariance)".  Unfortunately, the concept of the space has
not been well established in \cite{BLL}.  The rotation invariance is actually
replaced by an $\frak{so}(3)$ algebra, $[\vect{J}, \vect{J}]=\vect{J}$.
Obviously, this condition cannot guarantee that the geometry has spatial isotropic.
In fact, many geometries which are invariant under the transformations generated by
$\frak{so}(3)$ do not have spatial isotropic with respect to any point in the
manifolds.  The $P_{2+}$ is one of examples \cite{Huang}, which is split into
a 3d space-time and one 1d space.  The $NH_2$ geometry is another example, which
is split into 2d space and 2d space-time.

Therefore, the right requirements to pick up the genuine possible kinematics should
be that\\
(1) {\it space is isotropic with respect to any point on the manifold;}\\
(2) {\it parity and time-reversal are automorphisms of the kinematical groups;}\\
(3) {\it the geometry has Lorentz-like signature.}\\
Then, the geometries for genuine possible kinematics are only
3 relativistic geometries, \dS, \AdS, and $Min$;
3 absolute-time geometries, $NH_\pm$, $G$;
3 absolute-space geometries $E_{2-}$, $P_{2-}$, $C$;
and their time dualities, $P'$, $NH_+'$, $G'$, $HN_\pm$ and $C_2$.

In the viewpoint of differential geometry, the Minkowski space-time and
$P'$ geometry actually describe the same space-time because they have the
same metric tensor, the same topology and are diffeomorphic to each other.
The same identification can be made for other pairs of geometries on the
same reason.  Hence, the genuine possible kinematics from the viewpoint of
differential geometry are given in TABLE VI.
\begin{table}[h]
\caption{Geometries for the genuine possible kinematics}
\begin{tabular}{|l|c|c|c|}
\toprule[2pt]
 &  $>0$ &  $=0$ & $<0$ \\
 \toprule[2pt]
Relativistic &  \dS & $Min$  & \AdS\\
\hline
Absolute-time &  $NH_+$ & $G$ &  $NH_-$ \\
\hline
Absolute-space & $E_{2-}$ & $C$& $P_{2-}$ \\
\bottomrule[2pt]
\end{tabular}
\end{table}
Clearly, the geometries in the middle column have vanishing curvature.
\dS\ and \AdS\ have 4d positive and negative curvature. $NH_\pm$ have
conformal flat spaces and 1d ``curved" time in terms of Beltrami time.
$E_{2-}$ and $P_{2-}$ have 3d sphere and 3d hyperboloid space, and
1d time is conformal flat in Beltrami coordinates.

\begin{acknowledgments}\vskip -4mm
We would like to thank late Prof. H.-Y. Guo for helpful discussion.
This work is supported by NSFC under Grant Nos. 10775140, 10975141,
10705048, 10731080,
Knowledge Innovation Funds of CAS (KJCX3-SYW-S03) and the President Fund of
GUCAS.
\end{acknowledgments}

\end{document}